\documentclass[apjl,ip,twocolumn]{aastex61}
\usepackage{amsmath}
\usepackage{comment}
\usepackage{ifthen}
\usepackage[switch]{lineno}
\modulolinenumbers[5]

\newcommand{\forloop}[5][1]%
{%
\setcounter{#2}{#3}%
\ifthenelse{#4}%
	{%
	#5%
	\addtocounter{#2}{#1}%
	\forloop[#1]{#2}{\value{#2}}{#4}{#5}%
	}%
	{%
	}%
}%

\newcommand{\ctbd}[1]{}

\newcommand{\Lc}{Light curve}

\newcommand{\masy}{\ensuremath{\rm mas\,yr^{-1}}}
\newcommand{\kms}{\ensuremath{\rm km\,s^{-1}}}
\newcommand{\ms}{\ensuremath{\rm m\,s^{-1}}}

\newcommand{\gcmc}{\ensuremath{\rm g\,cm^{-3}}}

\newcommand{\feh}{\ensuremath{\rm [Fe/H]}}

\newcommand{\rsun}{\ensuremath{R_\sun}}
\newcommand{\msun}{\ensuremath{M_\sun}}
\newcommand{\lsun}{\ensuremath{L_\sun}}

\newcommand{\rstar}{\ensuremath{R_\star}}
\newcommand{\mstar}{\ensuremath{M_\star}}
\newcommand{\lstar}{\ensuremath{L_\star}}

\newcommand{\teffstar}{\ensuremath{T_{\rm eff\star}}}
\newcommand{\rhostar}{\ensuremath{\rho_\star}}
\newcommand{\loggstar}{\ensuremath{\log{g_{\star}}}}

\newcommand{\rpl}{\ensuremath{R_{p}}}
\newcommand{\mpl}{\ensuremath{M_{p}}}

\newcommand{\rhopl}{\ensuremath{\rho_{p}}}

\newcommand{\arstar}{\ensuremath{a/\rstar}}
\newcommand{\zrstar}{\ensuremath{\zeta/\rstar}}

\newcommand{\rjup}{\ensuremath{R_{\rm J}}}
\newcommand{\mjup}{\ensuremath{M_{\rm J}}}

\newcommand{\loopand}{\ifnum\value{planetcounter}=2 and \else\fi}
\newcommand{\loopcomma}{\ifnum\value{planetcounter}<2 ,\else. \fi}
\newcommand{\loopcommanoperiod}{\ifnum\value{planetcounter}<2 ,\else \space\fi}
\newcommand{\loopcommanospace}{\ifnum\value{planetcounter}<2 ,\else \fi}

\newcommand{\hatcurhtrxxxxxA}{TOI762}                           
\newcommand{\hatcurfieldxxxxxA}{\ensuremath{string}}            
\newcommand{\hatcurCCraxxxxxA}{\ensuremath{11^{\mathrm h}04^{\mathrm m}18.1831{\mathrm s}}}                   
\newcommand{\hatcurCCdecxxxxxA}{\ensuremath{-47{\arcdeg}49{\arcmin}17.0030{\arcsec}}}                 
\newcommand{\hatcurCCmagxxxxxA}{NULL}                           
\newcommand{\hatcurCCtwomassxxxxxA}{2MASS~11041818-4749169}     
\newcommand{\hatcurCCgscxxxxxA}{GSC~NULL}                       
\newcommand{\hatcurCCgaiaxxxxxA}{GAIA~5362352740204840832}      
\newcommand{\hatcurCCgaiadrtwoxxxxxA}{GAIA~DR2~5362352744504000256} 
\newcommand{\hatcurCCtassmvxxxxxA}{\ensuremath{nff\pmnff}}      
\newcommand{\hatcurCCtassmvshortxxxxxA}{\ensuremath{0.0}}       
\newcommand{\hatcurCCtassmBxxxxxA}{\ensuremath{nff\pmnff}}      
\newcommand{\hatcurCCtassmBshortxxxxxA}{\ensuremath{0.0}}       
\newcommand{\hatcurCCtassmIxxxxxA}{\ensuremath{nff\pmnff}}      
\newcommand{\hatcurCCtassmIshortxxxxxA}{\ensuremath{0.0}}       
\newcommand{\hatcurCCtassmgxxxxxA}{\ensuremath{nff\pmnff}}      
\newcommand{\hatcurCCtassmgshortxxxxxA}{\ensuremath{0.0}}       
\newcommand{\hatcurCCtassmrxxxxxA}{\ensuremath{nff\pmnff}}      
\newcommand{\hatcurCCtassmrshortxxxxxA}{\ensuremath{0.0}}       
\newcommand{\hatcurCCtassmixxxxxA}{\ensuremath{nff\pmnff}}      
\newcommand{\hatcurCCtassmishortxxxxxA}{\ensuremath{0.0}}       
\newcommand{\hatcurCCparallaxxxxxxA}{\ensuremath{10.118\pm0.023}} 
\newcommand{\hatcurCCgaiamGxxxxxA}{\ensuremath{14.9297\pm0.0028}} 
\newcommand{\hatcurCCgaiamBPxxxxxA}{\ensuremath{0\pm0}}         
\newcommand{\hatcurCCgaiamRPxxxxxA}{\ensuremath{0\pm0}}         
\newcommand{\hatcurCCtwomassJmagxxxxxA}{\ensuremath{12.099\pm0.080}} 
\newcommand{\hatcurCCtwomassHmagxxxxxA}{\ensuremath{11.444\pm0.077}} 
\newcommand{\hatcurCCtwomassKmagxxxxxA}{\ensuremath{11.187\pm0.082}} 
\newcommand{\hatcurCCcitJmagxxxxxA}{\ensuremath{11.987\pm0.027}} 
\newcommand{\hatcurCCcitHmagxxxxxA}{\ensuremath{11.338\pm0.029}} 
\newcommand{\hatcurCCcitKmagxxxxxA}{\ensuremath{11.108\pm0.028}} 
\newcommand{\hatcurCCbbJmagxxxxxA}{\ensuremath{12.081\pm0.030}} 
\newcommand{\hatcurCCbbHmagxxxxxA}{\ensuremath{11.365\pm0.030}} 
\newcommand{\hatcurCCbbKmagxxxxxA}{\ensuremath{11.128\pm0.028}} 
\newcommand{\hatcurCCesoJmagxxxxxA}{\ensuremath{12.090\pm0.034}} 
\newcommand{\hatcurCCesoHmagxxxxxA}{\ensuremath{11.366\pm0.048}} 
\newcommand{\hatcurCCesoKmagxxxxxA}{\ensuremath{11.124\pm0.029}} 
\newcommand{\hatcurCCesoJHmagxxxxxA}{\ensuremath{0.724\pm0.055}} 
\newcommand{\hatcurCCesoJKmagxxxxxA}{\ensuremath{0.966\pm0.043}} 
\newcommand{\hatcurCCesoHKmagxxxxxA}{\ensuremath{0.242\pm0.055}} 
\newcommand{\hatcurCCWonemagxxxxxA}{\ensuremath{10.991\pm0.088}} 
\newcommand{\hatcurCCWtwomagxxxxxA}{\ensuremath{10.890\pm0.094}} 
\newcommand{\hatcurCCWthreemagxxxxxA}{\ensuremath{10.71\pm0.12}} 
\newcommand{\hatcurCCWfourmagxxxxxA}{\ensuremath{nff\pmnff}}    
\newcommand{\hatcurLCdipxxxxxA}{\ensuremath{0.1}}               
\newcommand{\hatcurLCrprstarxxxxxA}{\ensuremath{0.17985\pm0.00095}} 
\newcommand{\hatcurLCbsqxxxxxA}{\ensuremath{0.5710_{-0.0101}^{+0.0086}}} 
\newcommand{\hatcurLCimpxxxxxA}{\ensuremath{0.7556_{-0.0067}^{+0.0057}}} 
\newcommand{\hatcurLCzetaxxxxxA}{\ensuremath{47.79\pm0.30}}     
\newcommand{\hatcurLCdurxxxxxA}{\ensuremath{0.05789\pm0.00031}} 
\newcommand{\hatcurLCdurshortxxxxxA}{\ensuremath{0.0579}}       
\newcommand{\hatcurLCdurhrxxxxxA}{\ensuremath{1.3895\pm0.0074}} 
\newcommand{\hatcurLCdurhrshortxxxxxA}{\ensuremath{1.389}}      
\newcommand{\hatcurLCqxxxxxA}{\ensuremath{0.016700\pm0.000098}} 
\newcommand{\hatcurLCqshortxxxxxA}{\ensuremath{0.017}}          
\newcommand{\hatcurLCingdurxxxxxA}{\ensuremath{0.01877\pm0.00048}} 
\newcommand{\hatcurLCPxxxxxA}{\ensuremath{3.47168260\pm0.00000072}} 
\newcommand{\hatcurLCPprecxxxxxA}{\ensuremath{3.4716826}}       
\newcommand{\hatcurLCPshortxxxxxA}{\ensuremath{3.4717}}         
\newcommand{\hatcurLCTxxxxxA}{\ensuremath{2459850.258470\pm0.000094}} 
\newcommand{\hatcurLCTAxxxxxA}{\ensuremath{2458572.67927\pm0.00027}} 
\newcommand{\hatcurLCTBxxxxxA}{\ensuremath{2460058.55942\pm0.00011}} 
\newcommand{\hatcurLChatnetmAxxxxxA}{\ensuremath{-6.77083\pm0.00010}} 
\newcommand{\hatcurLCiblendAxxxxxA}{\ensuremath{0.914\pm0.028}} 
\newcommand{\hatcurLChatnetmBxxxxxA}{\ensuremath{-6.80725\pm0.00011}} 
\newcommand{\hatcurLCiblendBxxxxxA}{\ensuremath{0.958\pm0.023}} 
\newcommand{\hatcurLChatnetmCxxxxxA}{\ensuremath{-6.69633\pm0.00011}} 
\newcommand{\hatcurLCiblendCxxxxxA}{\ensuremath{0.988\pm0.030}} 
\newcommand{\hatcurLChatnetmDxxxxxA}{\ensuremath{-6.844590\pm0.000095}} 
\newcommand{\hatcurLCiblendDxxxxxA}{\ensuremath{0.836\pm0.028}} 
\newcommand{\hatcurLCrhoxxxxxA}{\ensuremath{8.11\pm0.20}}       
\newcommand{\hatcurSMEiteffxxxxxA}{\ensuremath{3150\pm67}}      
\newcommand{\hatcurSMEizfehxxxxxA}{\ensuremath{0.24\pm0.10}}    
\newcommand{\hatcurSMEizfehshortxxxxxA}{\ensuremath{0.24}}      
\newcommand{\hatcurSMEiloggxxxxxA}{\ensuremath{5.00\pm0.50}}    
\newcommand{\hatcurSMEivsinxxxxxA}{\ensuremath{0\pm50}}         
\newcommand{\hatcurSMEivmacxxxxxA}{\ensuremath{nff\pmnff}}      
\newcommand{\hatcurSMEivmicxxxxxA}{\ensuremath{nff\pmnff}}      
\newcommand{\hatcurextraerrMJxxxxxA}{\ensuremath{0\pm0}}        
\newcommand{\hatcurextraerrMJtwosiglimxxxxxA}{\ensuremath{<0.0200}} 
\newcommand{\hatcurextraerrMHxxxxxA}{\ensuremath{0\pm0}}        
\newcommand{\hatcurextraerrMHtwosiglimxxxxxA}{\ensuremath{<0.0200}} 
\newcommand{\hatcurextraerrMKsxxxxxA}{\ensuremath{0\pm0}}       
\newcommand{\hatcurextraerrMKstwosiglimxxxxxA}{\ensuremath{<0.0200}} 
\newcommand{\hatcurextraerrMGxxxxxA}{\ensuremath{0\pm0}}        
\newcommand{\hatcurextraerrMGtwosiglimxxxxxA}{\ensuremath{<0.0200}} 
\newcommand{\hatcurextraerrMWonexxxxxA}{\ensuremath{0\pm0}}     
\newcommand{\hatcurextraerrMWonetwosiglimxxxxxA}{\ensuremath{<0.0200}} 
\newcommand{\hatcurextraerrMWtwoxxxxxA}{\ensuremath{0\pm0}}     
\newcommand{\hatcurextraerrMWtwotwosiglimxxxxxA}{\ensuremath{<0.0200}} 
\newcommand{\hatcurLBiBxxxxxA}{\ensuremath{0.5259}}             
\newcommand{\hatcurLBiiBxxxxxA}{\ensuremath{0.3296}}            
\newcommand{\hatcurLBiVxxxxxA}{\ensuremath{0.5154}}             
\newcommand{\hatcurLBiiVxxxxxA}{\ensuremath{0.3424}}            
\newcommand{\hatcurLBiRxxxxxA}{\ensuremath{0.4326}}             
\newcommand{\hatcurLBiiRxxxxxA}{\ensuremath{0.3323}}            
\newcommand{\hatcurLBiIxxxxxA}{\ensuremath{0.37\pm0.15}}        
\newcommand{\hatcurLBiiIxxxxxA}{\ensuremath{0.28_{-0.19}^{+0.14}}} 
\newcommand{\hatcurLBiuxxxxxA}{\ensuremath{0.4363}}             
\newcommand{\hatcurLBiiuxxxxxA}{\ensuremath{0.3435}}            
\newcommand{\hatcurLBigxxxxxA}{\ensuremath{0.65\pm0.14}}        
\newcommand{\hatcurLBiigxxxxxA}{\ensuremath{0.09\pm0.16}}       
\newcommand{\hatcurLBirxxxxxA}{\ensuremath{0.44\pm0.14}}        
\newcommand{\hatcurLBiirxxxxxA}{\ensuremath{0.30\pm0.17}}       
\newcommand{\hatcurLBiixxxxxA}{\ensuremath{0.23\pm0.12}}        
\newcommand{\hatcurLBiiixxxxxA}{\ensuremath{0.13\pm0.15}}       
\newcommand{\hatcurLBizxxxxxA}{\ensuremath{0.13\pm0.11}}        
\newcommand{\hatcurLBiizxxxxxA}{\ensuremath{0.10\pm0.12}}       
\newcommand{\hatcurLBiJxxxxxA}{\ensuremath{0.050_{-0.037}^{+0.066}}} 
\newcommand{\hatcurLBiiJxxxxxA}{\ensuremath{-0.000\pm0.071}}    
\newcommand{\hatcurLBiHxxxxxA}{\ensuremath{0.0936}}             
\newcommand{\hatcurLBiiHxxxxxA}{\ensuremath{0.2839}}            
\newcommand{\hatcurLBiKxxxxxA}{\ensuremath{0.0865}}             
\newcommand{\hatcurLBiiKxxxxxA}{\ensuremath{0.2452}}            
\newcommand{\hatcurLBiTxxxxxA}{\ensuremath{0.132\pm0.091}}      
\newcommand{\hatcurLBiiTxxxxxA}{\ensuremath{0.02\pm0.13}}       
\newcommand{\hatcurLBikepxxxxxA}{\ensuremath{0.2943}}           
\newcommand{\hatcurLBiikepxxxxxA}{\ensuremath{0.4341}}          
\newcommand{\hatcurLBiCxxxxxA}{\ensuremath{0.2393}}             
\newcommand{\hatcurLBiiCxxxxxA}{\ensuremath{0.4400}}            
\newcommand{\hatcurLBiMxxxxxA}{\ensuremath{0.4171}}             
\newcommand{\hatcurLBiiMxxxxxA}{\ensuremath{0.4013}}            
\newcommand{\hatcurLBiSonexxxxxA}{\ensuremath{0.0660}}            
\newcommand{\hatcurLBiiSonexxxxxA}{\ensuremath{0.1843}}           
\newcommand{\hatcurLBiStwoxxxxxA}{\ensuremath{0.0490}}            
\newcommand{\hatcurLBiiStwoxxxxxA}{\ensuremath{0.1551}}           
\newcommand{\hatcurLBiSthreexxxxxA}{\ensuremath{0.0533}}            
\newcommand{\hatcurLBiiSthreexxxxxA}{\ensuremath{0.1380}}           
\newcommand{\hatcurLBiSfourxxxxxA}{\ensuremath{0.0551}}            
\newcommand{\hatcurLBiiSfourxxxxxA}{\ensuremath{0.1075}}           
\newcommand{\hatcurISOmxxxxxA}{\ensuremath{0.442\pm0.025}}      
\newcommand{\hatcurISOmshortxxxxxA}{\ensuremath{0.44}}          
\newcommand{\hatcurISOmlongxxxxxA}{\ensuremath{0.442\pm0.025}}  
\newcommand{\hatcurISOrxxxxxA}{\ensuremath{0.4250\pm0.0091}}    
\newcommand{\hatcurISOrshortxxxxxA}{\ensuremath{0.42}}          
\newcommand{\hatcurISOrlongxxxxxA}{\ensuremath{0.4250\pm0.0091}} 
\newcommand{\hatcurISOrhoxxxxxA}{\ensuremath{8.11\pm0.20}}      
\newcommand{\hatcurISOrholongxxxxxA}{\ensuremath{8.11\pm0.20}}  
\newcommand{\hatcurISOloggxxxxxA}{\ensuremath{4.827\pm0.010}}   
\newcommand{\hatcurISOlumxxxxxA}{\ensuremath{0.0185\pm0.0011}}  
\newcommand{\hatcurISOlumshortxxxxxA}{\ensuremath{0.02}}        
\newcommand{\hatcurISOteffxxxxxA}{\ensuremath{3266\pm36}}       
\newcommand{\hatcurISOzfehxxxxxA}{\ensuremath{0.357\pm0.085}}   
\newcommand{\hatcurISOagexxxxxA}{\ensuremath{9.3\pm5.5}}        
\newcommand{\hatcurISOspecxxxxxA}{M}                            
\newcommand{\hatcurRVKxxxxxA}{\ensuremath{58.1\pm9.3}}          
\newcommand{\hatcurRVKtwosiglimxxxxxA}{\ensuremath{<73.5}}      
\newcommand{\hatcurRVrkxxxxxA}{\ensuremath{0\pm0}}              
\newcommand{\hatcurRVrhxxxxxA}{\ensuremath{0\pm0}}              
\newcommand{\hatcurRVkxxxxxA}{\ensuremath{0\pm0}}               
\newcommand{\hatcurRVhxxxxxA}{\ensuremath{0\pm0}}               
\newcommand{\hatcurRVtronexxxxxA}{\ensuremath{0\pm0}}           
\newcommand{\hatcurRVtrtwoxxxxxA}{\ensuremath{0\pm0}}           
\newcommand{\hatcurRVgammaAxxxxxA}{\ensuremath{48680.1\pm4.3}}  
\newcommand{\hatcurRVjitterAxxxxxA}{\ensuremath{0.1\pm4.1}}     
\newcommand{\hatcurRVjittertwosiglimAxxxxxA}{\ensuremath{<10.4}} 
\newcommand{\hatcurRVfitrmsAxxxxxA}{\ensuremath{0.0}}           
\newcommand{\hatcurRVgammaBxxxxxA}{\ensuremath{48607.7\pm6.0}}  
\newcommand{\hatcurRVjitterBxxxxxA}{\ensuremath{0.0\pm2.9}}     
\newcommand{\hatcurRVjittertwosiglimBxxxxxA}{\ensuremath{<8.0}} 
\newcommand{\hatcurRVfitrmsBxxxxxA}{\ensuremath{0.0}}           
\newcommand{\hatcurRVeccenxxxxxA}{\ensuremath{0\pm0}}           
\newcommand{\hatcurRVeccentwosiglimxxxxxA}{\ensuremath{<0.000}} 
\newcommand{\hatcurRVomegaxxxxxA}{\ensuremath{0\pm0}}           
\newcommand{\hatcurPPixxxxxA}{\ensuremath{87.500\pm0.039}}      
\newcommand{\hatcurPPgxxxxxA}{\ensuremath{11.3\pm1.9}}          
\newcommand{\hatcurPPloggxxxxxA}{\ensuremath{3.053\pm0.072}}    
\newcommand{\hatcurPParxxxxxA}{\ensuremath{17.29\pm0.14}}       
\newcommand{\hatcurPParelxxxxxA}{\ensuremath{0.03418\pm0.00065}} 
\newcommand{\hatcurPPrhoxxxxxA}{\ensuremath{0.76\pm0.13}}       
\newcommand{\hatcurPPmxxxxxA}{\ensuremath{0.251\pm0.042}}       
\newcommand{\hatcurPPmtwosiglimxxxxxA}{\ensuremath{<0.32}}      
\newcommand{\hatcurPPmshortxxxxxA}{\ensuremath{0.25}}           
\newcommand{\hatcurPPmlongxxxxxA}{\ensuremath{0.251\pm0.042}}   
\newcommand{\hatcurPPmexxxxxA}{\ensuremath{80\pm13}}            
\newcommand{\hatcurPPmeshortxxxxxA}{\ensuremath{79.8}}          
\newcommand{\hatcurPPmelongxxxxxA}{\ensuremath{80\pm13}}        
\newcommand{\hatcurPPrxxxxxA}{\ensuremath{0.744\pm0.017}}       
\newcommand{\hatcurPPrshortxxxxxA}{\ensuremath{0.74}}           
\newcommand{\hatcurPPrlongxxxxxA}{\ensuremath{0.744\pm0.017}}   
\newcommand{\hatcurPPrexxxxxA}{\ensuremath{8.34\pm0.19}}        
\newcommand{\hatcurPPreshortxxxxxA}{\ensuremath{8.3}}           
\newcommand{\hatcurPPrelongxxxxxA}{\ensuremath{8.34\pm0.19}}    
\newcommand{\hatcurPPmrcorrxxxxxA}{\ensuremath{0.17}}           
\newcommand{\hatcurPPteffxxxxxA}{\ensuremath{555.4\pm6.4}}      
\newcommand{\hatcurPPthetaxxxxxA}{\ensuremath{0.0523\pm0.0085}} 
\newcommand{\hatcurPPfluxperixxxxxA}{\ensuremath{2.156\pm0.099}} 
\newcommand{\hatcurPPfluxperidimxxxxxA}{\ensuremath{7}}         
\newcommand{\hatcurPPfluxapxxxxxA}{\ensuremath{2.156\pm0.099}}  
\newcommand{\hatcurPPfluxapdimxxxxxA}{\ensuremath{7}}           
\newcommand{\hatcurPPfluxavgxxxxxA}{\ensuremath{2.156\pm0.099}} 
\newcommand{\hatcurPPfluxavgdimxxxxxA}{\ensuremath{7}}          
\newcommand{\hatcurPPfluxavglogxxxxxA}{\ensuremath{7.334\pm0.020}} 
\newcommand{\hatcurXsecphasexxxxxA}{\ensuremath{0\pm0}}         
\newcommand{\hatcurXsecondaryxxxxxA}{\ensuremath{2459851.994310\pm0.000094}} 
\newcommand{\hatcurXsecdurxxxxxA}{\ensuremath{0.05789\pm0.00031}} 
\newcommand{\hatcurXsecingdurxxxxxA}{\ensuremath{0.01877\pm0.00048}} 
\newcommand{\hatcurPPphiconjxxxxxA}{\ensuremath{0\pm0}}         
\newcommand{\hatcurPPperixxxxxA}{\ensuremath{2459849.390545\pm0.000093}} 
\newcommand{\hatcurPPaequivxxxxxA}{\ensuremath{0.2512\pm0.0058}} 
\newcommand{\hatcurPPtcircxxxxxA}{\ensuremath{432\pm76}}        
\newcommand{\hatcurPPtinfallxxxxxA}{\ensuremath{91000\pm15000}} 
\newcommand{\hatcurXdistxxxxxA}{\ensuremath{98.80\pm0.23}}      
\newcommand{\hatcurXAvxxxxxA}{\ensuremath{0.1740\pm0.0052}}     
\newcommand{\hatcurXdistredxxxxxA}{\ensuremath{98.78\pm0.22}}   
\newcommand{\hatcurXEBVxxxxxA}{\ensuremath{0.0560_{-0.0020}^{+0.0010}}} 
\newcommand{\hatcurCCpmraxxxxxA}{\ensuremath{-159.174\pm0.020}} 
\newcommand{\hatcurCCpmdecxxxxxA}{\ensuremath{-24.780\pm0.020}} 
\newcommand{\hatcurCCpmxxxxxA}{\ensuremath{161.091\pm0.028}}    
\newcommand{\hatcurhtrxxxxxB}{TIC46432937}                      
\newcommand{\hatcurfieldxxxxxB}{\ensuremath{string}}            
\newcommand{\hatcurCCraxxxxxB}{\ensuremath{05^{\mathrm h}35^{\mathrm m}28.5693{\mathrm s}}}                   
\newcommand{\hatcurCCdecxxxxxB}{\ensuremath{-14{\arcdeg}35{\arcmin}50.4600{\arcsec}}}                 
\newcommand{\hatcurCCmagxxxxxB}{14.310}                         
\newcommand{\hatcurCCtwomassxxxxxB}{2MASS~11041818-4749169}     
\newcommand{\hatcurCCgscxxxxxB}{GSC~NULL}                       
\newcommand{\hatcurCCgaiaxxxxxB}{GAIA~}                         
\newcommand{\hatcurCCgaiadrtwoxxxxxB}{GAIA~DR2~2984391358868786816} 
\newcommand{\hatcurCCtassmvxxxxxB}{\ensuremath{14.310\pm0.020}} 
\newcommand{\hatcurCCtassmvshortxxxxxB}{\ensuremath{14.3}}      
\newcommand{\hatcurCCtassmBxxxxxB}{\ensuremath{15.707\pm0.050}} 
\newcommand{\hatcurCCtassmBshortxxxxxB}{\ensuremath{15.7}}      
\newcommand{\hatcurCCtassmIxxxxxB}{\ensuremath{nff\pmnff}}      
\newcommand{\hatcurCCtassmIshortxxxxxB}{\ensuremath{0.0}}       
\newcommand{\hatcurCCtassmgxxxxxB}{\ensuremath{15.040\pm0.060}} 
\newcommand{\hatcurCCtassmgshortxxxxxB}{\ensuremath{15.0}}      
\newcommand{\hatcurCCtassmrxxxxxB}{\ensuremath{13.740\pm0.020}} 
\newcommand{\hatcurCCtassmrshortxxxxxB}{\ensuremath{13.7}}      
\newcommand{\hatcurCCtassmixxxxxB}{\ensuremath{12.792\pm0.040}} 
\newcommand{\hatcurCCtassmishortxxxxxB}{\ensuremath{12.8}}      
\newcommand{\hatcurCCparallaxxxxxxB}{\ensuremath{11.031\pm0.013}} 
\newcommand{\hatcurCCgaiamGxxxxxB}{\ensuremath{13.4172\pm0.0028}} 
\newcommand{\hatcurCCgaiamBPxxxxxB}{\ensuremath{14.4962\pm0.0033}} 
\newcommand{\hatcurCCgaiamRPxxxxxB}{\ensuremath{12.3785\pm0.0039}} 
\newcommand{\hatcurCCtwomassJmagxxxxxB}{\ensuremath{11.011\pm0.022}} 
\newcommand{\hatcurCCtwomassHmagxxxxxB}{\ensuremath{10.427\pm0.023}} 
\newcommand{\hatcurCCtwomassKmagxxxxxB}{\ensuremath{10.195\pm0.020}} 
\newcommand{\hatcurCCcitJmagxxxxxB}{\ensuremath{11.004\pm0.023}} 
\newcommand{\hatcurCCcitHmagxxxxxB}{\ensuremath{10.418\pm0.024}} 
\newcommand{\hatcurCCcitKmagxxxxxB}{\ensuremath{10.219\pm0.021}} 
\newcommand{\hatcurCCbbJmagxxxxxB}{\ensuremath{11.090\pm0.025}} 
\newcommand{\hatcurCCbbHmagxxxxxB}{\ensuremath{10.444\pm0.025}} 
\newcommand{\hatcurCCbbKmagxxxxxB}{\ensuremath{10.239\pm0.021}} 
\newcommand{\hatcurCCesoJmagxxxxxB}{\ensuremath{11.098\pm0.030}} 
\newcommand{\hatcurCCesoHmagxxxxxB}{\ensuremath{10.444\pm0.040}} 
\newcommand{\hatcurCCesoKmagxxxxxB}{\ensuremath{10.236\pm0.023}} 
\newcommand{\hatcurCCesoJHmagxxxxxB}{\ensuremath{0.654\pm0.047}} 
\newcommand{\hatcurCCesoJKmagxxxxxB}{\ensuremath{0.862\pm0.035}} 
\newcommand{\hatcurCCesoHKmagxxxxxB}{\ensuremath{0.208\pm0.045}} 
\newcommand{\hatcurCCWonemagxxxxxB}{\ensuremath{10.114\pm0.023}} 
\newcommand{\hatcurCCWtwomagxxxxxB}{\ensuremath{10.063\pm0.020}} 
\newcommand{\hatcurCCWthreemagxxxxxB}{\ensuremath{9.910\pm0.055}} 
\newcommand{\hatcurCCWfourmagxxxxxB}{\ensuremath{nff\pmnff}}    
\newcommand{\hatcurLCdipxxxxxB}{\ensuremath{0.1}}               
\newcommand{\hatcurLCrprstarxxxxxB}{\ensuremath{0.2301\pm0.0037}} 
\newcommand{\hatcurLCbsqxxxxxB}{\ensuremath{0.681_{-0.012}^{+0.026}}} 
\newcommand{\hatcurLCimpxxxxxB}{\ensuremath{0.8250_{-0.0073}^{+0.0157}}} 
\newcommand{\hatcurLCzetaxxxxxB}{\ensuremath{64.73_{-0.89}^{+2.37}}} 
\newcommand{\hatcurLCdurxxxxxB}{\ensuremath{0.04981\pm0.00026}} 
\newcommand{\hatcurLCdurshortxxxxxB}{\ensuremath{0.0498}}       
\newcommand{\hatcurLCdurhrxxxxxB}{\ensuremath{1.1955\pm0.0062}} 
\newcommand{\hatcurLCdurhrshortxxxxxB}{\ensuremath{1.196}}      
\newcommand{\hatcurLCqxxxxxB}{\ensuremath{0.03460\pm0.00018}}   
\newcommand{\hatcurLCqshortxxxxxB}{\ensuremath{0.035}}          
\newcommand{\hatcurLCingdurxxxxxB}{\ensuremath{0.05227\pm0.00026}} 
\newcommand{\hatcurLCPxxxxxB}{\ensuremath{1.440445270\pm0.000000087}} 
\newcommand{\hatcurLCPprecxxxxxB}{\ensuremath{1.4404453}}       
\newcommand{\hatcurLCPshortxxxxxB}{\ensuremath{1.4404}}         
\newcommand{\hatcurLCTxxxxxB}{\ensuremath{2459952.288940\pm0.000039}} 
\newcommand{\hatcurLCTAxxxxxB}{\ensuremath{2458468.630306\pm0.000094}} 
\newcommand{\hatcurLCTBxxxxxB}{\ensuremath{2460316.721590\pm0.000046}} 
\newcommand{\hatcurLChatnetmAxxxxxB}{\ensuremath{-8.116450\pm0.000040}} 
\newcommand{\hatcurLCiblendAxxxxxB}{\ensuremath{1\pm0}}         
\newcommand{\hatcurLChatnetmBxxxxxB}{\ensuremath{-8.019870\pm0.000041}} 
\newcommand{\hatcurLCiblendBxxxxxB}{\ensuremath{1\pm0}}         
\newcommand{\hatcurLCrhoxxxxxB}{\ensuremath{5.34\pm0.11}}       
\newcommand{\hatcurSMEiteffxxxxxB}{\ensuremath{3535\pm65}}      
\newcommand{\hatcurSMEizfehxxxxxB}{\ensuremath{0.03\pm0.10}}    
\newcommand{\hatcurSMEizfehshortxxxxxB}{\ensuremath{0.03}}      
\newcommand{\hatcurSMEiloggxxxxxB}{\ensuremath{5.00\pm0.50}}    
\newcommand{\hatcurSMEivsinxxxxxB}{\ensuremath{0\pm50}}         
\newcommand{\hatcurSMEivmacxxxxxB}{\ensuremath{nff\pmnff}}      
\newcommand{\hatcurSMEivmicxxxxxB}{\ensuremath{nff\pmnff}}      
\newcommand{\hatcurextraerrMJxxxxxB}{\ensuremath{0\pm0}}        
\newcommand{\hatcurextraerrMJtwosiglimxxxxxB}{\ensuremath{<0.0200}} 
\newcommand{\hatcurextraerrMHxxxxxB}{\ensuremath{0\pm0}}        
\newcommand{\hatcurextraerrMHtwosiglimxxxxxB}{\ensuremath{<0.0200}} 
\newcommand{\hatcurextraerrMKsxxxxxB}{\ensuremath{0\pm0}}       
\newcommand{\hatcurextraerrMKstwosiglimxxxxxB}{\ensuremath{<0.0200}} 
\newcommand{\hatcurextraerrMGxxxxxB}{\ensuremath{0\pm0}}        
\newcommand{\hatcurextraerrMGtwosiglimxxxxxB}{\ensuremath{<0.0200}} 
\newcommand{\hatcurextraerrMRPxxxxxB}{\ensuremath{0\pm0}}       
\newcommand{\hatcurextraerrMRPtwosiglimxxxxxB}{\ensuremath{<0.0200}} 
\newcommand{\hatcurextraerrMgxxxxxB}{\ensuremath{0\pm0}}        
\newcommand{\hatcurextraerrMgtwosiglimxxxxxB}{\ensuremath{<0.0200}} 
\newcommand{\hatcurextraerrMrxxxxxB}{\ensuremath{0\pm0}}        
\newcommand{\hatcurextraerrMrtwosiglimxxxxxB}{\ensuremath{<0.0200}} 
\newcommand{\hatcurextraerrMixxxxxB}{\ensuremath{0\pm0}}        
\newcommand{\hatcurextraerrMitwosiglimxxxxxB}{\ensuremath{<0.0200}} 
\newcommand{\hatcurextraerrMWonexxxxxB}{\ensuremath{0\pm0}}     
\newcommand{\hatcurextraerrMWonetwosiglimxxxxxB}{\ensuremath{<0.0200}} 
\newcommand{\hatcurextraerrMWtwoxxxxxB}{\ensuremath{0\pm0}}     
\newcommand{\hatcurextraerrMWtwotwosiglimxxxxxB}{\ensuremath{<0.0200}} 
\newcommand{\hatcurLBiBxxxxxB}{\ensuremath{0.4448}}             
\newcommand{\hatcurLBiiBxxxxxB}{\ensuremath{0.3380}}            
\newcommand{\hatcurLBiVxxxxxB}{\ensuremath{0.4071}}             
\newcommand{\hatcurLBiiVxxxxxB}{\ensuremath{0.3615}}            
\newcommand{\hatcurLBiRxxxxxB}{\ensuremath{0.3575}}             
\newcommand{\hatcurLBiiRxxxxxB}{\ensuremath{0.3335}}            
\newcommand{\hatcurLBiIxxxxxB}{\ensuremath{0.108_{-0.071}^{+0.097}}} 
\newcommand{\hatcurLBiiIxxxxxB}{\ensuremath{0.06\pm0.12}}       
\newcommand{\hatcurLBiuxxxxxB}{\ensuremath{0.4132}}             
\newcommand{\hatcurLBiiuxxxxxB}{\ensuremath{0.3279}}            
\newcommand{\hatcurLBigxxxxxB}{\ensuremath{0.4098}}             
\newcommand{\hatcurLBiigxxxxxB}{\ensuremath{0.3420}}            
\newcommand{\hatcurLBirxxxxxB}{\ensuremath{0.4207}}             
\newcommand{\hatcurLBiirxxxxxB}{\ensuremath{0.3178}}            
\newcommand{\hatcurLBiixxxxxB}{\ensuremath{0.2336}}             
\newcommand{\hatcurLBiiixxxxxB}{\ensuremath{0.3384}}            
\newcommand{\hatcurLBizxxxxxB}{\ensuremath{0.143\pm0.096}}      
\newcommand{\hatcurLBiizxxxxxB}{\ensuremath{0.15\pm0.14}}       
\newcommand{\hatcurLBiJxxxxxB}{\ensuremath{0.142\pm0.088}}      
\newcommand{\hatcurLBiiJxxxxxB}{\ensuremath{0.09\pm0.13}}       
\newcommand{\hatcurLBiHxxxxxB}{\ensuremath{0.20\pm0.11}}        
\newcommand{\hatcurLBiiHxxxxxB}{\ensuremath{0.08\pm0.15}}       
\newcommand{\hatcurLBiKxxxxxB}{\ensuremath{0.0759}}             
\newcommand{\hatcurLBiiKxxxxxB}{\ensuremath{0.2304}}            
\newcommand{\hatcurLBiTxxxxxB}{\ensuremath{0.20\pm0.10}}        
\newcommand{\hatcurLBiiTxxxxxB}{\ensuremath{0.17\pm0.14}}       
\newcommand{\hatcurLBikepxxxxxB}{\ensuremath{0.2854}}           
\newcommand{\hatcurLBiikepxxxxxB}{\ensuremath{0.4148}}          
\newcommand{\hatcurLBiCxxxxxB}{\ensuremath{0.2416}}             
\newcommand{\hatcurLBiiCxxxxxB}{\ensuremath{0.4173}}            
\newcommand{\hatcurLBiMxxxxxB}{\ensuremath{0.3749}}             
\newcommand{\hatcurLBiiMxxxxxB}{\ensuremath{0.4051}}            
\newcommand{\hatcurLBiSonexxxxxB}{\ensuremath{0.0625}}            
\newcommand{\hatcurLBiiSonexxxxxB}{\ensuremath{0.1665}}           
\newcommand{\hatcurLBiStwoxxxxxB}{\ensuremath{0.0462}}            
\newcommand{\hatcurLBiiStwoxxxxxB}{\ensuremath{0.1375}}           
\newcommand{\hatcurLBiSthreexxxxxB}{\ensuremath{0.0495}}            
\newcommand{\hatcurLBiiSthreexxxxxB}{\ensuremath{0.1196}}           
\newcommand{\hatcurLBiSfourxxxxxB}{\ensuremath{0.0533}}            
\newcommand{\hatcurLBiiSfourxxxxxB}{\ensuremath{0.0911}}           
\newcommand{\hatcurISOmxxxxxB}{\ensuremath{0.563\pm0.029}}      
\newcommand{\hatcurISOmshortxxxxxB}{\ensuremath{0.56}}          
\newcommand{\hatcurISOmlongxxxxxB}{\ensuremath{0.563\pm0.029}}  
\newcommand{\hatcurISOrxxxxxB}{\ensuremath{0.5299\pm0.0091}}    
\newcommand{\hatcurISOrshortxxxxxB}{\ensuremath{0.53}}          
\newcommand{\hatcurISOrlongxxxxxB}{\ensuremath{0.5299\pm0.0091}} 
\newcommand{\hatcurISOrhoxxxxxB}{\ensuremath{5.34\pm0.11}}      
\newcommand{\hatcurISOrholongxxxxxB}{\ensuremath{5.34\pm0.11}}  
\newcommand{\hatcurISOloggxxxxxB}{\ensuremath{4.740\pm0.010}}   
\newcommand{\hatcurISOlumxxxxxB}{\ensuremath{0.0412\pm0.0031}}  
\newcommand{\hatcurISOlumshortxxxxxB}{\ensuremath{0.04}}        
\newcommand{\hatcurISOteffxxxxxB}{\ensuremath{3572\pm57}}       
\newcommand{\hatcurISOzfehxxxxxB}{\ensuremath{0.323\pm0.081}}   
\newcommand{\hatcurISOagexxxxxB}{\ensuremath{7.4\pm5.1}}        
\newcommand{\hatcurISOspecxxxxxB}{M}                            
\newcommand{\hatcurRVKxxxxxB}{\ensuremath{837.1\pm4.4}}         
\newcommand{\hatcurRVKtwosiglimxxxxxB}{\ensuremath{<843.4}}     
\newcommand{\hatcurRVrkxxxxxB}{\ensuremath{0\pm0}}              
\newcommand{\hatcurRVrhxxxxxB}{\ensuremath{0\pm0}}              
\newcommand{\hatcurRVkxxxxxB}{\ensuremath{0\pm0}}               
\newcommand{\hatcurRVhxxxxxB}{\ensuremath{0\pm0}}               
\newcommand{\hatcurRVtronexxxxxB}{\ensuremath{0\pm0}}           
\newcommand{\hatcurRVtrtwoxxxxxB}{\ensuremath{0\pm0}}           
\newcommand{\hatcurRVgammaxxxxxB}{\ensuremath{102279.1\pm2.5}}  
\newcommand{\hatcurRVjitterxxxxxB}{\ensuremath{5.3\pm2.5}}      
\newcommand{\hatcurRVjittertwosiglimxxxxxB}{\ensuremath{<10.4}} 
\newcommand{\hatcurRVfitrmsxxxxxB}{\ensuremath{.1fym}}          %
\newcommand{\hatcurRVeccenxxxxxB}{\ensuremath{0\pm0}}           
\newcommand{\hatcurRVeccentwosiglimxxxxxB}{\ensuremath{<0.000}} 
\newcommand{\hatcurRVomegaxxxxxB}{\ensuremath{0\pm0}}           
\newcommand{\hatcurPPixxxxxB}{\ensuremath{84.350_{-0.140}^{+0.090}}} 
\newcommand{\hatcurPPgxxxxxB}{\ensuremath{56.4_{-2.9}^{+2.1}}}  
\newcommand{\hatcurPPloggxxxxxB}{\ensuremath{3.751_{-0.023}^{+0.016}}} 
\newcommand{\hatcurPParxxxxxB}{\ensuremath{8.381\pm0.056}}      
\newcommand{\hatcurPParelxxxxxB}{\ensuremath{0.02065\pm0.00035}} 
\newcommand{\hatcurPPrhoxxxxxB}{\ensuremath{2.37\pm0.15}}       
\newcommand{\hatcurPPmxxxxxB}{\ensuremath{3.20\pm0.11}}         
\newcommand{\hatcurPPmtwosiglimxxxxxB}{\ensuremath{<3.38}}      
\newcommand{\hatcurPPmshortxxxxxB}{\ensuremath{3.20}}           
\newcommand{\hatcurPPmlongxxxxxB}{\ensuremath{3.20\pm0.11}}     
\newcommand{\hatcurPPmexxxxxB}{\ensuremath{1016\pm35}}          
\newcommand{\hatcurPPmeshortxxxxxB}{\ensuremath{1016.4}}        
\newcommand{\hatcurPPmelongxxxxxB}{\ensuremath{1016\pm35}}      
\newcommand{\hatcurPPrxxxxxB}{\ensuremath{1.188\pm0.030}}       
\newcommand{\hatcurPPrshortxxxxxB}{\ensuremath{1.19}}           
\newcommand{\hatcurPPrlongxxxxxB}{\ensuremath{1.188\pm0.030}}   
\newcommand{\hatcurPPrexxxxxB}{\ensuremath{13.32\pm0.34}}       
\newcommand{\hatcurPPreshortxxxxxB}{\ensuremath{13.3}}          
\newcommand{\hatcurPPrelongxxxxxB}{\ensuremath{13.32\pm0.34}}   
\newcommand{\hatcurPPmrcorrxxxxxB}{\ensuremath{0.53}}           
\newcommand{\hatcurPPteffxxxxxB}{\ensuremath{872\pm14}}         
\newcommand{\hatcurPPthetaxxxxxB}{\ensuremath{0.1951\pm0.0051}} 
\newcommand{\hatcurPPfluxperixxxxxB}{\ensuremath{1.314\pm0.086}} 
\newcommand{\hatcurPPfluxperidimxxxxxB}{\ensuremath{8}}         
\newcommand{\hatcurPPfluxapxxxxxB}{\ensuremath{1.314\pm0.086}}  
\newcommand{\hatcurPPfluxapdimxxxxxB}{\ensuremath{8}}           
\newcommand{\hatcurPPfluxavgxxxxxB}{\ensuremath{1.314\pm0.086}} 
\newcommand{\hatcurPPfluxavgdimxxxxxB}{\ensuremath{8}}          
\newcommand{\hatcurPPfluxavglogxxxxxB}{\ensuremath{8.119\pm0.028}} 
\newcommand{\hatcurXsecphasexxxxxB}{\ensuremath{0\pm0}}         
\newcommand{\hatcurXsecondaryxxxxxB}{\ensuremath{2459953.009160\pm0.000039}} 
\newcommand{\hatcurXsecdurxxxxxB}{\ensuremath{0.05248\pm0.00074}} 
\newcommand{\hatcurXsecingdurxxxxxB}{\ensuremath{0.02624\pm0.00037}} 
\newcommand{\hatcurPPphiconjxxxxxB}{\ensuremath{0\pm0}}         
\newcommand{\hatcurPPperixxxxxB}{\ensuremath{2459951.928824\pm0.000038}} 
\newcommand{\hatcurPPaequivxxxxxB}{\ensuremath{0.1018\pm0.0033}} 
\newcommand{\hatcurPPtcircxxxxxB}{\ensuremath{13.9_{-1.7}^{+1.3}}} 
\newcommand{\hatcurPPtinfallxxxxxB}{\ensuremath{99.7\pm4.0}}    
\newcommand{\hatcurXdistxxxxxB}{\ensuremath{90.60\pm0.12}}      
\newcommand{\hatcurXAvxxxxxB}{\ensuremath{0.023\pm0.012}}       
\newcommand{\hatcurXdistredxxxxxB}{\ensuremath{90.64\pm0.10}}   
\newcommand{\hatcurXEBVxxxxxB}{\ensuremath{0.0070\pm0.0040}}    
\newcommand{\hatcurCCpmraxxxxxB}{\ensuremath{-13.365\pm0.013}}  
\newcommand{\hatcurCCpmdecxxxxxB}{\ensuremath{36.962\pm0.012}}  
\newcommand{\hatcurCCpmxxxxxB}{\ensuremath{39.304\pm0.018}}     
\newcommand{\hatcurCCbbHmag}[1]{\ifnum#1=762 %
\hatcurCCbbHmagxxxxxA
\else
\ifnum#1=4643 %
\hatcurCCbbHmagxxxxxB
\else
??????\fi
\fi
}
\newcommand{\hatcurCCbbJmag}[1]{\ifnum#1=762 %
\hatcurCCbbJmagxxxxxA
\else
\ifnum#1=4643 %
\hatcurCCbbJmagxxxxxB
\else
??????\fi
\fi
}
\newcommand{\hatcurCCbbKmag}[1]{\ifnum#1=762 %
\hatcurCCbbKmagxxxxxA
\else
\ifnum#1=4643 %
\hatcurCCbbKmagxxxxxB
\else
??????\fi
\fi
}
\newcommand{\hatcurCCcitHmag}[1]{\ifnum#1=762 %
\hatcurCCcitHmagxxxxxA
\else
\ifnum#1=4643 %
\hatcurCCcitHmagxxxxxB
\else
??????\fi
\fi
}
\newcommand{\hatcurCCcitJmag}[1]{\ifnum#1=762 %
\hatcurCCcitJmagxxxxxA
\else
\ifnum#1=4643 %
\hatcurCCcitJmagxxxxxB
\else
??????\fi
\fi
}
\newcommand{\hatcurCCcitKmag}[1]{\ifnum#1=762 %
\hatcurCCcitKmagxxxxxA
\else
\ifnum#1=4643 %
\hatcurCCcitKmagxxxxxB
\else
??????\fi
\fi
}
\newcommand{\hatcurCCdec}[1]{\ifnum#1=762 %
\hatcurCCdecxxxxxA
\else
\ifnum#1=4643 %
\hatcurCCdecxxxxxB
\else
??????\fi
\fi
}
\newcommand{\hatcurCCesoHKmag}[1]{\ifnum#1=762 %
\hatcurCCesoHKmagxxxxxA
\else
\ifnum#1=4643 %
\hatcurCCesoHKmagxxxxxB
\else
??????\fi
\fi
}
\newcommand{\hatcurCCesoHmag}[1]{\ifnum#1=762 %
\hatcurCCesoHmagxxxxxA
\else
\ifnum#1=4643 %
\hatcurCCesoHmagxxxxxB
\else
??????\fi
\fi
}
\newcommand{\hatcurCCesoJHmag}[1]{\ifnum#1=762 %
\hatcurCCesoJHmagxxxxxA
\else
\ifnum#1=4643 %
\hatcurCCesoJHmagxxxxxB
\else
??????\fi
\fi
}
\newcommand{\hatcurCCesoJKmag}[1]{\ifnum#1=762 %
\hatcurCCesoJKmagxxxxxA
\else
\ifnum#1=4643 %
\hatcurCCesoJKmagxxxxxB
\else
??????\fi
\fi
}
\newcommand{\hatcurCCesoJmag}[1]{\ifnum#1=762 %
\hatcurCCesoJmagxxxxxA
\else
\ifnum#1=4643 %
\hatcurCCesoJmagxxxxxB
\else
??????\fi
\fi
}
\newcommand{\hatcurCCesoKmag}[1]{\ifnum#1=762 %
\hatcurCCesoKmagxxxxxA
\else
\ifnum#1=4643 %
\hatcurCCesoKmagxxxxxB
\else
??????\fi
\fi
}
\newcommand{\hatcurCCgaia}[1]{\ifnum#1=762 %
\hatcurCCgaiaxxxxxA
\else
\ifnum#1=4643 %
\hatcurCCgaiaxxxxxB
\else
??????\fi
\fi
}
\newcommand{\hatcurCCgaiadrtwo}[1]{\ifnum#1=762 %
\hatcurCCgaiadrtwoxxxxxA
\else
\ifnum#1=4643 %
\hatcurCCgaiadrtwoxxxxxB
\else
??????\fi
\fi
}
\newcommand{\hatcurCCgaiamBP}[1]{\ifnum#1=762 %
\hatcurCCgaiamBPxxxxxA
\else
\ifnum#1=4643 %
\hatcurCCgaiamBPxxxxxB
\else
??????\fi
\fi
}
\newcommand{\hatcurCCgaiamG}[1]{\ifnum#1=762 %
\hatcurCCgaiamGxxxxxA
\else
\ifnum#1=4643 %
\hatcurCCgaiamGxxxxxB
\else
??????\fi
\fi
}
\newcommand{\hatcurCCgaiamRP}[1]{\ifnum#1=762 %
\hatcurCCgaiamRPxxxxxA
\else
\ifnum#1=4643 %
\hatcurCCgaiamRPxxxxxB
\else
??????\fi
\fi
}
\newcommand{\hatcurCCgsc}[1]{\ifnum#1=762 %
\hatcurCCgscxxxxxA
\else
\ifnum#1=4643 %
\hatcurCCgscxxxxxB
\else
??????\fi
\fi
}
\newcommand{\hatcurCCmag}[1]{\ifnum#1=762 %
\hatcurCCmagxxxxxA
\else
\ifnum#1=4643 %
\hatcurCCmagxxxxxB
\else
??????\fi
\fi
}
\newcommand{\hatcurCCparallax}[1]{\ifnum#1=762 %
\hatcurCCparallaxxxxxxA
\else
\ifnum#1=4643 %
\hatcurCCparallaxxxxxxB
\else
??????\fi
\fi
}
\newcommand{\hatcurCCpm}[1]{\ifnum#1=762 %
\hatcurCCpmxxxxxA
\else
\ifnum#1=4643 %
\hatcurCCpmxxxxxB
\else
??????\fi
\fi
}
\newcommand{\hatcurCCpmdec}[1]{\ifnum#1=762 %
\hatcurCCpmdecxxxxxA
\else
\ifnum#1=4643 %
\hatcurCCpmdecxxxxxB
\else
??????\fi
\fi
}
\newcommand{\hatcurCCpmra}[1]{\ifnum#1=762 %
\hatcurCCpmraxxxxxA
\else
\ifnum#1=4643 %
\hatcurCCpmraxxxxxB
\else
??????\fi
\fi
}
\newcommand{\hatcurCCra}[1]{\ifnum#1=762 %
\hatcurCCraxxxxxA
\else
\ifnum#1=4643 %
\hatcurCCraxxxxxB
\else
??????\fi
\fi
}
\newcommand{\hatcurCCtassmB}[1]{\ifnum#1=762 %
\hatcurCCtassmBxxxxxA
\else
\ifnum#1=4643 %
\hatcurCCtassmBxxxxxB
\else
??????\fi
\fi
}
\newcommand{\hatcurCCtassmBshort}[1]{\ifnum#1=762 %
\hatcurCCtassmBshortxxxxxA
\else
\ifnum#1=4643 %
\hatcurCCtassmBshortxxxxxB
\else
??????\fi
\fi
}
\newcommand{\hatcurCCtassmg}[1]{\ifnum#1=762 %
\hatcurCCtassmgxxxxxA
\else
\ifnum#1=4643 %
\hatcurCCtassmgxxxxxB
\else
??????\fi
\fi
}
\newcommand{\hatcurCCtassmgshort}[1]{\ifnum#1=762 %
\hatcurCCtassmgshortxxxxxA
\else
\ifnum#1=4643 %
\hatcurCCtassmgshortxxxxxB
\else
??????\fi
\fi
}
\newcommand{\hatcurCCtassmi}[1]{\ifnum#1=762 %
\hatcurCCtassmixxxxxA
\else
\ifnum#1=4643 %
\hatcurCCtassmixxxxxB
\else
??????\fi
\fi
}
\newcommand{\hatcurCCtassmI}[1]{\ifnum#1=762 %
\hatcurCCtassmIxxxxxA
\else
\ifnum#1=4643 %
\hatcurCCtassmIxxxxxB
\else
??????\fi
\fi
}
\newcommand{\hatcurCCtassmishort}[1]{\ifnum#1=762 %
\hatcurCCtassmishortxxxxxA
\else
\ifnum#1=4643 %
\hatcurCCtassmishortxxxxxB
\else
??????\fi
\fi
}
\newcommand{\hatcurCCtassmIshort}[1]{\ifnum#1=762 %
\hatcurCCtassmIshortxxxxxA
\else
\ifnum#1=4643 %
\hatcurCCtassmIshortxxxxxB
\else
??????\fi
\fi
}
\newcommand{\hatcurCCtassmr}[1]{\ifnum#1=762 %
\hatcurCCtassmrxxxxxA
\else
\ifnum#1=4643 %
\hatcurCCtassmrxxxxxB
\else
??????\fi
\fi
}
\newcommand{\hatcurCCtassmrshort}[1]{\ifnum#1=762 %
\hatcurCCtassmrshortxxxxxA
\else
\ifnum#1=4643 %
\hatcurCCtassmrshortxxxxxB
\else
??????\fi
\fi
}
\newcommand{\hatcurCCtassmv}[1]{\ifnum#1=762 %
\hatcurCCtassmvxxxxxA
\else
\ifnum#1=4643 %
\hatcurCCtassmvxxxxxB
\else
??????\fi
\fi
}
\newcommand{\hatcurCCtassmvshort}[1]{\ifnum#1=762 %
\hatcurCCtassmvshortxxxxxA
\else
\ifnum#1=4643 %
\hatcurCCtassmvshortxxxxxB
\else
??????\fi
\fi
}
\newcommand{\hatcurCCtwomass}[1]{\ifnum#1=762 %
\hatcurCCtwomassxxxxxA
\else
\ifnum#1=4643 %
\hatcurCCtwomassxxxxxB
\else
??????\fi
\fi
}
\newcommand{\hatcurCCtwomassHmag}[1]{\ifnum#1=762 %
\hatcurCCtwomassHmagxxxxxA
\else
\ifnum#1=4643 %
\hatcurCCtwomassHmagxxxxxB
\else
??????\fi
\fi
}
\newcommand{\hatcurCCtwomassJmag}[1]{\ifnum#1=762 %
\hatcurCCtwomassJmagxxxxxA
\else
\ifnum#1=4643 %
\hatcurCCtwomassJmagxxxxxB
\else
??????\fi
\fi
}
\newcommand{\hatcurCCtwomassKmag}[1]{\ifnum#1=762 %
\hatcurCCtwomassKmagxxxxxA
\else
\ifnum#1=4643 %
\hatcurCCtwomassKmagxxxxxB
\else
??????\fi
\fi
}
\newcommand{\hatcurCCWfourmag}[1]{\ifnum#1=762 %
\hatcurCCWfourmagxxxxxA
\else
\ifnum#1=4643 %
\hatcurCCWfourmagxxxxxB
\else
??????\fi
\fi
}
\newcommand{\hatcurCCWonemag}[1]{\ifnum#1=762 %
\hatcurCCWonemagxxxxxA
\else
\ifnum#1=4643 %
\hatcurCCWonemagxxxxxB
\else
??????\fi
\fi
}
\newcommand{\hatcurCCWthreemag}[1]{\ifnum#1=762 %
\hatcurCCWthreemagxxxxxA
\else
\ifnum#1=4643 %
\hatcurCCWthreemagxxxxxB
\else
??????\fi
\fi
}
\newcommand{\hatcurCCWtwomag}[1]{\ifnum#1=762 %
\hatcurCCWtwomagxxxxxA
\else
\ifnum#1=4643 %
\hatcurCCWtwomagxxxxxB
\else
??????\fi
\fi
}
\newcommand{\hatcurextraerrMg}[1]{\ifnum#1=4643 %
\hatcurextraerrMgxxxxxB
\else
??????\fi
}
\newcommand{\hatcurextraerrMG}[1]{\ifnum#1=762 %
\hatcurextraerrMGxxxxxA
\else
\ifnum#1=4643 %
\hatcurextraerrMGxxxxxB
\else
??????\fi
\fi
}
\newcommand{\hatcurextraerrMgtwosiglim}[1]{\ifnum#1=4643 %
\hatcurextraerrMgtwosiglimxxxxxB
\else
??????\fi
}
\newcommand{\hatcurextraerrMGtwosiglim}[1]{\ifnum#1=762 %
\hatcurextraerrMGtwosiglimxxxxxA
\else
\ifnum#1=4643 %
\hatcurextraerrMGtwosiglimxxxxxB
\else
??????\fi
\fi
}
\newcommand{\hatcurextraerrMH}[1]{\ifnum#1=762 %
\hatcurextraerrMHxxxxxA
\else
\ifnum#1=4643 %
\hatcurextraerrMHxxxxxB
\else
??????\fi
\fi
}
\newcommand{\hatcurextraerrMHtwosiglim}[1]{\ifnum#1=762 %
\hatcurextraerrMHtwosiglimxxxxxA
\else
\ifnum#1=4643 %
\hatcurextraerrMHtwosiglimxxxxxB
\else
??????\fi
\fi
}
\newcommand{\hatcurextraerrMi}[1]{\ifnum#1=4643 %
\hatcurextraerrMixxxxxB
\else
??????\fi
}
\newcommand{\hatcurextraerrMitwosiglim}[1]{\ifnum#1=4643 %
\hatcurextraerrMitwosiglimxxxxxB
\else
??????\fi
}
\newcommand{\hatcurextraerrMJ}[1]{\ifnum#1=762 %
\hatcurextraerrMJxxxxxA
\else
\ifnum#1=4643 %
\hatcurextraerrMJxxxxxB
\else
??????\fi
\fi
}
\newcommand{\hatcurextraerrMJtwosiglim}[1]{\ifnum#1=762 %
\hatcurextraerrMJtwosiglimxxxxxA
\else
\ifnum#1=4643 %
\hatcurextraerrMJtwosiglimxxxxxB
\else
??????\fi
\fi
}
\newcommand{\hatcurextraerrMKs}[1]{\ifnum#1=762 %
\hatcurextraerrMKsxxxxxA
\else
\ifnum#1=4643 %
\hatcurextraerrMKsxxxxxB
\else
??????\fi
\fi
}
\newcommand{\hatcurextraerrMKstwosiglim}[1]{\ifnum#1=762 %
\hatcurextraerrMKstwosiglimxxxxxA
\else
\ifnum#1=4643 %
\hatcurextraerrMKstwosiglimxxxxxB
\else
??????\fi
\fi
}
\newcommand{\hatcurextraerrMr}[1]{\ifnum#1=4643 %
\hatcurextraerrMrxxxxxB
\else
??????\fi
}
\newcommand{\hatcurextraerrMRP}[1]{\ifnum#1=762 %
\hatcurextraerrMRPxxxxxA
\else
\ifnum#1=4643 %
\hatcurextraerrMRPxxxxxB
\else
??????\fi
\fi
}
\newcommand{\hatcurextraerrMRPtwosiglim}[1]{\ifnum#1=762 %
\hatcurextraerrMRPtwosiglimxxxxxA
\else
\ifnum#1=4643 %
\hatcurextraerrMRPtwosiglimxxxxxB
\else
??????\fi
\fi
}
\newcommand{\hatcurextraerrMrtwosiglim}[1]{\ifnum#1=4643 %
\hatcurextraerrMrtwosiglimxxxxxB
\else
??????\fi
}
\newcommand{\hatcurextraerrMWone}[1]{\ifnum#1=762 %
\hatcurextraerrMWonexxxxxA
\else
\ifnum#1=4643 %
\hatcurextraerrMWonexxxxxB
\else
??????\fi
\fi
}
\newcommand{\hatcurextraerrMWonetwosiglim}[1]{\ifnum#1=762 %
\hatcurextraerrMWonetwosiglimxxxxxA
\else
\ifnum#1=4643 %
\hatcurextraerrMWonetwosiglimxxxxxB
\else
??????\fi
\fi
}
\newcommand{\hatcurextraerrMWtwo}[1]{\ifnum#1=762 %
\hatcurextraerrMWtwoxxxxxA
\else
\ifnum#1=4643 %
\hatcurextraerrMWtwoxxxxxB
\else
??????\fi
\fi
}
\newcommand{\hatcurextraerrMWtwotwosiglim}[1]{\ifnum#1=762 %
\hatcurextraerrMWtwotwosiglimxxxxxA
\else
\ifnum#1=4643 %
\hatcurextraerrMWtwotwosiglimxxxxxB
\else
??????\fi
\fi
}
\newcommand{\hatcurfield}[1]{\ifnum#1=762 %
\hatcurfieldxxxxxA
\else
\ifnum#1=4643 %
\hatcurfieldxxxxxB
\else
??????\fi
\fi
}
\newcommand{\hatcurhtr}[1]{\ifnum#1=762 %
\hatcurhtrxxxxxA
\else
\ifnum#1=4643 %
\hatcurhtrxxxxxB
\else
??????\fi
\fi
}
\newcommand{\hatcurISOage}[1]{\ifnum#1=762 %
\hatcurISOagexxxxxA
\else
\ifnum#1=4643 %
\hatcurISOagexxxxxB
\else
??????\fi
\fi
}
\newcommand{\hatcurISOlogg}[1]{\ifnum#1=762 %
\hatcurISOloggxxxxxA
\else
\ifnum#1=4643 %
\hatcurISOloggxxxxxB
\else
??????\fi
\fi
}
\newcommand{\hatcurISOlum}[1]{\ifnum#1=762 %
\hatcurISOlumxxxxxA
\else
\ifnum#1=4643 %
\hatcurISOlumxxxxxB
\else
??????\fi
\fi
}
\newcommand{\hatcurISOlumshort}[1]{\ifnum#1=762 %
\hatcurISOlumshortxxxxxA
\else
\ifnum#1=4643 %
\hatcurISOlumshortxxxxxB
\else
??????\fi
\fi
}
\newcommand{\hatcurISOm}[1]{\ifnum#1=762 %
\hatcurISOmxxxxxA
\else
\ifnum#1=4643 %
\hatcurISOmxxxxxB
\else
??????\fi
\fi
}
\newcommand{\hatcurISOmlong}[1]{\ifnum#1=762 %
\hatcurISOmlongxxxxxA
\else
\ifnum#1=4643 %
\hatcurISOmlongxxxxxB
\else
??????\fi
\fi
}
\newcommand{\hatcurISOmshort}[1]{\ifnum#1=762 %
\hatcurISOmshortxxxxxA
\else
\ifnum#1=4643 %
\hatcurISOmshortxxxxxB
\else
??????\fi
\fi
}
\newcommand{\hatcurISOr}[1]{\ifnum#1=762 %
\hatcurISOrxxxxxA
\else
\ifnum#1=4643 %
\hatcurISOrxxxxxB
\else
??????\fi
\fi
}
\newcommand{\hatcurISOrho}[1]{\ifnum#1=762 %
\hatcurISOrhoxxxxxA
\else
\ifnum#1=4643 %
\hatcurISOrhoxxxxxB
\else
??????\fi
\fi
}
\newcommand{\hatcurISOrholong}[1]{\ifnum#1=762 %
\hatcurISOrholongxxxxxA
\else
\ifnum#1=4643 %
\hatcurISOrholongxxxxxB
\else
??????\fi
\fi
}
\newcommand{\hatcurISOrlong}[1]{\ifnum#1=762 %
\hatcurISOrlongxxxxxA
\else
\ifnum#1=4643 %
\hatcurISOrlongxxxxxB
\else
??????\fi
\fi
}
\newcommand{\hatcurISOrshort}[1]{\ifnum#1=762 %
\hatcurISOrshortxxxxxA
\else
\ifnum#1=4643 %
\hatcurISOrshortxxxxxB
\else
??????\fi
\fi
}
\newcommand{\hatcurISOspec}[1]{\ifnum#1=762 %
\hatcurISOspecxxxxxA
\else
\ifnum#1=4643 %
\hatcurISOspecxxxxxB
\else
??????\fi
\fi
}
\newcommand{\hatcurISOteff}[1]{\ifnum#1=762 %
\hatcurISOteffxxxxxA
\else
\ifnum#1=4643 %
\hatcurISOteffxxxxxB
\else
??????\fi
\fi
}
\newcommand{\hatcurISOzfeh}[1]{\ifnum#1=762 %
\hatcurISOzfehxxxxxA
\else
\ifnum#1=4643 %
\hatcurISOzfehxxxxxB
\else
??????\fi
\fi
}
\newcommand{\hatcurLBiB}[1]{\ifnum#1=762 %
\hatcurLBiBxxxxxA
\else
\ifnum#1=4643 %
\hatcurLBiBxxxxxB
\else
??????\fi
\fi
}
\newcommand{\hatcurLBiC}[1]{\ifnum#1=762 %
\hatcurLBiCxxxxxA
\else
\ifnum#1=4643 %
\hatcurLBiCxxxxxB
\else
??????\fi
\fi
}
\newcommand{\hatcurLBig}[1]{\ifnum#1=762 %
\hatcurLBigxxxxxA
\else
\ifnum#1=4643 %
\hatcurLBigxxxxxB
\else
??????\fi
\fi
}
\newcommand{\hatcurLBiH}[1]{\ifnum#1=762 %
\hatcurLBiHxxxxxA
\else
\ifnum#1=4643 %
\hatcurLBiHxxxxxB
\else
??????\fi
\fi
}
\newcommand{\hatcurLBii}[1]{\ifnum#1=762 %
\hatcurLBiixxxxxA
\else
\ifnum#1=4643 %
\hatcurLBiixxxxxB
\else
??????\fi
\fi
}
\newcommand{\hatcurLBiI}[1]{\ifnum#1=762 %
\hatcurLBiIxxxxxA
\else
\ifnum#1=4643 %
\hatcurLBiIxxxxxB
\else
??????\fi
\fi
}
\newcommand{\hatcurLBiiB}[1]{\ifnum#1=762 %
\hatcurLBiiBxxxxxA
\else
\ifnum#1=4643 %
\hatcurLBiiBxxxxxB
\else
??????\fi
\fi
}
\newcommand{\hatcurLBiiC}[1]{\ifnum#1=762 %
\hatcurLBiiCxxxxxA
\else
\ifnum#1=4643 %
\hatcurLBiiCxxxxxB
\else
??????\fi
\fi
}
\newcommand{\hatcurLBiig}[1]{\ifnum#1=762 %
\hatcurLBiigxxxxxA
\else
\ifnum#1=4643 %
\hatcurLBiigxxxxxB
\else
??????\fi
\fi
}
\newcommand{\hatcurLBiiH}[1]{\ifnum#1=762 %
\hatcurLBiiHxxxxxA
\else
\ifnum#1=4643 %
\hatcurLBiiHxxxxxB
\else
??????\fi
\fi
}
\newcommand{\hatcurLBiii}[1]{\ifnum#1=762 %
\hatcurLBiiixxxxxA
\else
\ifnum#1=4643 %
\hatcurLBiiixxxxxB
\else
??????\fi
\fi
}
\newcommand{\hatcurLBiiI}[1]{\ifnum#1=762 %
\hatcurLBiiIxxxxxA
\else
\ifnum#1=4643 %
\hatcurLBiiIxxxxxB
\else
??????\fi
\fi
}
\newcommand{\hatcurLBiiJ}[1]{\ifnum#1=762 %
\hatcurLBiiJxxxxxA
\else
\ifnum#1=4643 %
\hatcurLBiiJxxxxxB
\else
??????\fi
\fi
}
\newcommand{\hatcurLBiiK}[1]{\ifnum#1=762 %
\hatcurLBiiKxxxxxA
\else
\ifnum#1=4643 %
\hatcurLBiiKxxxxxB
\else
??????\fi
\fi
}
\newcommand{\hatcurLBiikep}[1]{\ifnum#1=762 %
\hatcurLBiikepxxxxxA
\else
\ifnum#1=4643 %
\hatcurLBiikepxxxxxB
\else
??????\fi
\fi
}
\newcommand{\hatcurLBiiM}[1]{\ifnum#1=762 %
\hatcurLBiiMxxxxxA
\else
\ifnum#1=4643 %
\hatcurLBiiMxxxxxB
\else
??????\fi
\fi
}
\newcommand{\hatcurLBiir}[1]{\ifnum#1=762 %
\hatcurLBiirxxxxxA
\else
\ifnum#1=4643 %
\hatcurLBiirxxxxxB
\else
??????\fi
\fi
}
\newcommand{\hatcurLBiiR}[1]{\ifnum#1=762 %
\hatcurLBiiRxxxxxA
\else
\ifnum#1=4643 %
\hatcurLBiiRxxxxxB
\else
??????\fi
\fi
}
\newcommand{\hatcurLBiiSfour}[1]{\ifnum#1=762 %
\hatcurLBiiSfourxxxxxA
\else
\ifnum#1=4643 %
\hatcurLBiiSfourxxxxxB
\else
??????\fi
\fi
}
\newcommand{\hatcurLBiiSone}[1]{\ifnum#1=762 %
\hatcurLBiiSonexxxxxA
\else
\ifnum#1=4643 %
\hatcurLBiiSonexxxxxB
\else
??????\fi
\fi
}
\newcommand{\hatcurLBiiSthree}[1]{\ifnum#1=762 %
\hatcurLBiiSthreexxxxxA
\else
\ifnum#1=4643 %
\hatcurLBiiSthreexxxxxB
\else
??????\fi
\fi
}
\newcommand{\hatcurLBiiStwo}[1]{\ifnum#1=762 %
\hatcurLBiiStwoxxxxxA
\else
\ifnum#1=4643 %
\hatcurLBiiStwoxxxxxB
\else
??????\fi
\fi
}
\newcommand{\hatcurLBiiT}[1]{\ifnum#1=762 %
\hatcurLBiiTxxxxxA
\else
\ifnum#1=4643 %
\hatcurLBiiTxxxxxB
\else
??????\fi
\fi
}
\newcommand{\hatcurLBiiu}[1]{\ifnum#1=762 %
\hatcurLBiiuxxxxxA
\else
\ifnum#1=4643 %
\hatcurLBiiuxxxxxB
\else
??????\fi
\fi
}
\newcommand{\hatcurLBiiV}[1]{\ifnum#1=762 %
\hatcurLBiiVxxxxxA
\else
\ifnum#1=4643 %
\hatcurLBiiVxxxxxB
\else
??????\fi
\fi
}
\newcommand{\hatcurLBiiz}[1]{\ifnum#1=762 %
\hatcurLBiizxxxxxA
\else
\ifnum#1=4643 %
\hatcurLBiizxxxxxB
\else
??????\fi
\fi
}
\newcommand{\hatcurLBiJ}[1]{\ifnum#1=762 %
\hatcurLBiJxxxxxA
\else
\ifnum#1=4643 %
\hatcurLBiJxxxxxB
\else
??????\fi
\fi
}
\newcommand{\hatcurLBiK}[1]{\ifnum#1=762 %
\hatcurLBiKxxxxxA
\else
\ifnum#1=4643 %
\hatcurLBiKxxxxxB
\else
??????\fi
\fi
}
\newcommand{\hatcurLBikep}[1]{\ifnum#1=762 %
\hatcurLBikepxxxxxA
\else
\ifnum#1=4643 %
\hatcurLBikepxxxxxB
\else
??????\fi
\fi
}
\newcommand{\hatcurLBiM}[1]{\ifnum#1=762 %
\hatcurLBiMxxxxxA
\else
\ifnum#1=4643 %
\hatcurLBiMxxxxxB
\else
??????\fi
\fi
}
\newcommand{\hatcurLBir}[1]{\ifnum#1=762 %
\hatcurLBirxxxxxA
\else
\ifnum#1=4643 %
\hatcurLBirxxxxxB
\else
??????\fi
\fi
}
\newcommand{\hatcurLBiR}[1]{\ifnum#1=762 %
\hatcurLBiRxxxxxA
\else
\ifnum#1=4643 %
\hatcurLBiRxxxxxB
\else
??????\fi
\fi
}
\newcommand{\hatcurLBiSfour}[1]{\ifnum#1=762 %
\hatcurLBiSfourxxxxxA
\else
\ifnum#1=4643 %
\hatcurLBiSfourxxxxxB
\else
??????\fi
\fi
}
\newcommand{\hatcurLBiSone}[1]{\ifnum#1=762 %
\hatcurLBiSonexxxxxA
\else
\ifnum#1=4643 %
\hatcurLBiSonexxxxxB
\else
??????\fi
\fi
}
\newcommand{\hatcurLBiSthree}[1]{\ifnum#1=762 %
\hatcurLBiSthreexxxxxA
\else
\ifnum#1=4643 %
\hatcurLBiSthreexxxxxB
\else
??????\fi
\fi
}
\newcommand{\hatcurLBiStwo}[1]{\ifnum#1=762 %
\hatcurLBiStwoxxxxxA
\else
\ifnum#1=4643 %
\hatcurLBiStwoxxxxxB
\else
??????\fi
\fi
}
\newcommand{\hatcurLBiT}[1]{\ifnum#1=762 %
\hatcurLBiTxxxxxA
\else
\ifnum#1=4643 %
\hatcurLBiTxxxxxB
\else
??????\fi
\fi
}
\newcommand{\hatcurLBiu}[1]{\ifnum#1=762 %
\hatcurLBiuxxxxxA
\else
\ifnum#1=4643 %
\hatcurLBiuxxxxxB
\else
??????\fi
\fi
}
\newcommand{\hatcurLBiV}[1]{\ifnum#1=762 %
\hatcurLBiVxxxxxA
\else
\ifnum#1=4643 %
\hatcurLBiVxxxxxB
\else
??????\fi
\fi
}
\newcommand{\hatcurLBiz}[1]{\ifnum#1=762 %
\hatcurLBizxxxxxA
\else
\ifnum#1=4643 %
\hatcurLBizxxxxxB
\else
??????\fi
\fi
}
\newcommand{\hatcurLCbsq}[1]{\ifnum#1=762 %
\hatcurLCbsqxxxxxA
\else
\ifnum#1=4643 %
\hatcurLCbsqxxxxxB
\else
??????\fi
\fi
}
\newcommand{\hatcurLCdip}[1]{\ifnum#1=762 %
\hatcurLCdipxxxxxA
\else
\ifnum#1=4643 %
\hatcurLCdipxxxxxB
\else
??????\fi
\fi
}
\newcommand{\hatcurLCdur}[1]{\ifnum#1=762 %
\hatcurLCdurxxxxxA
\else
\ifnum#1=4643 %
\hatcurLCdurxxxxxB
\else
??????\fi
\fi
}
\newcommand{\hatcurLCdurhr}[1]{\ifnum#1=762 %
\hatcurLCdurhrxxxxxA
\else
\ifnum#1=4643 %
\hatcurLCdurhrxxxxxB
\else
??????\fi
\fi
}
\newcommand{\hatcurLCdurhrshort}[1]{\ifnum#1=762 %
\hatcurLCdurhrshortxxxxxA
\else
\ifnum#1=4643 %
\hatcurLCdurhrshortxxxxxB
\else
??????\fi
\fi
}
\newcommand{\hatcurLCdurshort}[1]{\ifnum#1=762 %
\hatcurLCdurshortxxxxxA
\else
\ifnum#1=4643 %
\hatcurLCdurshortxxxxxB
\else
??????\fi
\fi
}
\newcommand{\hatcurLChatnetmA}[1]{\ifnum#1=762 %
\hatcurLChatnetmAxxxxxA
\else
\ifnum#1=4643 %
\hatcurLChatnetmAxxxxxB
\else
??????\fi
\fi
}
\newcommand{\hatcurLChatnetmB}[1]{\ifnum#1=762 %
\hatcurLChatnetmBxxxxxA
\else
\ifnum#1=4643 %
\hatcurLChatnetmBxxxxxB
\else
??????\fi
\fi
}
\newcommand{\hatcurLChatnetmC}[1]{\ifnum#1=762 %
\hatcurLChatnetmCxxxxxA
\else
??????\fi
}
\newcommand{\hatcurLChatnetmD}[1]{\ifnum#1=762 %
\hatcurLChatnetmDxxxxxA
\else
??????\fi
}
\newcommand{\hatcurLCiblendA}[1]{\ifnum#1=762 %
\hatcurLCiblendAxxxxxA
\else
\ifnum#1=4643 %
\hatcurLCiblendAxxxxxB
\else
??????\fi
\fi
}
\newcommand{\hatcurLCiblendB}[1]{\ifnum#1=762 %
\hatcurLCiblendBxxxxxA
\else
\ifnum#1=4643 %
\hatcurLCiblendBxxxxxB
\else
??????\fi
\fi
}
\newcommand{\hatcurLCiblendC}[1]{\ifnum#1=762 %
\hatcurLCiblendCxxxxxA
\else
??????\fi
}
\newcommand{\hatcurLCiblendD}[1]{\ifnum#1=762 %
\hatcurLCiblendDxxxxxA
\else
??????\fi
}
\newcommand{\hatcurLCimp}[1]{\ifnum#1=762 %
\hatcurLCimpxxxxxA
\else
\ifnum#1=4643 %
\hatcurLCimpxxxxxB
\else
??????\fi
\fi
}
\newcommand{\hatcurLCingdur}[1]{\ifnum#1=762 %
\hatcurLCingdurxxxxxA
\else
\ifnum#1=4643 %
\hatcurLCingdurxxxxxB
\else
??????\fi
\fi
}
\newcommand{\hatcurLCP}[1]{\ifnum#1=762 %
\hatcurLCPxxxxxA
\else
\ifnum#1=4643 %
\hatcurLCPxxxxxB
\else
??????\fi
\fi
}
\newcommand{\hatcurLCPprec}[1]{\ifnum#1=762 %
\hatcurLCPprecxxxxxA
\else
\ifnum#1=4643 %
\hatcurLCPprecxxxxxB
\else
??????\fi
\fi
}
\newcommand{\hatcurLCPshort}[1]{\ifnum#1=762 %
\hatcurLCPshortxxxxxA
\else
\ifnum#1=4643 %
\hatcurLCPshortxxxxxB
\else
??????\fi
\fi
}
\newcommand{\hatcurLCq}[1]{\ifnum#1=762 %
\hatcurLCqxxxxxA
\else
\ifnum#1=4643 %
\hatcurLCqxxxxxB
\else
??????\fi
\fi
}
\newcommand{\hatcurLCqshort}[1]{\ifnum#1=762 %
\hatcurLCqshortxxxxxA
\else
\ifnum#1=4643 %
\hatcurLCqshortxxxxxB
\else
??????\fi
\fi
}
\newcommand{\hatcurLCrho}[1]{\ifnum#1=762 %
\hatcurLCrhoxxxxxA
\else
\ifnum#1=4643 %
\hatcurLCrhoxxxxxB
\else
??????\fi
\fi
}
\newcommand{\hatcurLCrprstar}[1]{\ifnum#1=762 %
\hatcurLCrprstarxxxxxA
\else
\ifnum#1=4643 %
\hatcurLCrprstarxxxxxB
\else
??????\fi
\fi
}
\newcommand{\hatcurLCT}[1]{\ifnum#1=762 %
\hatcurLCTxxxxxA
\else
\ifnum#1=4643 %
\hatcurLCTxxxxxB
\else
??????\fi
\fi
}
\newcommand{\hatcurLCTA}[1]{\ifnum#1=762 %
\hatcurLCTAxxxxxA
\else
\ifnum#1=4643 %
\hatcurLCTAxxxxxB
\else
??????\fi
\fi
}
\newcommand{\hatcurLCTB}[1]{\ifnum#1=762 %
\hatcurLCTBxxxxxA
\else
\ifnum#1=4643 %
\hatcurLCTBxxxxxB
\else
??????\fi
\fi
}
\newcommand{\hatcurLCzeta}[1]{\ifnum#1=762 %
\hatcurLCzetaxxxxxA
\else
\ifnum#1=4643 %
\hatcurLCzetaxxxxxB
\else
??????\fi
\fi
}
\newcommand{\hatcurPPaequiv}[1]{\ifnum#1=762 %
\hatcurPPaequivxxxxxA
\else
\ifnum#1=4643 %
\hatcurPPaequivxxxxxB
\else
??????\fi
\fi
}
\newcommand{\hatcurPPar}[1]{\ifnum#1=762 %
\hatcurPParxxxxxA
\else
\ifnum#1=4643 %
\hatcurPParxxxxxB
\else
??????\fi
\fi
}
\newcommand{\hatcurPParel}[1]{\ifnum#1=762 %
\hatcurPParelxxxxxA
\else
\ifnum#1=4643 %
\hatcurPParelxxxxxB
\else
??????\fi
\fi
}
\newcommand{\hatcurPPfluxap}[1]{\ifnum#1=762 %
\hatcurPPfluxapxxxxxA
\else
\ifnum#1=4643 %
\hatcurPPfluxapxxxxxB
\else
??????\fi
\fi
}
\newcommand{\hatcurPPfluxapdim}[1]{\ifnum#1=762 %
\hatcurPPfluxapdimxxxxxA
\else
\ifnum#1=4643 %
\hatcurPPfluxapdimxxxxxB
\else
??????\fi
\fi
}
\newcommand{\hatcurPPfluxavg}[1]{\ifnum#1=762 %
\hatcurPPfluxavgxxxxxA
\else
\ifnum#1=4643 %
\hatcurPPfluxavgxxxxxB
\else
??????\fi
\fi
}
\newcommand{\hatcurPPfluxavgdim}[1]{\ifnum#1=762 %
\hatcurPPfluxavgdimxxxxxA
\else
\ifnum#1=4643 %
\hatcurPPfluxavgdimxxxxxB
\else
??????\fi
\fi
}
\newcommand{\hatcurPPfluxavglog}[1]{\ifnum#1=762 %
\hatcurPPfluxavglogxxxxxA
\else
\ifnum#1=4643 %
\hatcurPPfluxavglogxxxxxB
\else
??????\fi
\fi
}
\newcommand{\hatcurPPfluxperi}[1]{\ifnum#1=762 %
\hatcurPPfluxperixxxxxA
\else
\ifnum#1=4643 %
\hatcurPPfluxperixxxxxB
\else
??????\fi
\fi
}
\newcommand{\hatcurPPfluxperidim}[1]{\ifnum#1=762 %
\hatcurPPfluxperidimxxxxxA
\else
\ifnum#1=4643 %
\hatcurPPfluxperidimxxxxxB
\else
??????\fi
\fi
}
\newcommand{\hatcurPPg}[1]{\ifnum#1=762 %
\hatcurPPgxxxxxA
\else
\ifnum#1=4643 %
\hatcurPPgxxxxxB
\else
??????\fi
\fi
}
\newcommand{\hatcurPPi}[1]{\ifnum#1=762 %
\hatcurPPixxxxxA
\else
\ifnum#1=4643 %
\hatcurPPixxxxxB
\else
??????\fi
\fi
}
\newcommand{\hatcurPPlogg}[1]{\ifnum#1=762 %
\hatcurPPloggxxxxxA
\else
\ifnum#1=4643 %
\hatcurPPloggxxxxxB
\else
??????\fi
\fi
}
\newcommand{\hatcurPPm}[1]{\ifnum#1=762 %
\hatcurPPmxxxxxA
\else
\ifnum#1=4643 %
\hatcurPPmxxxxxB
\else
??????\fi
\fi
}
\newcommand{\hatcurPPme}[1]{\ifnum#1=762 %
\hatcurPPmexxxxxA
\else
\ifnum#1=4643 %
\hatcurPPmexxxxxB
\else
??????\fi
\fi
}
\newcommand{\hatcurPPmelong}[1]{\ifnum#1=762 %
\hatcurPPmelongxxxxxA
\else
\ifnum#1=4643 %
\hatcurPPmelongxxxxxB
\else
??????\fi
\fi
}
\newcommand{\hatcurPPmeshort}[1]{\ifnum#1=762 %
\hatcurPPmeshortxxxxxA
\else
\ifnum#1=4643 %
\hatcurPPmeshortxxxxxB
\else
??????\fi
\fi
}
\newcommand{\hatcurPPmlong}[1]{\ifnum#1=762 %
\hatcurPPmlongxxxxxA
\else
\ifnum#1=4643 %
\hatcurPPmlongxxxxxB
\else
??????\fi
\fi
}
\newcommand{\hatcurPPmrcorr}[1]{\ifnum#1=762 %
\hatcurPPmrcorrxxxxxA
\else
\ifnum#1=4643 %
\hatcurPPmrcorrxxxxxB
\else
??????\fi
\fi
}
\newcommand{\hatcurPPmshort}[1]{\ifnum#1=762 %
\hatcurPPmshortxxxxxA
\else
\ifnum#1=4643 %
\hatcurPPmshortxxxxxB
\else
??????\fi
\fi
}
\newcommand{\hatcurPPmtwosiglim}[1]{\ifnum#1=762 %
\hatcurPPmtwosiglimxxxxxA
\else
\ifnum#1=4643 %
\hatcurPPmtwosiglimxxxxxB
\else
??????\fi
\fi
}
\newcommand{\hatcurPPperi}[1]{\ifnum#1=762 %
\hatcurPPperixxxxxA
\else
\ifnum#1=4643 %
\hatcurPPperixxxxxB
\else
??????\fi
\fi
}
\newcommand{\hatcurPPphiconj}[1]{\ifnum#1=762 %
\hatcurPPphiconjxxxxxA
\else
\ifnum#1=4643 %
\hatcurPPphiconjxxxxxB
\else
??????\fi
\fi
}
\newcommand{\hatcurPPr}[1]{\ifnum#1=762 %
\hatcurPPrxxxxxA
\else
\ifnum#1=4643 %
\hatcurPPrxxxxxB
\else
??????\fi
\fi
}
\newcommand{\hatcurPPre}[1]{\ifnum#1=762 %
\hatcurPPrexxxxxA
\else
\ifnum#1=4643 %
\hatcurPPrexxxxxB
\else
??????\fi
\fi
}
\newcommand{\hatcurPPrelong}[1]{\ifnum#1=762 %
\hatcurPPrelongxxxxxA
\else
\ifnum#1=4643 %
\hatcurPPrelongxxxxxB
\else
??????\fi
\fi
}
\newcommand{\hatcurPPreshort}[1]{\ifnum#1=762 %
\hatcurPPreshortxxxxxA
\else
\ifnum#1=4643 %
\hatcurPPreshortxxxxxB
\else
??????\fi
\fi
}
\newcommand{\hatcurPPrho}[1]{\ifnum#1=762 %
\hatcurPPrhoxxxxxA
\else
\ifnum#1=4643 %
\hatcurPPrhoxxxxxB
\else
??????\fi
\fi
}
\newcommand{\hatcurPPrlong}[1]{\ifnum#1=762 %
\hatcurPPrlongxxxxxA
\else
\ifnum#1=4643 %
\hatcurPPrlongxxxxxB
\else
??????\fi
\fi
}
\newcommand{\hatcurPPrshort}[1]{\ifnum#1=762 %
\hatcurPPrshortxxxxxA
\else
\ifnum#1=4643 %
\hatcurPPrshortxxxxxB
\else
??????\fi
\fi
}
\newcommand{\hatcurPPtcirc}[1]{\ifnum#1=762 %
\hatcurPPtcircxxxxxA
\else
\ifnum#1=4643 %
\hatcurPPtcircxxxxxB
\else
??????\fi
\fi
}
\newcommand{\hatcurPPteff}[1]{\ifnum#1=762 %
\hatcurPPteffxxxxxA
\else
\ifnum#1=4643 %
\hatcurPPteffxxxxxB
\else
??????\fi
\fi
}
\newcommand{\hatcurPPtheta}[1]{\ifnum#1=762 %
\hatcurPPthetaxxxxxA
\else
\ifnum#1=4643 %
\hatcurPPthetaxxxxxB
\else
??????\fi
\fi
}
\newcommand{\hatcurPPtinfall}[1]{\ifnum#1=762 %
\hatcurPPtinfallxxxxxA
\else
\ifnum#1=4643 %
\hatcurPPtinfallxxxxxB
\else
??????\fi
\fi
}
\newcommand{\hatcurRVeccen}[1]{\ifnum#1=762 %
\hatcurRVeccenxxxxxA
\else
\ifnum#1=4643 %
\hatcurRVeccenxxxxxB
\else
??????\fi
\fi
}
\newcommand{\hatcurRVeccentwosiglim}[1]{\ifnum#1=762 %
\hatcurRVeccentwosiglimxxxxxA
\else
\ifnum#1=4643 %
\hatcurRVeccentwosiglimxxxxxB
\else
??????\fi
\fi
}
\newcommand{\hatcurRVfitrms}[1]{\ifnum#1=4643 %
\hatcurRVfitrmsxxxxxB
\else
??????\fi
}
\newcommand{\hatcurRVfitrmsA}[1]{\ifnum#1=762 %
\hatcurRVfitrmsAxxxxxA
\else
??????\fi
}
\newcommand{\hatcurRVfitrmsB}[1]{\ifnum#1=762 %
\hatcurRVfitrmsBxxxxxA
\else
??????\fi
}
\newcommand{\hatcurRVgamma}[1]{\ifnum#1=4643 %
\hatcurRVgammaxxxxxB
\else
??????\fi
}
\newcommand{\hatcurRVgammaA}[1]{\ifnum#1=762 %
\hatcurRVgammaAxxxxxA
\else
??????\fi
}
\newcommand{\hatcurRVgammaB}[1]{\ifnum#1=762 %
\hatcurRVgammaBxxxxxA
\else
??????\fi
}
\newcommand{\hatcurRVh}[1]{\ifnum#1=762 %
\hatcurRVhxxxxxA
\else
\ifnum#1=4643 %
\hatcurRVhxxxxxB
\else
??????\fi
\fi
}
\newcommand{\hatcurRVjitter}[1]{\ifnum#1=4643 %
\hatcurRVjitterxxxxxB
\else
??????\fi
}
\newcommand{\hatcurRVjitterA}[1]{\ifnum#1=762 %
\hatcurRVjitterAxxxxxA
\else
??????\fi
}
\newcommand{\hatcurRVjitterB}[1]{\ifnum#1=762 %
\hatcurRVjitterBxxxxxA
\else
??????\fi
}
\newcommand{\hatcurRVjittertwosiglim}[1]{\ifnum#1=4643 %
\hatcurRVjittertwosiglimxxxxxB
\else
??????\fi
}
\newcommand{\hatcurRVjittertwosiglimA}[1]{\ifnum#1=762 %
\hatcurRVjittertwosiglimAxxxxxA
\else
??????\fi
}
\newcommand{\hatcurRVjittertwosiglimB}[1]{\ifnum#1=762 %
\hatcurRVjittertwosiglimBxxxxxA
\else
??????\fi
}
\newcommand{\hatcurRVk}[1]{\ifnum#1=762 %
\hatcurRVkxxxxxA
\else
\ifnum#1=4643 %
\hatcurRVkxxxxxB
\else
??????\fi
\fi
}
\newcommand{\hatcurRVK}[1]{\ifnum#1=762 %
\hatcurRVKxxxxxA
\else
\ifnum#1=4643 %
\hatcurRVKxxxxxB
\else
??????\fi
\fi
}
\newcommand{\hatcurRVKtwosiglim}[1]{\ifnum#1=762 %
\hatcurRVKtwosiglimxxxxxA
\else
\ifnum#1=4643 %
\hatcurRVKtwosiglimxxxxxB
\else
??????\fi
\fi
}
\newcommand{\hatcurRVomega}[1]{\ifnum#1=762 %
\hatcurRVomegaxxxxxA
\else
\ifnum#1=4643 %
\hatcurRVomegaxxxxxB
\else
??????\fi
\fi
}
\newcommand{\hatcurRVrh}[1]{\ifnum#1=762 %
\hatcurRVrhxxxxxA
\else
\ifnum#1=4643 %
\hatcurRVrhxxxxxB
\else
??????\fi
\fi
}
\newcommand{\hatcurRVrk}[1]{\ifnum#1=762 %
\hatcurRVrkxxxxxA
\else
\ifnum#1=4643 %
\hatcurRVrkxxxxxB
\else
??????\fi
\fi
}
\newcommand{\hatcurRVtrone}[1]{\ifnum#1=762 %
\hatcurRVtronexxxxxA
\else
\ifnum#1=4643 %
\hatcurRVtronexxxxxB
\else
??????\fi
\fi
}
\newcommand{\hatcurRVtrtwo}[1]{\ifnum#1=762 %
\hatcurRVtrtwoxxxxxA
\else
\ifnum#1=4643 %
\hatcurRVtrtwoxxxxxB
\else
??????\fi
\fi
}
\newcommand{\hatcurSMEilogg}[1]{\ifnum#1=762 %
\hatcurSMEiloggxxxxxA
\else
\ifnum#1=4643 %
\hatcurSMEiloggxxxxxB
\else
??????\fi
\fi
}
\newcommand{\hatcurSMEiteff}[1]{\ifnum#1=762 %
\hatcurSMEiteffxxxxxA
\else
\ifnum#1=4643 %
\hatcurSMEiteffxxxxxB
\else
??????\fi
\fi
}
\newcommand{\hatcurSMEivmac}[1]{\ifnum#1=762 %
\hatcurSMEivmacxxxxxA
\else
\ifnum#1=4643 %
\hatcurSMEivmacxxxxxB
\else
??????\fi
\fi
}
\newcommand{\hatcurSMEivmic}[1]{\ifnum#1=762 %
\hatcurSMEivmicxxxxxA
\else
\ifnum#1=4643 %
\hatcurSMEivmicxxxxxB
\else
??????\fi
\fi
}
\newcommand{\hatcurSMEivsin}[1]{\ifnum#1=762 %
\hatcurSMEivsinxxxxxA
\else
\ifnum#1=4643 %
\hatcurSMEivsinxxxxxB
\else
??????\fi
\fi
}
\newcommand{\hatcurSMEizfeh}[1]{\ifnum#1=762 %
\hatcurSMEizfehxxxxxA
\else
\ifnum#1=4643 %
\hatcurSMEizfehxxxxxB
\else
??????\fi
\fi
}
\newcommand{\hatcurSMEizfehshort}[1]{\ifnum#1=762 %
\hatcurSMEizfehshortxxxxxA
\else
\ifnum#1=4643 %
\hatcurSMEizfehshortxxxxxB
\else
??????\fi
\fi
}
\newcommand{\hatcurXAv}[1]{\ifnum#1=762 %
\hatcurXAvxxxxxA
\else
\ifnum#1=4643 %
\hatcurXAvxxxxxB
\else
??????\fi
\fi
}
\newcommand{\hatcurXdist}[1]{\ifnum#1=762 %
\hatcurXdistxxxxxA
\else
\ifnum#1=4643 %
\hatcurXdistxxxxxB
\else
??????\fi
\fi
}
\newcommand{\hatcurXdistred}[1]{\ifnum#1=762 %
\hatcurXdistredxxxxxA
\else
\ifnum#1=4643 %
\hatcurXdistredxxxxxB
\else
??????\fi
\fi
}
\newcommand{\hatcurXEBV}[1]{\ifnum#1=762 %
\hatcurXEBVxxxxxA
\else
\ifnum#1=4643 %
\hatcurXEBVxxxxxB
\else
??????\fi
\fi
}
\newcommand{\hatcurXsecdur}[1]{\ifnum#1=762 %
\hatcurXsecdurxxxxxA
\else
\ifnum#1=4643 %
\hatcurXsecdurxxxxxB
\else
??????\fi
\fi
}
\newcommand{\hatcurXsecingdur}[1]{\ifnum#1=762 %
\hatcurXsecingdurxxxxxA
\else
\ifnum#1=4643 %
\hatcurXsecingdurxxxxxB
\else
??????\fi
\fi
}
\newcommand{\hatcurXsecondary}[1]{\ifnum#1=762 %
\hatcurXsecondaryxxxxxA
\else
\ifnum#1=4643 %
\hatcurXsecondaryxxxxxB
\else
??????\fi
\fi
}
\newcommand{\hatcurXsecphase}[1]{\ifnum#1=762 %
\hatcurXsecphasexxxxxA
\else
\ifnum#1=4643 %
\hatcurXsecphasexxxxxB
\else
??????\fi
\fi
}
\newcommand{\hatcurhtreccenxxxxxA}{TOI762}                             
\newcommand{\hatcurfieldeccenxxxxxA}{\ensuremath{string}}              
\newcommand{\hatcurCCraeccenxxxxxA}{\ensuremath{11^{\mathrm h}04^{\mathrm m}18.1831{\mathrm s}}}                     
\newcommand{\hatcurCCdececcenxxxxxA}{\ensuremath{-47{\arcdeg}49{\arcmin}17.0030{\arcsec}}}                   
\newcommand{\hatcurCCmageccenxxxxxA}{NULL}                             
\newcommand{\hatcurCCtwomasseccenxxxxxA}{2MASS~11041818-4749169}       
\newcommand{\hatcurCCgsceccenxxxxxA}{GSC~NULL}                         
\newcommand{\hatcurCCgaiaeccenxxxxxA}{GAIA~5362352740204840832}        
\newcommand{\hatcurCCgaiadrtwoeccenxxxxxA}{GAIA~DR2~5362352744504000256} 
\newcommand{\hatcurCCtassmveccenxxxxxA}{\ensuremath{nff\pmnff}}        
\newcommand{\hatcurCCtassmvshorteccenxxxxxA}{\ensuremath{0.0}}         
\newcommand{\hatcurCCtassmBeccenxxxxxA}{\ensuremath{nff\pmnff}}        
\newcommand{\hatcurCCtassmBshorteccenxxxxxA}{\ensuremath{0.0}}         
\newcommand{\hatcurCCtassmIeccenxxxxxA}{\ensuremath{nff\pmnff}}        
\newcommand{\hatcurCCtassmIshorteccenxxxxxA}{\ensuremath{0.0}}         
\newcommand{\hatcurCCtassmgeccenxxxxxA}{\ensuremath{nff\pmnff}}        
\newcommand{\hatcurCCtassmgshorteccenxxxxxA}{\ensuremath{0.0}}         
\newcommand{\hatcurCCtassmreccenxxxxxA}{\ensuremath{nff\pmnff}}        
\newcommand{\hatcurCCtassmrshorteccenxxxxxA}{\ensuremath{0.0}}         
\newcommand{\hatcurCCtassmieccenxxxxxA}{\ensuremath{nff\pmnff}}        
\newcommand{\hatcurCCtassmishorteccenxxxxxA}{\ensuremath{0.0}}         
\newcommand{\hatcurCCparallaxeccenxxxxxA}{\ensuremath{10.118\pm0.023}} 
\newcommand{\hatcurCCgaiamGeccenxxxxxA}{\ensuremath{14.9297\pm0.0028}} 
\newcommand{\hatcurCCgaiamBPeccenxxxxxA}{\ensuremath{0\pm0}}           
\newcommand{\hatcurCCgaiamRPeccenxxxxxA}{\ensuremath{0\pm0}}           
\newcommand{\hatcurCCtwomassJmageccenxxxxxA}{\ensuremath{12.099\pm0.080}} 
\newcommand{\hatcurCCtwomassHmageccenxxxxxA}{\ensuremath{11.444\pm0.077}} 
\newcommand{\hatcurCCtwomassKmageccenxxxxxA}{\ensuremath{11.187\pm0.082}} 
\newcommand{\hatcurCCcitJmageccenxxxxxA}{\ensuremath{11.987\pm0.027}}  
\newcommand{\hatcurCCcitHmageccenxxxxxA}{\ensuremath{11.338\pm0.029}}  
\newcommand{\hatcurCCcitKmageccenxxxxxA}{\ensuremath{11.108\pm0.028}}  
\newcommand{\hatcurCCbbJmageccenxxxxxA}{\ensuremath{12.081\pm0.030}}   
\newcommand{\hatcurCCbbHmageccenxxxxxA}{\ensuremath{11.365\pm0.030}}   
\newcommand{\hatcurCCbbKmageccenxxxxxA}{\ensuremath{11.128\pm0.028}}   
\newcommand{\hatcurCCesoJmageccenxxxxxA}{\ensuremath{12.090\pm0.034}}  
\newcommand{\hatcurCCesoHmageccenxxxxxA}{\ensuremath{11.366\pm0.048}}  
\newcommand{\hatcurCCesoKmageccenxxxxxA}{\ensuremath{11.124\pm0.029}}  
\newcommand{\hatcurCCesoJHmageccenxxxxxA}{\ensuremath{0.724\pm0.055}}  
\newcommand{\hatcurCCesoJKmageccenxxxxxA}{\ensuremath{0.966\pm0.043}}  
\newcommand{\hatcurCCesoHKmageccenxxxxxA}{\ensuremath{0.242\pm0.055}}  
\newcommand{\hatcurCCWonemageccenxxxxxA}{\ensuremath{10.991\pm0.088}}  
\newcommand{\hatcurCCWtwomageccenxxxxxA}{\ensuremath{10.890\pm0.094}}  
\newcommand{\hatcurCCWthreemageccenxxxxxA}{\ensuremath{10.71\pm0.12}}  
\newcommand{\hatcurCCWfourmageccenxxxxxA}{\ensuremath{nff\pmnff}}      
\newcommand{\hatcurLCdipeccenxxxxxA}{\ensuremath{0.1}}                 
\newcommand{\hatcurLCrprstareccenxxxxxA}{\ensuremath{0.1795\pm0.0011}} 
\newcommand{\hatcurLCbsqeccenxxxxxA}{\ensuremath{0.573_{-0.011}^{+0.010}}} 
\newcommand{\hatcurLCimpeccenxxxxxA}{\ensuremath{0.7566_{-0.0074}^{+0.0066}}} 
\newcommand{\hatcurLCzetaeccenxxxxxA}{\ensuremath{47.57\pm0.31}}       
\newcommand{\hatcurLCdureccenxxxxxA}{\ensuremath{0.05819\pm0.00039}}   
\newcommand{\hatcurLCdurshorteccenxxxxxA}{\ensuremath{0.0582}}         
\newcommand{\hatcurLCdurhreccenxxxxxA}{\ensuremath{1.3966\pm0.0093}}   
\newcommand{\hatcurLCdurhrshorteccenxxxxxA}{\ensuremath{1.397}}        
\newcommand{\hatcurLCqeccenxxxxxA}{\ensuremath{0.01680\pm0.00012}}     
\newcommand{\hatcurLCqshorteccenxxxxxA}{\ensuremath{0.017}}            
\newcommand{\hatcurLCingdureccenxxxxxA}{\ensuremath{0.01892\pm0.00058}} 
\newcommand{\hatcurLCPeccenxxxxxA}{\ensuremath{3.47168232\pm0.00000074}} 
\newcommand{\hatcurLCPprececcenxxxxxA}{\ensuremath{3.4716823}}         
\newcommand{\hatcurLCPshorteccenxxxxxA}{\ensuremath{3.4717}}           
\newcommand{\hatcurLCTeccenxxxxxA}{\ensuremath{2459874.560250\pm0.000090}} 
\newcommand{\hatcurLCTAeccenxxxxxA}{\ensuremath{2458572.67937\pm0.00028}} 
\newcommand{\hatcurLCTBeccenxxxxxA}{\ensuremath{2460058.55942\pm0.00010}} 
\newcommand{\hatcurLChatnetmAeccenxxxxxA}{\ensuremath{-6.77083\pm0.00010}} 
\newcommand{\hatcurLCiblendAeccenxxxxxA}{\ensuremath{0.926\pm0.028}}   
\newcommand{\hatcurLChatnetmBeccenxxxxxA}{\ensuremath{-6.80726\pm0.00010}} 
\newcommand{\hatcurLCiblendBeccenxxxxxA}{\ensuremath{0.974\pm0.021}}   
\newcommand{\hatcurLChatnetmCeccenxxxxxA}{\ensuremath{-6.69633\pm0.00011}} 
\newcommand{\hatcurLCiblendCeccenxxxxxA}{\ensuremath{1.004\pm0.032}}   
\newcommand{\hatcurLChatnetmDeccenxxxxxA}{\ensuremath{-6.844570\pm0.000093}} 
\newcommand{\hatcurLCiblendDeccenxxxxxA}{\ensuremath{0.854\pm0.029}}   
\newcommand{\hatcurLCrhoeccenxxxxxA}{\ensuremath{8.31_{-0.31}^{+0.41}}} 
\newcommand{\hatcurSMEiteffeccenxxxxxA}{\ensuremath{3150\pm67}}        
\newcommand{\hatcurSMEizfeheccenxxxxxA}{\ensuremath{0.24\pm0.10}}      
\newcommand{\hatcurSMEizfehshorteccenxxxxxA}{\ensuremath{0.24}}        
\newcommand{\hatcurSMEiloggeccenxxxxxA}{\ensuremath{5.00\pm0.50}}      
\newcommand{\hatcurSMEivsineccenxxxxxA}{\ensuremath{0\pm50}}           
\newcommand{\hatcurSMEivmaceccenxxxxxA}{\ensuremath{nff\pmnff}}        
\newcommand{\hatcurSMEivmiceccenxxxxxA}{\ensuremath{nff\pmnff}}        
\newcommand{\hatcurextraerrMJeccenxxxxxA}{\ensuremath{0\pm0}}          
\newcommand{\hatcurextraerrMJtwosiglimeccenxxxxxA}{\ensuremath{<0.0200}} 
\newcommand{\hatcurextraerrMHeccenxxxxxA}{\ensuremath{0\pm0}}          
\newcommand{\hatcurextraerrMHtwosiglimeccenxxxxxA}{\ensuremath{<0.0200}} 
\newcommand{\hatcurextraerrMKseccenxxxxxA}{\ensuremath{0\pm0}}         
\newcommand{\hatcurextraerrMKstwosiglimeccenxxxxxA}{\ensuremath{<0.0200}} 
\newcommand{\hatcurextraerrMGeccenxxxxxA}{\ensuremath{0\pm0}}          
\newcommand{\hatcurextraerrMGtwosiglimeccenxxxxxA}{\ensuremath{<0.0200}} 
\newcommand{\hatcurextraerrMWoneeccenxxxxxA}{\ensuremath{0\pm0}}       
\newcommand{\hatcurextraerrMWonetwosiglimeccenxxxxxA}{\ensuremath{<0.0200}} 
\newcommand{\hatcurextraerrMWtwoeccenxxxxxA}{\ensuremath{0\pm0}}       
\newcommand{\hatcurextraerrMWtwotwosiglimeccenxxxxxA}{\ensuremath{<0.0200}} 
\newcommand{\hatcurLBiBeccenxxxxxA}{\ensuremath{0.5259}}               
\newcommand{\hatcurLBiiBeccenxxxxxA}{\ensuremath{0.3296}}              
\newcommand{\hatcurLBiVeccenxxxxxA}{\ensuremath{0.5154}}               
\newcommand{\hatcurLBiiVeccenxxxxxA}{\ensuremath{0.3424}}              
\newcommand{\hatcurLBiReccenxxxxxA}{\ensuremath{0.4326}}               
\newcommand{\hatcurLBiiReccenxxxxxA}{\ensuremath{0.3323}}              
\newcommand{\hatcurLBiIeccenxxxxxA}{\ensuremath{0.42\pm0.14}}          
\newcommand{\hatcurLBiiIeccenxxxxxA}{\ensuremath{0.18\pm0.15}}         
\newcommand{\hatcurLBiueccenxxxxxA}{\ensuremath{0.4363}}               
\newcommand{\hatcurLBiiueccenxxxxxA}{\ensuremath{0.3435}}              
\newcommand{\hatcurLBigeccenxxxxxA}{\ensuremath{0.57\pm0.13}}          
\newcommand{\hatcurLBiigeccenxxxxxA}{\ensuremath{0.20\pm0.15}}         
\newcommand{\hatcurLBireccenxxxxxA}{\ensuremath{0.44\pm0.12}}          
\newcommand{\hatcurLBiireccenxxxxxA}{\ensuremath{0.30\pm0.14}}         
\newcommand{\hatcurLBiieccenxxxxxA}{\ensuremath{0.15\pm0.10}}          
\newcommand{\hatcurLBiiieccenxxxxxA}{\ensuremath{0.18\pm0.12}}         
\newcommand{\hatcurLBizeccenxxxxxA}{\ensuremath{0.16\pm0.10}}          
\newcommand{\hatcurLBiizeccenxxxxxA}{\ensuremath{0.13\pm0.13}}         
\newcommand{\hatcurLBiJeccenxxxxxA}{\ensuremath{0.057_{-0.042}^{+0.083}}} 
\newcommand{\hatcurLBiiJeccenxxxxxA}{\ensuremath{-0.001\pm0.085}}      
\newcommand{\hatcurLBiHeccenxxxxxA}{\ensuremath{0.0936}}               
\newcommand{\hatcurLBiiHeccenxxxxxA}{\ensuremath{0.2839}}              
\newcommand{\hatcurLBiKeccenxxxxxA}{\ensuremath{0.0865}}               
\newcommand{\hatcurLBiiKeccenxxxxxA}{\ensuremath{0.2452}}              
\newcommand{\hatcurLBiTeccenxxxxxA}{\ensuremath{0.15\pm0.10}}          
\newcommand{\hatcurLBiiTeccenxxxxxA}{\ensuremath{0.06\pm0.13}}         
\newcommand{\hatcurLBikepeccenxxxxxA}{\ensuremath{0.2943}}             
\newcommand{\hatcurLBiikepeccenxxxxxA}{\ensuremath{0.4341}}            
\newcommand{\hatcurLBiCeccenxxxxxA}{\ensuremath{0.2393}}               
\newcommand{\hatcurLBiiCeccenxxxxxA}{\ensuremath{0.4400}}              
\newcommand{\hatcurLBiMeccenxxxxxA}{\ensuremath{0.4171}}               
\newcommand{\hatcurLBiiMeccenxxxxxA}{\ensuremath{0.4013}}              
\newcommand{\hatcurLBiSoneeccenxxxxxA}{\ensuremath{0.0660}}              
\newcommand{\hatcurLBiiSoneeccenxxxxxA}{\ensuremath{0.1843}}             
\newcommand{\hatcurLBiStwoeccenxxxxxA}{\ensuremath{0.0490}}              
\newcommand{\hatcurLBiiStwoeccenxxxxxA}{\ensuremath{0.1551}}             
\newcommand{\hatcurLBiSthreeeccenxxxxxA}{\ensuremath{0.0533}}              
\newcommand{\hatcurLBiiSthreeeccenxxxxxA}{\ensuremath{0.1380}}             
\newcommand{\hatcurLBiSfoureccenxxxxxA}{\ensuremath{0.0551}}              
\newcommand{\hatcurLBiiSfoureccenxxxxxA}{\ensuremath{0.1075}}             
\newcommand{\hatcurISOmeccenxxxxxA}{\ensuremath{0.430\pm0.020}}        
\newcommand{\hatcurISOmshorteccenxxxxxA}{\ensuremath{0.43}}            
\newcommand{\hatcurISOmlongeccenxxxxxA}{\ensuremath{0.430\pm0.020}}    
\newcommand{\hatcurISOreccenxxxxxA}{\ensuremath{0.4169\pm0.0087}}      
\newcommand{\hatcurISOrshorteccenxxxxxA}{\ensuremath{0.42}}            
\newcommand{\hatcurISOrlongeccenxxxxxA}{\ensuremath{0.4169\pm0.0087}}  
\newcommand{\hatcurISOrhoeccenxxxxxA}{\ensuremath{8.31_{-0.31}^{+0.41}}} 
\newcommand{\hatcurISOrholongeccenxxxxxA}{\ensuremath{8.31_{-0.31}^{+0.41}}} 
\newcommand{\hatcurISOloggeccenxxxxxA}{\ensuremath{4.829\pm0.014}}     
\newcommand{\hatcurISOlumeccenxxxxxA}{\ensuremath{0.01762\pm0.00085}}  
\newcommand{\hatcurISOlumshorteccenxxxxxA}{\ensuremath{0.02}}          
\newcommand{\hatcurISOteffeccenxxxxxA}{\ensuremath{3255.7_{-9.9}^{+14.1}}} 
\newcommand{\hatcurISOzfeheccenxxxxxA}{\ensuremath{0.316\pm0.072}}     
\newcommand{\hatcurISOageeccenxxxxxA}{\ensuremath{5.9_{-5.1}^{+8.5}}}  
\newcommand{\hatcurISOspececcenxxxxxA}{M}                              
\newcommand{\hatcurRVKeccenxxxxxA}{\ensuremath{60.2\pm9.2}}            
\newcommand{\hatcurRVKtwosiglimeccenxxxxxA}{\ensuremath{<76.0}}        
\newcommand{\hatcurRVrkeccenxxxxxA}{\ensuremath{0.12\pm0.11}}          
\newcommand{\hatcurRVrheccenxxxxxA}{\ensuremath{-0.085\pm0.082}}       
\newcommand{\hatcurRVkeccenxxxxxA}{\ensuremath{0.021_{-0.020}^{+0.030}}} 
\newcommand{\hatcurRVheccenxxxxxA}{\ensuremath{-0.014_{-0.021}^{+0.015}}} 
\newcommand{\hatcurRVtroneeccenxxxxxA}{\ensuremath{0\pm0}}             
\newcommand{\hatcurRVtrtwoeccenxxxxxA}{\ensuremath{0\pm0}}             
\newcommand{\hatcurRVecceneccenxxxxxA}{\ensuremath{0.032\pm0.026}}     
\newcommand{\hatcurRVeccentwosiglimeccenxxxxxA}{\ensuremath{<0.083}}   
\newcommand{\hatcurRVomegaeccenxxxxxA}{\ensuremath{310\pm120}}         
\newcommand{\hatcurPPieccenxxxxxA}{\ensuremath{87.550\pm0.069}}        
\newcommand{\hatcurPPgeccenxxxxxA}{\ensuremath{11.9\pm1.9}}            
\newcommand{\hatcurPPloggeccenxxxxxA}{\ensuremath{3.077\pm0.069}}      
\newcommand{\hatcurPPareccenxxxxxA}{\ensuremath{17.43_{-0.22}^{+0.28}}} 
\newcommand{\hatcurPPareleccenxxxxxA}{\ensuremath{0.03388\pm0.00052}}  
\newcommand{\hatcurPPrhoeccenxxxxxA}{\ensuremath{0.82\pm0.13}}         
\newcommand{\hatcurPPmeccenxxxxxA}{\ensuremath{0.255\pm0.040}}         
\newcommand{\hatcurPPmtwosiglimeccenxxxxxA}{\ensuremath{<0.32}}        
\newcommand{\hatcurPPmshorteccenxxxxxA}{\ensuremath{0.26}}             
\newcommand{\hatcurPPmlongeccenxxxxxA}{\ensuremath{0.255\pm0.040}}     
\newcommand{\hatcurPPmeeccenxxxxxA}{\ensuremath{81\pm13}}              
\newcommand{\hatcurPPmeshorteccenxxxxxA}{\ensuremath{81.1}}            
\newcommand{\hatcurPPmelongeccenxxxxxA}{\ensuremath{81\pm13}}          
\newcommand{\hatcurPPreccenxxxxxA}{\ensuremath{0.728\pm0.015}}         
\newcommand{\hatcurPPrshorteccenxxxxxA}{\ensuremath{0.73}}             
\newcommand{\hatcurPPrlongeccenxxxxxA}{\ensuremath{0.728\pm0.015}}     
\newcommand{\hatcurPPreeccenxxxxxA}{\ensuremath{8.16\pm0.17}}          
\newcommand{\hatcurPPreshorteccenxxxxxA}{\ensuremath{8.2}}             
\newcommand{\hatcurPPrelongeccenxxxxxA}{\ensuremath{8.16\pm0.17}}      
\newcommand{\hatcurPPmrcorreccenxxxxxA}{\ensuremath{0.13}}             
\newcommand{\hatcurPPteffeccenxxxxxA}{\ensuremath{551.6\pm4.5}}        
\newcommand{\hatcurPPthetaeccenxxxxxA}{\ensuremath{0.0553\pm0.0085}}   
\newcommand{\hatcurPPfluxperieccenxxxxxA}{\ensuremath{2.232_{-0.077}^{+0.128}}} 
\newcommand{\hatcurPPfluxperidimeccenxxxxxA}{\ensuremath{7}}           
\newcommand{\hatcurPPfluxapeccenxxxxxA}{\ensuremath{1.97\pm0.13}}      
\newcommand{\hatcurPPfluxapdimeccenxxxxxA}{\ensuremath{7}}             
\newcommand{\hatcurPPfluxavgeccenxxxxxA}{\ensuremath{2.098\pm0.069}}   
\newcommand{\hatcurPPfluxavgdimeccenxxxxxA}{\ensuremath{7}}            
\newcommand{\hatcurPPfluxavglogeccenxxxxxA}{\ensuremath{7.322\pm0.014}} 
\newcommand{\hatcurXsecphaseeccenxxxxxA}{\ensuremath{0.513\pm0.018}}   
\newcommand{\hatcurXsecondaryeccenxxxxxA}{\ensuremath{2459876.341\pm0.062}} 
\newcommand{\hatcurXsecdureccenxxxxxA}{\ensuremath{0.05759\pm0.00067}} 
\newcommand{\hatcurXsecingdureccenxxxxxA}{\ensuremath{0.0174\pm0.0014}} 
\newcommand{\hatcurPPphiconjeccenxxxxxA}{\ensuremath{0.30_{-0.29}^{+0.10}}} 
\newcommand{\hatcurPPperieccenxxxxxA}{\ensuremath{2459873.53\pm0.91}}  
\newcommand{\hatcurPPaequiveccenxxxxxA}{\ensuremath{0.2548\pm0.0042}}  
\newcommand{\hatcurPPtcirceccenxxxxxA}{\ensuremath{465\pm83}}          
\newcommand{\hatcurPPtinfalleccenxxxxxA}{\ensuremath{90000\pm16000}}   
\newcommand{\hatcurXdisteccenxxxxxA}{\ensuremath{98.80\pm0.20}}        
\newcommand{\hatcurXAveccenxxxxxA}{\ensuremath{0.1820_{-0.0050}^{+0.0030}}} 
\newcommand{\hatcurXdistredeccenxxxxxA}{\ensuremath{98.83\pm0.20}}     
\newcommand{\hatcurXEBVeccenxxxxxA}{\ensuremath{0.0590_{-0.0020}^{+0.0010}}} 
\newcommand{\hatcurCCpmraeccenxxxxxA}{\ensuremath{-159.174\pm0.020}}   
\newcommand{\hatcurCCpmdececcenxxxxxA}{\ensuremath{-24.780\pm0.020}}   
\newcommand{\hatcurCCpmeccenxxxxxA}{\ensuremath{161.091\pm0.028}}      
\newcommand{\hatcurhtreccenxxxxxB}{TIC46432937}                        
\newcommand{\hatcurfieldeccenxxxxxB}{\ensuremath{string}}              
\newcommand{\hatcurCCraeccenxxxxxB}{\ensuremath{05^{\mathrm h}35^{\mathrm m}28.5693{\mathrm s}}}                     
\newcommand{\hatcurCCdececcenxxxxxB}{\ensuremath{-14{\arcdeg}35{\arcmin}50.4600{\arcsec}}}                   
\newcommand{\hatcurCCmageccenxxxxxB}{14.310}                           
\newcommand{\hatcurCCtwomasseccenxxxxxB}{2MASS~11041818-4749169}       
\newcommand{\hatcurCCgsceccenxxxxxB}{GSC~NULL}                         
\newcommand{\hatcurCCgaiaeccenxxxxxB}{GAIA~}                           
\newcommand{\hatcurCCgaiadrtwoeccenxxxxxB}{GAIA~DR2~2984391358868786816} 
\newcommand{\hatcurCCtassmveccenxxxxxB}{\ensuremath{14.310\pm0.020}}   
\newcommand{\hatcurCCtassmvshorteccenxxxxxB}{\ensuremath{14.3}}        
\newcommand{\hatcurCCtassmBeccenxxxxxB}{\ensuremath{15.707\pm0.050}}   
\newcommand{\hatcurCCtassmBshorteccenxxxxxB}{\ensuremath{15.7}}        
\newcommand{\hatcurCCtassmIeccenxxxxxB}{\ensuremath{nff\pmnff}}        
\newcommand{\hatcurCCtassmIshorteccenxxxxxB}{\ensuremath{0.0}}         
\newcommand{\hatcurCCtassmgeccenxxxxxB}{\ensuremath{15.040\pm0.060}}   
\newcommand{\hatcurCCtassmgshorteccenxxxxxB}{\ensuremath{15.0}}        
\newcommand{\hatcurCCtassmreccenxxxxxB}{\ensuremath{13.740\pm0.020}}   
\newcommand{\hatcurCCtassmrshorteccenxxxxxB}{\ensuremath{13.7}}        
\newcommand{\hatcurCCtassmieccenxxxxxB}{\ensuremath{12.792\pm0.040}}   
\newcommand{\hatcurCCtassmishorteccenxxxxxB}{\ensuremath{12.8}}        
\newcommand{\hatcurCCparallaxeccenxxxxxB}{\ensuremath{11.031\pm0.013}} 
\newcommand{\hatcurCCgaiamGeccenxxxxxB}{\ensuremath{13.4172\pm0.0028}} 
\newcommand{\hatcurCCgaiamBPeccenxxxxxB}{\ensuremath{14.4962\pm0.0033}} 
\newcommand{\hatcurCCgaiamRPeccenxxxxxB}{\ensuremath{12.3785\pm0.0039}} 
\newcommand{\hatcurCCtwomassJmageccenxxxxxB}{\ensuremath{11.011\pm0.022}} 
\newcommand{\hatcurCCtwomassHmageccenxxxxxB}{\ensuremath{10.427\pm0.023}} 
\newcommand{\hatcurCCtwomassKmageccenxxxxxB}{\ensuremath{10.195\pm0.020}} 
\newcommand{\hatcurCCcitJmageccenxxxxxB}{\ensuremath{11.004\pm0.023}}  
\newcommand{\hatcurCCcitHmageccenxxxxxB}{\ensuremath{10.418\pm0.024}}  
\newcommand{\hatcurCCcitKmageccenxxxxxB}{\ensuremath{10.219\pm0.021}}  
\newcommand{\hatcurCCbbJmageccenxxxxxB}{\ensuremath{11.090\pm0.025}}   
\newcommand{\hatcurCCbbHmageccenxxxxxB}{\ensuremath{10.444\pm0.025}}   
\newcommand{\hatcurCCbbKmageccenxxxxxB}{\ensuremath{10.239\pm0.021}}   
\newcommand{\hatcurCCesoJmageccenxxxxxB}{\ensuremath{11.098\pm0.030}}  
\newcommand{\hatcurCCesoHmageccenxxxxxB}{\ensuremath{10.444\pm0.040}}  
\newcommand{\hatcurCCesoKmageccenxxxxxB}{\ensuremath{10.236\pm0.023}}  
\newcommand{\hatcurCCesoJHmageccenxxxxxB}{\ensuremath{0.654\pm0.047}}  
\newcommand{\hatcurCCesoJKmageccenxxxxxB}{\ensuremath{0.862\pm0.035}}  
\newcommand{\hatcurCCesoHKmageccenxxxxxB}{\ensuremath{0.208\pm0.045}}  
\newcommand{\hatcurCCWonemageccenxxxxxB}{\ensuremath{10.114\pm0.023}}  
\newcommand{\hatcurCCWtwomageccenxxxxxB}{\ensuremath{10.063\pm0.020}}  
\newcommand{\hatcurCCWthreemageccenxxxxxB}{\ensuremath{9.910\pm0.055}} 
\newcommand{\hatcurCCWfourmageccenxxxxxB}{\ensuremath{nff\pmnff}}      
\newcommand{\hatcurLCdipeccenxxxxxB}{\ensuremath{0.1}}                 
\newcommand{\hatcurLCrprstareccenxxxxxB}{\ensuremath{0.2298\pm0.0036}} 
\newcommand{\hatcurLCbsqeccenxxxxxB}{\ensuremath{0.679_{-0.011}^{+0.027}}} 
\newcommand{\hatcurLCimpeccenxxxxxB}{\ensuremath{0.8240_{-0.0065}^{+0.0159}}} 
\newcommand{\hatcurLCzetaeccenxxxxxB}{\ensuremath{64.54_{-0.72}^{+2.49}}} 
\newcommand{\hatcurLCdureccenxxxxxB}{\ensuremath{0.04981\pm0.00025}}   
\newcommand{\hatcurLCdurshorteccenxxxxxB}{\ensuremath{0.0498}}         
\newcommand{\hatcurLCdurhreccenxxxxxB}{\ensuremath{1.1955\pm0.0061}}   
\newcommand{\hatcurLCdurhrshorteccenxxxxxB}{\ensuremath{1.196}}        
\newcommand{\hatcurLCqeccenxxxxxB}{\ensuremath{0.03460\pm0.00018}}     
\newcommand{\hatcurLCqshorteccenxxxxxB}{\ensuremath{0.035}}            
\newcommand{\hatcurLCingdureccenxxxxxB}{\ensuremath{0.05226\pm0.00025}} 
\newcommand{\hatcurLCPeccenxxxxxB}{\ensuremath{1.440445270\pm0.000000091}} 
\newcommand{\hatcurLCPprececcenxxxxxB}{\ensuremath{1.4404453}}         
\newcommand{\hatcurLCPshorteccenxxxxxB}{\ensuremath{1.4404}}           
\newcommand{\hatcurLCTeccenxxxxxB}{\ensuremath{2460008.466300\pm0.000038}} 
\newcommand{\hatcurLCTAeccenxxxxxB}{\ensuremath{2458468.630306\pm0.000100}} 
\newcommand{\hatcurLCTBeccenxxxxxB}{\ensuremath{2460316.721588\pm0.000045}} 
\newcommand{\hatcurLChatnetmAeccenxxxxxB}{\ensuremath{-8.116450\pm0.000040}} 
\newcommand{\hatcurLCiblendAeccenxxxxxB}{\ensuremath{1\pm0}}           
\newcommand{\hatcurLChatnetmBeccenxxxxxB}{\ensuremath{-8.019870\pm0.000041}} 
\newcommand{\hatcurLCiblendBeccenxxxxxB}{\ensuremath{1\pm0}}           
\newcommand{\hatcurLCrhoeccenxxxxxB}{\ensuremath{5.36\pm0.11}}         
\newcommand{\hatcurSMEiteffeccenxxxxxB}{\ensuremath{3535\pm65}}        
\newcommand{\hatcurSMEizfeheccenxxxxxB}{\ensuremath{0.03\pm0.10}}      
\newcommand{\hatcurSMEizfehshorteccenxxxxxB}{\ensuremath{0.03}}        
\newcommand{\hatcurSMEiloggeccenxxxxxB}{\ensuremath{5.00\pm0.50}}      
\newcommand{\hatcurSMEivsineccenxxxxxB}{\ensuremath{0\pm50}}           
\newcommand{\hatcurSMEivmaceccenxxxxxB}{\ensuremath{nff\pmnff}}        
\newcommand{\hatcurSMEivmiceccenxxxxxB}{\ensuremath{nff\pmnff}}        
\newcommand{\hatcurextraerrMJeccenxxxxxB}{\ensuremath{0\pm0}}          
\newcommand{\hatcurextraerrMJtwosiglimeccenxxxxxB}{\ensuremath{<0.0200}} 
\newcommand{\hatcurextraerrMHeccenxxxxxB}{\ensuremath{0\pm0}}          
\newcommand{\hatcurextraerrMHtwosiglimeccenxxxxxB}{\ensuremath{<0.0200}} 
\newcommand{\hatcurextraerrMKseccenxxxxxB}{\ensuremath{0\pm0}}         
\newcommand{\hatcurextraerrMKstwosiglimeccenxxxxxB}{\ensuremath{<0.0200}} 
\newcommand{\hatcurextraerrMGeccenxxxxxB}{\ensuremath{0\pm0}}          
\newcommand{\hatcurextraerrMGtwosiglimeccenxxxxxB}{\ensuremath{<0.0200}} 
\newcommand{\hatcurextraerrMRPeccenxxxxxB}{\ensuremath{0\pm0}}         
\newcommand{\hatcurextraerrMRPtwosiglimeccenxxxxxB}{\ensuremath{<0.0200}} 
\newcommand{\hatcurextraerrMgeccenxxxxxB}{\ensuremath{0\pm0}}          
\newcommand{\hatcurextraerrMgtwosiglimeccenxxxxxB}{\ensuremath{<0.0200}} 
\newcommand{\hatcurextraerrMreccenxxxxxB}{\ensuremath{0\pm0}}          
\newcommand{\hatcurextraerrMrtwosiglimeccenxxxxxB}{\ensuremath{<0.0200}} 
\newcommand{\hatcurextraerrMieccenxxxxxB}{\ensuremath{0\pm0}}          
\newcommand{\hatcurextraerrMitwosiglimeccenxxxxxB}{\ensuremath{<0.0200}} 
\newcommand{\hatcurextraerrMWoneeccenxxxxxB}{\ensuremath{0\pm0}}       
\newcommand{\hatcurextraerrMWonetwosiglimeccenxxxxxB}{\ensuremath{<0.0200}} 
\newcommand{\hatcurextraerrMWtwoeccenxxxxxB}{\ensuremath{0\pm0}}       
\newcommand{\hatcurextraerrMWtwotwosiglimeccenxxxxxB}{\ensuremath{<0.0200}} 
\newcommand{\hatcurLBiBeccenxxxxxB}{\ensuremath{0.4448}}               
\newcommand{\hatcurLBiiBeccenxxxxxB}{\ensuremath{0.3380}}              
\newcommand{\hatcurLBiVeccenxxxxxB}{\ensuremath{0.4071}}               
\newcommand{\hatcurLBiiVeccenxxxxxB}{\ensuremath{0.3615}}              
\newcommand{\hatcurLBiReccenxxxxxB}{\ensuremath{0.3575}}               
\newcommand{\hatcurLBiiReccenxxxxxB}{\ensuremath{0.3335}}              
\newcommand{\hatcurLBiIeccenxxxxxB}{\ensuremath{0.109_{-0.070}^{+0.099}}} 
\newcommand{\hatcurLBiiIeccenxxxxxB}{\ensuremath{0.06\pm0.12}}         
\newcommand{\hatcurLBiueccenxxxxxB}{\ensuremath{0.4132}}               
\newcommand{\hatcurLBiiueccenxxxxxB}{\ensuremath{0.3279}}              
\newcommand{\hatcurLBigeccenxxxxxB}{\ensuremath{0.4098}}               
\newcommand{\hatcurLBiigeccenxxxxxB}{\ensuremath{0.3420}}              
\newcommand{\hatcurLBireccenxxxxxB}{\ensuremath{0.4207}}               
\newcommand{\hatcurLBiireccenxxxxxB}{\ensuremath{0.3178}}              
\newcommand{\hatcurLBiieccenxxxxxB}{\ensuremath{0.2336}}               
\newcommand{\hatcurLBiiieccenxxxxxB}{\ensuremath{0.3384}}              
\newcommand{\hatcurLBizeccenxxxxxB}{\ensuremath{0.158\pm0.092}}        
\newcommand{\hatcurLBiizeccenxxxxxB}{\ensuremath{0.13\pm0.14}}         
\newcommand{\hatcurLBiJeccenxxxxxB}{\ensuremath{0.143\pm0.090}}        
\newcommand{\hatcurLBiiJeccenxxxxxB}{\ensuremath{0.09\pm0.13}}         
\newcommand{\hatcurLBiHeccenxxxxxB}{\ensuremath{0.20\pm0.11}}          
\newcommand{\hatcurLBiiHeccenxxxxxB}{\ensuremath{0.09\pm0.14}}         
\newcommand{\hatcurLBiKeccenxxxxxB}{\ensuremath{0.0759}}               
\newcommand{\hatcurLBiiKeccenxxxxxB}{\ensuremath{0.2304}}              
\newcommand{\hatcurLBiTeccenxxxxxB}{\ensuremath{0.20\pm0.10}}          
\newcommand{\hatcurLBiiTeccenxxxxxB}{\ensuremath{0.18\pm0.14}}         
\newcommand{\hatcurLBikepeccenxxxxxB}{\ensuremath{0.2854}}             
\newcommand{\hatcurLBiikepeccenxxxxxB}{\ensuremath{0.4148}}            
\newcommand{\hatcurLBiCeccenxxxxxB}{\ensuremath{0.2416}}               
\newcommand{\hatcurLBiiCeccenxxxxxB}{\ensuremath{0.4173}}              
\newcommand{\hatcurLBiMeccenxxxxxB}{\ensuremath{0.3749}}               
\newcommand{\hatcurLBiiMeccenxxxxxB}{\ensuremath{0.4051}}              
\newcommand{\hatcurLBiSoneeccenxxxxxB}{\ensuremath{0.0625}}              
\newcommand{\hatcurLBiiSoneeccenxxxxxB}{\ensuremath{0.1665}}             
\newcommand{\hatcurLBiStwoeccenxxxxxB}{\ensuremath{0.0462}}              
\newcommand{\hatcurLBiiStwoeccenxxxxxB}{\ensuremath{0.1375}}             
\newcommand{\hatcurLBiSthreeeccenxxxxxB}{\ensuremath{0.0495}}              
\newcommand{\hatcurLBiiSthreeeccenxxxxxB}{\ensuremath{0.1196}}             
\newcommand{\hatcurLBiSfoureccenxxxxxB}{\ensuremath{0.0533}}              
\newcommand{\hatcurLBiiSfoureccenxxxxxB}{\ensuremath{0.0911}}             
\newcommand{\hatcurISOmeccenxxxxxB}{\ensuremath{0.559\pm0.029}}        
\newcommand{\hatcurISOmshorteccenxxxxxB}{\ensuremath{0.56}}            
\newcommand{\hatcurISOmlongeccenxxxxxB}{\ensuremath{0.559\pm0.029}}    
\newcommand{\hatcurISOreccenxxxxxB}{\ensuremath{0.5280\pm0.0093}}      
\newcommand{\hatcurISOrshorteccenxxxxxB}{\ensuremath{0.53}}            
\newcommand{\hatcurISOrlongeccenxxxxxB}{\ensuremath{0.5280\pm0.0093}}  
\newcommand{\hatcurISOrhoeccenxxxxxB}{\ensuremath{5.36\pm0.11}}        
\newcommand{\hatcurISOrholongeccenxxxxxB}{\ensuremath{5.36\pm0.11}}    
\newcommand{\hatcurISOloggeccenxxxxxB}{\ensuremath{4.740\pm0.010}}     
\newcommand{\hatcurISOlumeccenxxxxxB}{\ensuremath{0.0410\pm0.0030}}    
\newcommand{\hatcurISOlumshorteccenxxxxxB}{\ensuremath{0.04}}          
\newcommand{\hatcurISOteffeccenxxxxxB}{\ensuremath{3573\pm55}}         
\newcommand{\hatcurISOzfeheccenxxxxxB}{\ensuremath{0.320\pm0.082}}     
\newcommand{\hatcurISOageeccenxxxxxB}{\ensuremath{6.9_{-4.6}^{+6.2}}}  
\newcommand{\hatcurISOspececcenxxxxxB}{M}                              
\newcommand{\hatcurRVKeccenxxxxxB}{\ensuremath{837.9\pm4.2}}           
\newcommand{\hatcurRVKtwosiglimeccenxxxxxB}{\ensuremath{<844.1}}       
\newcommand{\hatcurRVrkeccenxxxxxB}{\ensuremath{0.022\pm0.023}}        
\newcommand{\hatcurRVrheccenxxxxxB}{\ensuremath{-0.012\pm0.047}}       
\newcommand{\hatcurRVkeccenxxxxxB}{\ensuremath{0.0010_{-0.0011}^{+0.0015}}} 
\newcommand{\hatcurRVheccenxxxxxB}{\ensuremath{-0.0004_{-0.0036}^{+0.0024}}} 
\newcommand{\hatcurRVtroneeccenxxxxxB}{\ensuremath{0\pm0}}             
\newcommand{\hatcurRVtrtwoeccenxxxxxB}{\ensuremath{0\pm0}}             
\newcommand{\hatcurRVgammaeccenxxxxxB}{\ensuremath{102279.0\pm2.5}}    
\newcommand{\hatcurRVjittereccenxxxxxB}{\ensuremath{5.1\pm2.6}}        
\newcommand{\hatcurRVjittertwosiglimeccenxxxxxB}{\ensuremath{<10.4}}   
\newcommand{\hatcurRVfitrmseccenxxxxxB}{\ensuremath{.1fym}}            %
\newcommand{\hatcurRVecceneccenxxxxxB}{\ensuremath{0.0030\pm0.0027}}   
\newcommand{\hatcurRVeccentwosiglimeccenxxxxxB}{\ensuremath{<0.009}}   
\newcommand{\hatcurRVomegaeccenxxxxxB}{\ensuremath{260\pm140}}         
\newcommand{\hatcurPPieccenxxxxxB}{\ensuremath{84.360_{-0.130}^{+0.090}}} 
\newcommand{\hatcurPPgeccenxxxxxB}{\ensuremath{56.7_{-2.8}^{+2.0}}}    
\newcommand{\hatcurPPloggeccenxxxxxB}{\ensuremath{3.753_{-0.022}^{+0.015}}} 
\newcommand{\hatcurPPareccenxxxxxB}{\ensuremath{8.391\pm0.055}}        
\newcommand{\hatcurPPareleccenxxxxxB}{\ensuremath{0.02060\pm0.00036}}  
\newcommand{\hatcurPPrhoeccenxxxxxB}{\ensuremath{2.40\pm0.15}}         
\newcommand{\hatcurPPmeccenxxxxxB}{\ensuremath{3.19\pm0.11}}           
\newcommand{\hatcurPPmtwosiglimeccenxxxxxB}{\ensuremath{<3.38}}        
\newcommand{\hatcurPPmshorteccenxxxxxB}{\ensuremath{3.19}}             
\newcommand{\hatcurPPmlongeccenxxxxxB}{\ensuremath{3.19\pm0.11}}       
\newcommand{\hatcurPPmeeccenxxxxxB}{\ensuremath{1013\pm36}}            
\newcommand{\hatcurPPmeshorteccenxxxxxB}{\ensuremath{1013.0}}          
\newcommand{\hatcurPPmelongeccenxxxxxB}{\ensuremath{1013\pm36}}        
\newcommand{\hatcurPPreccenxxxxxB}{\ensuremath{1.182\pm0.030}}         
\newcommand{\hatcurPPrshorteccenxxxxxB}{\ensuremath{1.18}}             
\newcommand{\hatcurPPrlongeccenxxxxxB}{\ensuremath{1.182\pm0.030}}     
\newcommand{\hatcurPPreeccenxxxxxB}{\ensuremath{13.25\pm0.33}}         
\newcommand{\hatcurPPreshorteccenxxxxxB}{\ensuremath{13.3}}            
\newcommand{\hatcurPPrelongeccenxxxxxB}{\ensuremath{13.25\pm0.33}}     
\newcommand{\hatcurPPmrcorreccenxxxxxB}{\ensuremath{0.58}}             
\newcommand{\hatcurPPteffeccenxxxxxB}{\ensuremath{872\pm14}}           
\newcommand{\hatcurPPthetaeccenxxxxxB}{\ensuremath{0.1963\pm0.0050}}   
\newcommand{\hatcurPPfluxperieccenxxxxxB}{\ensuremath{1.321\pm0.085}}  
\newcommand{\hatcurPPfluxperidimeccenxxxxxB}{\ensuremath{8}}           
\newcommand{\hatcurPPfluxapeccenxxxxxB}{\ensuremath{1.304\pm0.084}}    
\newcommand{\hatcurPPfluxapdimeccenxxxxxB}{\ensuremath{8}}             
\newcommand{\hatcurPPfluxavgeccenxxxxxB}{\ensuremath{1.313\pm0.084}}   
\newcommand{\hatcurPPfluxavgdimeccenxxxxxB}{\ensuremath{8}}            
\newcommand{\hatcurPPfluxavglogeccenxxxxxB}{\ensuremath{8.118\pm0.028}} 
\newcommand{\hatcurXsecphaseeccenxxxxxB}{\ensuremath{0.50064\pm0.00093}} 
\newcommand{\hatcurXsecondaryeccenxxxxxB}{\ensuremath{2460009.1874\pm0.0013}} 
\newcommand{\hatcurXsecdureccenxxxxxB}{\ensuremath{0.05237\pm0.00073}} 
\newcommand{\hatcurXsecingdureccenxxxxxB}{\ensuremath{0.02619\pm0.00037}} 
\newcommand{\hatcurPPphiconjeccenxxxxxB}{\ensuremath{0.21_{-0.32}^{+0.24}}} 
\newcommand{\hatcurPPperieccenxxxxxB}{\ensuremath{2460008.17\pm0.42}}  
\newcommand{\hatcurPPaequiveccenxxxxxB}{\ensuremath{0.1018\pm0.0033}}  
\newcommand{\hatcurPPtcirceccenxxxxxB}{\ensuremath{14.0\pm1.4}}        
\newcommand{\hatcurPPtinfalleccenxxxxxB}{\ensuremath{100.0\pm3.9}}     
\newcommand{\hatcurXdisteccenxxxxxB}{\ensuremath{90.600_{-0.100}^{+0.200}}} 
\newcommand{\hatcurXAveccenxxxxxB}{\ensuremath{0.023\pm0.012}}         
\newcommand{\hatcurXdistredeccenxxxxxB}{\ensuremath{90.65\pm0.10}}     
\newcommand{\hatcurXEBVeccenxxxxxB}{\ensuremath{0.0070_{-0.0030}^{+0.0050}}} 
\newcommand{\hatcurCCpmraeccenxxxxxB}{\ensuremath{-13.365\pm0.013}}    
\newcommand{\hatcurCCpmdececcenxxxxxB}{\ensuremath{36.962\pm0.012}}    
\newcommand{\hatcurCCpmeccenxxxxxB}{\ensuremath{39.304\pm0.018}}       
\newcommand{\hatcurCCbbHmageccen}[1]{\ifnum#1=762 %
\hatcurCCbbHmageccenxxxxxA
\else
\ifnum#1=4643 %
\hatcurCCbbHmageccenxxxxxB
\else
??????\fi
\fi
}
\newcommand{\hatcurCCbbJmageccen}[1]{\ifnum#1=762 %
\hatcurCCbbJmageccenxxxxxA
\else
\ifnum#1=4643 %
\hatcurCCbbJmageccenxxxxxB
\else
??????\fi
\fi
}
\newcommand{\hatcurCCbbKmageccen}[1]{\ifnum#1=762 %
\hatcurCCbbKmageccenxxxxxA
\else
\ifnum#1=4643 %
\hatcurCCbbKmageccenxxxxxB
\else
??????\fi
\fi
}
\newcommand{\hatcurCCcitHmageccen}[1]{\ifnum#1=762 %
\hatcurCCcitHmageccenxxxxxA
\else
\ifnum#1=4643 %
\hatcurCCcitHmageccenxxxxxB
\else
??????\fi
\fi
}
\newcommand{\hatcurCCcitJmageccen}[1]{\ifnum#1=762 %
\hatcurCCcitJmageccenxxxxxA
\else
\ifnum#1=4643 %
\hatcurCCcitJmageccenxxxxxB
\else
??????\fi
\fi
}
\newcommand{\hatcurCCcitKmageccen}[1]{\ifnum#1=762 %
\hatcurCCcitKmageccenxxxxxA
\else
\ifnum#1=4643 %
\hatcurCCcitKmageccenxxxxxB
\else
??????\fi
\fi
}
\newcommand{\hatcurCCdececcen}[1]{\ifnum#1=762 %
\hatcurCCdececcenxxxxxA
\else
\ifnum#1=4643 %
\hatcurCCdececcenxxxxxB
\else
??????\fi
\fi
}
\newcommand{\hatcurCCesoHKmageccen}[1]{\ifnum#1=762 %
\hatcurCCesoHKmageccenxxxxxA
\else
\ifnum#1=4643 %
\hatcurCCesoHKmageccenxxxxxB
\else
??????\fi
\fi
}
\newcommand{\hatcurCCesoHmageccen}[1]{\ifnum#1=762 %
\hatcurCCesoHmageccenxxxxxA
\else
\ifnum#1=4643 %
\hatcurCCesoHmageccenxxxxxB
\else
??????\fi
\fi
}
\newcommand{\hatcurCCesoJHmageccen}[1]{\ifnum#1=762 %
\hatcurCCesoJHmageccenxxxxxA
\else
\ifnum#1=4643 %
\hatcurCCesoJHmageccenxxxxxB
\else
??????\fi
\fi
}
\newcommand{\hatcurCCesoJKmageccen}[1]{\ifnum#1=762 %
\hatcurCCesoJKmageccenxxxxxA
\else
\ifnum#1=4643 %
\hatcurCCesoJKmageccenxxxxxB
\else
??????\fi
\fi
}
\newcommand{\hatcurCCesoJmageccen}[1]{\ifnum#1=762 %
\hatcurCCesoJmageccenxxxxxA
\else
\ifnum#1=4643 %
\hatcurCCesoJmageccenxxxxxB
\else
??????\fi
\fi
}
\newcommand{\hatcurCCesoKmageccen}[1]{\ifnum#1=762 %
\hatcurCCesoKmageccenxxxxxA
\else
\ifnum#1=4643 %
\hatcurCCesoKmageccenxxxxxB
\else
??????\fi
\fi
}
\newcommand{\hatcurCCgaiadrtwoeccen}[1]{\ifnum#1=762 %
\hatcurCCgaiadrtwoeccenxxxxxA
\else
\ifnum#1=4643 %
\hatcurCCgaiadrtwoeccenxxxxxB
\else
??????\fi
\fi
}
\newcommand{\hatcurCCgaiaeccen}[1]{\ifnum#1=762 %
\hatcurCCgaiaeccenxxxxxA
\else
\ifnum#1=4643 %
\hatcurCCgaiaeccenxxxxxB
\else
??????\fi
\fi
}
\newcommand{\hatcurCCgaiamBPeccen}[1]{\ifnum#1=762 %
\hatcurCCgaiamBPeccenxxxxxA
\else
\ifnum#1=4643 %
\hatcurCCgaiamBPeccenxxxxxB
\else
??????\fi
\fi
}
\newcommand{\hatcurCCgaiamGeccen}[1]{\ifnum#1=762 %
\hatcurCCgaiamGeccenxxxxxA
\else
\ifnum#1=4643 %
\hatcurCCgaiamGeccenxxxxxB
\else
??????\fi
\fi
}
\newcommand{\hatcurCCgaiamRPeccen}[1]{\ifnum#1=762 %
\hatcurCCgaiamRPeccenxxxxxA
\else
\ifnum#1=4643 %
\hatcurCCgaiamRPeccenxxxxxB
\else
??????\fi
\fi
}
\newcommand{\hatcurCCgsceccen}[1]{\ifnum#1=762 %
\hatcurCCgsceccenxxxxxA
\else
\ifnum#1=4643 %
\hatcurCCgsceccenxxxxxB
\else
??????\fi
\fi
}
\newcommand{\hatcurCCmageccen}[1]{\ifnum#1=762 %
\hatcurCCmageccenxxxxxA
\else
\ifnum#1=4643 %
\hatcurCCmageccenxxxxxB
\else
??????\fi
\fi
}
\newcommand{\hatcurCCparallaxeccen}[1]{\ifnum#1=762 %
\hatcurCCparallaxeccenxxxxxA
\else
\ifnum#1=4643 %
\hatcurCCparallaxeccenxxxxxB
\else
??????\fi
\fi
}
\newcommand{\hatcurCCpmdececcen}[1]{\ifnum#1=762 %
\hatcurCCpmdececcenxxxxxA
\else
\ifnum#1=4643 %
\hatcurCCpmdececcenxxxxxB
\else
??????\fi
\fi
}
\newcommand{\hatcurCCpmeccen}[1]{\ifnum#1=762 %
\hatcurCCpmeccenxxxxxA
\else
\ifnum#1=4643 %
\hatcurCCpmeccenxxxxxB
\else
??????\fi
\fi
}
\newcommand{\hatcurCCpmraeccen}[1]{\ifnum#1=762 %
\hatcurCCpmraeccenxxxxxA
\else
\ifnum#1=4643 %
\hatcurCCpmraeccenxxxxxB
\else
??????\fi
\fi
}
\newcommand{\hatcurCCraeccen}[1]{\ifnum#1=762 %
\hatcurCCraeccenxxxxxA
\else
\ifnum#1=4643 %
\hatcurCCraeccenxxxxxB
\else
??????\fi
\fi
}
\newcommand{\hatcurCCtassmBeccen}[1]{\ifnum#1=762 %
\hatcurCCtassmBeccenxxxxxA
\else
\ifnum#1=4643 %
\hatcurCCtassmBeccenxxxxxB
\else
??????\fi
\fi
}
\newcommand{\hatcurCCtassmBshorteccen}[1]{\ifnum#1=762 %
\hatcurCCtassmBshorteccenxxxxxA
\else
\ifnum#1=4643 %
\hatcurCCtassmBshorteccenxxxxxB
\else
??????\fi
\fi
}
\newcommand{\hatcurCCtassmgeccen}[1]{\ifnum#1=762 %
\hatcurCCtassmgeccenxxxxxA
\else
\ifnum#1=4643 %
\hatcurCCtassmgeccenxxxxxB
\else
??????\fi
\fi
}
\newcommand{\hatcurCCtassmgshorteccen}[1]{\ifnum#1=762 %
\hatcurCCtassmgshorteccenxxxxxA
\else
\ifnum#1=4643 %
\hatcurCCtassmgshorteccenxxxxxB
\else
??????\fi
\fi
}
\newcommand{\hatcurCCtassmieccen}[1]{\ifnum#1=762 %
\hatcurCCtassmieccenxxxxxA
\else
\ifnum#1=4643 %
\hatcurCCtassmieccenxxxxxB
\else
??????\fi
\fi
}
\newcommand{\hatcurCCtassmIeccen}[1]{\ifnum#1=762 %
\hatcurCCtassmIeccenxxxxxA
\else
\ifnum#1=4643 %
\hatcurCCtassmIeccenxxxxxB
\else
??????\fi
\fi
}
\newcommand{\hatcurCCtassmishorteccen}[1]{\ifnum#1=762 %
\hatcurCCtassmishorteccenxxxxxA
\else
\ifnum#1=4643 %
\hatcurCCtassmishorteccenxxxxxB
\else
??????\fi
\fi
}
\newcommand{\hatcurCCtassmIshorteccen}[1]{\ifnum#1=762 %
\hatcurCCtassmIshorteccenxxxxxA
\else
\ifnum#1=4643 %
\hatcurCCtassmIshorteccenxxxxxB
\else
??????\fi
\fi
}
\newcommand{\hatcurCCtassmreccen}[1]{\ifnum#1=762 %
\hatcurCCtassmreccenxxxxxA
\else
\ifnum#1=4643 %
\hatcurCCtassmreccenxxxxxB
\else
??????\fi
\fi
}
\newcommand{\hatcurCCtassmrshorteccen}[1]{\ifnum#1=762 %
\hatcurCCtassmrshorteccenxxxxxA
\else
\ifnum#1=4643 %
\hatcurCCtassmrshorteccenxxxxxB
\else
??????\fi
\fi
}
\newcommand{\hatcurCCtassmveccen}[1]{\ifnum#1=762 %
\hatcurCCtassmveccenxxxxxA
\else
\ifnum#1=4643 %
\hatcurCCtassmveccenxxxxxB
\else
??????\fi
\fi
}
\newcommand{\hatcurCCtassmvshorteccen}[1]{\ifnum#1=762 %
\hatcurCCtassmvshorteccenxxxxxA
\else
\ifnum#1=4643 %
\hatcurCCtassmvshorteccenxxxxxB
\else
??????\fi
\fi
}
\newcommand{\hatcurCCtwomasseccen}[1]{\ifnum#1=762 %
\hatcurCCtwomasseccenxxxxxA
\else
\ifnum#1=4643 %
\hatcurCCtwomasseccenxxxxxB
\else
??????\fi
\fi
}
\newcommand{\hatcurCCtwomassHmageccen}[1]{\ifnum#1=762 %
\hatcurCCtwomassHmageccenxxxxxA
\else
\ifnum#1=4643 %
\hatcurCCtwomassHmageccenxxxxxB
\else
??????\fi
\fi
}
\newcommand{\hatcurCCtwomassJmageccen}[1]{\ifnum#1=762 %
\hatcurCCtwomassJmageccenxxxxxA
\else
\ifnum#1=4643 %
\hatcurCCtwomassJmageccenxxxxxB
\else
??????\fi
\fi
}
\newcommand{\hatcurCCtwomassKmageccen}[1]{\ifnum#1=762 %
\hatcurCCtwomassKmageccenxxxxxA
\else
\ifnum#1=4643 %
\hatcurCCtwomassKmageccenxxxxxB
\else
??????\fi
\fi
}
\newcommand{\hatcurCCWfourmageccen}[1]{\ifnum#1=762 %
\hatcurCCWfourmageccenxxxxxA
\else
\ifnum#1=4643 %
\hatcurCCWfourmageccenxxxxxB
\else
??????\fi
\fi
}
\newcommand{\hatcurCCWonemageccen}[1]{\ifnum#1=762 %
\hatcurCCWonemageccenxxxxxA
\else
\ifnum#1=4643 %
\hatcurCCWonemageccenxxxxxB
\else
??????\fi
\fi
}
\newcommand{\hatcurCCWthreemageccen}[1]{\ifnum#1=762 %
\hatcurCCWthreemageccenxxxxxA
\else
\ifnum#1=4643 %
\hatcurCCWthreemageccenxxxxxB
\else
??????\fi
\fi
}
\newcommand{\hatcurCCWtwomageccen}[1]{\ifnum#1=762 %
\hatcurCCWtwomageccenxxxxxA
\else
\ifnum#1=4643 %
\hatcurCCWtwomageccenxxxxxB
\else
??????\fi
\fi
}
\newcommand{\hatcurextraerrMgeccen}[1]{\ifnum#1=762 %
\hatcurextraerrMgeccenxxxxxA
\else
\ifnum#1=4643 %
\hatcurextraerrMgeccenxxxxxB
\else
??????\fi
\fi
}
\newcommand{\hatcurextraerrMGeccen}[1]{\ifnum#1=762 %
\hatcurextraerrMGeccenxxxxxA
\else
\ifnum#1=4643 %
\hatcurextraerrMGeccenxxxxxB
\else
??????\fi
\fi
}
\newcommand{\hatcurextraerrMgtwosiglimeccen}[1]{\ifnum#1=762 %
\hatcurextraerrMgtwosiglimeccenxxxxxA
\else
\ifnum#1=4643 %
\hatcurextraerrMgtwosiglimeccenxxxxxB
\else
??????\fi
\fi
}
\newcommand{\hatcurextraerrMGtwosiglimeccen}[1]{\ifnum#1=762 %
\hatcurextraerrMGtwosiglimeccenxxxxxA
\else
\ifnum#1=4643 %
\hatcurextraerrMGtwosiglimeccenxxxxxB
\else
??????\fi
\fi
}
\newcommand{\hatcurextraerrMHeccen}[1]{\ifnum#1=762 %
\hatcurextraerrMHeccenxxxxxA
\else
\ifnum#1=4643 %
\hatcurextraerrMHeccenxxxxxB
\else
??????\fi
\fi
}
\newcommand{\hatcurextraerrMHtwosiglimeccen}[1]{\ifnum#1=762 %
\hatcurextraerrMHtwosiglimeccenxxxxxA
\else
\ifnum#1=4643 %
\hatcurextraerrMHtwosiglimeccenxxxxxB
\else
??????\fi
\fi
}
\newcommand{\hatcurextraerrMieccen}[1]{\ifnum#1=762 %
\hatcurextraerrMieccenxxxxxA
\else
\ifnum#1=4643 %
\hatcurextraerrMieccenxxxxxB
\else
??????\fi
\fi
}
\newcommand{\hatcurextraerrMitwosiglimeccen}[1]{\ifnum#1=762 %
\hatcurextraerrMitwosiglimeccenxxxxxA
\else
\ifnum#1=4643 %
\hatcurextraerrMitwosiglimeccenxxxxxB
\else
??????\fi
\fi
}
\newcommand{\hatcurextraerrMJeccen}[1]{\ifnum#1=762 %
\hatcurextraerrMJeccenxxxxxA
\else
\ifnum#1=4643 %
\hatcurextraerrMJeccenxxxxxB
\else
??????\fi
\fi
}
\newcommand{\hatcurextraerrMJtwosiglimeccen}[1]{\ifnum#1=762 %
\hatcurextraerrMJtwosiglimeccenxxxxxA
\else
\ifnum#1=4643 %
\hatcurextraerrMJtwosiglimeccenxxxxxB
\else
??????\fi
\fi
}
\newcommand{\hatcurextraerrMKseccen}[1]{\ifnum#1=762 %
\hatcurextraerrMKseccenxxxxxA
\else
\ifnum#1=4643 %
\hatcurextraerrMKseccenxxxxxB
\else
??????\fi
\fi
}
\newcommand{\hatcurextraerrMKstwosiglimeccen}[1]{\ifnum#1=762 %
\hatcurextraerrMKstwosiglimeccenxxxxxA
\else
\ifnum#1=4643 %
\hatcurextraerrMKstwosiglimeccenxxxxxB
\else
??????\fi
\fi
}
\newcommand{\hatcurextraerrMreccen}[1]{\ifnum#1=762 %
\hatcurextraerrMreccenxxxxxA
\else
\ifnum#1=4643 %
\hatcurextraerrMreccenxxxxxB
\else
??????\fi
\fi
}
\newcommand{\hatcurextraerrMRPeccen}[1]{\ifnum#1=762 %
\hatcurextraerrMRPeccenxxxxxA
\else
\ifnum#1=4643 %
\hatcurextraerrMRPeccenxxxxxB
\else
??????\fi
\fi
}
\newcommand{\hatcurextraerrMRPtwosiglimeccen}[1]{\ifnum#1=762 %
\hatcurextraerrMRPtwosiglimeccenxxxxxA
\else
\ifnum#1=4643 %
\hatcurextraerrMRPtwosiglimeccenxxxxxB
\else
??????\fi
\fi
}
\newcommand{\hatcurextraerrMrtwosiglimeccen}[1]{\ifnum#1=762 %
\hatcurextraerrMrtwosiglimeccenxxxxxA
\else
\ifnum#1=4643 %
\hatcurextraerrMrtwosiglimeccenxxxxxB
\else
??????\fi
\fi
}
\newcommand{\hatcurextraerrMWoneeccen}[1]{\ifnum#1=762 %
\hatcurextraerrMWoneeccenxxxxxA
\else
\ifnum#1=4643 %
\hatcurextraerrMWoneeccenxxxxxB
\else
??????\fi
\fi
}
\newcommand{\hatcurextraerrMWonetwosiglimeccen}[1]{\ifnum#1=762 %
\hatcurextraerrMWonetwosiglimeccenxxxxxA
\else
\ifnum#1=4643 %
\hatcurextraerrMWonetwosiglimeccenxxxxxB
\else
??????\fi
\fi
}
\newcommand{\hatcurextraerrMWtwoeccen}[1]{\ifnum#1=762 %
\hatcurextraerrMWtwoeccenxxxxxA
\else
\ifnum#1=4643 %
\hatcurextraerrMWtwoeccenxxxxxB
\else
??????\fi
\fi
}
\newcommand{\hatcurextraerrMWtwotwosiglimeccen}[1]{\ifnum#1=762 %
\hatcurextraerrMWtwotwosiglimeccenxxxxxA
\else
\ifnum#1=4643 %
\hatcurextraerrMWtwotwosiglimeccenxxxxxB
\else
??????\fi
\fi
}
\newcommand{\hatcurfieldeccen}[1]{\ifnum#1=762 %
\hatcurfieldeccenxxxxxA
\else
\ifnum#1=4643 %
\hatcurfieldeccenxxxxxB
\else
??????\fi
\fi
}
\newcommand{\hatcurhtreccen}[1]{\ifnum#1=762 %
\hatcurhtreccenxxxxxA
\else
\ifnum#1=4643 %
\hatcurhtreccenxxxxxB
\else
??????\fi
\fi
}
\newcommand{\hatcurISOageeccen}[1]{\ifnum#1=762 %
\hatcurISOageeccenxxxxxA
\else
\ifnum#1=4643 %
\hatcurISOageeccenxxxxxB
\else
??????\fi
\fi
}
\newcommand{\hatcurISOloggeccen}[1]{\ifnum#1=762 %
\hatcurISOloggeccenxxxxxA
\else
\ifnum#1=4643 %
\hatcurISOloggeccenxxxxxB
\else
??????\fi
\fi
}
\newcommand{\hatcurISOlumeccen}[1]{\ifnum#1=762 %
\hatcurISOlumeccenxxxxxA
\else
\ifnum#1=4643 %
\hatcurISOlumeccenxxxxxB
\else
??????\fi
\fi
}
\newcommand{\hatcurISOlumshorteccen}[1]{\ifnum#1=762 %
\hatcurISOlumshorteccenxxxxxA
\else
\ifnum#1=4643 %
\hatcurISOlumshorteccenxxxxxB
\else
??????\fi
\fi
}
\newcommand{\hatcurISOmeccen}[1]{\ifnum#1=762 %
\hatcurISOmeccenxxxxxA
\else
\ifnum#1=4643 %
\hatcurISOmeccenxxxxxB
\else
??????\fi
\fi
}
\newcommand{\hatcurISOmlongeccen}[1]{\ifnum#1=762 %
\hatcurISOmlongeccenxxxxxA
\else
\ifnum#1=4643 %
\hatcurISOmlongeccenxxxxxB
\else
??????\fi
\fi
}
\newcommand{\hatcurISOmshorteccen}[1]{\ifnum#1=762 %
\hatcurISOmshorteccenxxxxxA
\else
\ifnum#1=4643 %
\hatcurISOmshorteccenxxxxxB
\else
??????\fi
\fi
}
\newcommand{\hatcurISOreccen}[1]{\ifnum#1=762 %
\hatcurISOreccenxxxxxA
\else
\ifnum#1=4643 %
\hatcurISOreccenxxxxxB
\else
??????\fi
\fi
}
\newcommand{\hatcurISOrhoeccen}[1]{\ifnum#1=762 %
\hatcurISOrhoeccenxxxxxA
\else
\ifnum#1=4643 %
\hatcurISOrhoeccenxxxxxB
\else
??????\fi
\fi
}
\newcommand{\hatcurISOrholongeccen}[1]{\ifnum#1=762 %
\hatcurISOrholongeccenxxxxxA
\else
\ifnum#1=4643 %
\hatcurISOrholongeccenxxxxxB
\else
??????\fi
\fi
}
\newcommand{\hatcurISOrlongeccen}[1]{\ifnum#1=762 %
\hatcurISOrlongeccenxxxxxA
\else
\ifnum#1=4643 %
\hatcurISOrlongeccenxxxxxB
\else
??????\fi
\fi
}
\newcommand{\hatcurISOrshorteccen}[1]{\ifnum#1=762 %
\hatcurISOrshorteccenxxxxxA
\else
\ifnum#1=4643 %
\hatcurISOrshorteccenxxxxxB
\else
??????\fi
\fi
}
\newcommand{\hatcurISOspececcen}[1]{\ifnum#1=762 %
\hatcurISOspececcenxxxxxA
\else
\ifnum#1=4643 %
\hatcurISOspececcenxxxxxB
\else
??????\fi
\fi
}
\newcommand{\hatcurISOteffeccen}[1]{\ifnum#1=762 %
\hatcurISOteffeccenxxxxxA
\else
\ifnum#1=4643 %
\hatcurISOteffeccenxxxxxB
\else
??????\fi
\fi
}
\newcommand{\hatcurISOzfeheccen}[1]{\ifnum#1=762 %
\hatcurISOzfeheccenxxxxxA
\else
\ifnum#1=4643 %
\hatcurISOzfeheccenxxxxxB
\else
??????\fi
\fi
}
\newcommand{\hatcurLBiBeccen}[1]{\ifnum#1=762 %
\hatcurLBiBeccenxxxxxA
\else
\ifnum#1=4643 %
\hatcurLBiBeccenxxxxxB
\else
??????\fi
\fi
}
\newcommand{\hatcurLBiCeccen}[1]{\ifnum#1=762 %
\hatcurLBiCeccenxxxxxA
\else
\ifnum#1=4643 %
\hatcurLBiCeccenxxxxxB
\else
??????\fi
\fi
}
\newcommand{\hatcurLBigeccen}[1]{\ifnum#1=762 %
\hatcurLBigeccenxxxxxA
\else
\ifnum#1=4643 %
\hatcurLBigeccenxxxxxB
\else
??????\fi
\fi
}
\newcommand{\hatcurLBiHeccen}[1]{\ifnum#1=762 %
\hatcurLBiHeccenxxxxxA
\else
\ifnum#1=4643 %
\hatcurLBiHeccenxxxxxB
\else
??????\fi
\fi
}
\newcommand{\hatcurLBiiBeccen}[1]{\ifnum#1=762 %
\hatcurLBiiBeccenxxxxxA
\else
\ifnum#1=4643 %
\hatcurLBiiBeccenxxxxxB
\else
??????\fi
\fi
}
\newcommand{\hatcurLBiiCeccen}[1]{\ifnum#1=762 %
\hatcurLBiiCeccenxxxxxA
\else
\ifnum#1=4643 %
\hatcurLBiiCeccenxxxxxB
\else
??????\fi
\fi
}
\newcommand{\hatcurLBiieccen}[1]{\ifnum#1=762 %
\hatcurLBiieccenxxxxxA
\else
\ifnum#1=4643 %
\hatcurLBiieccenxxxxxB
\else
??????\fi
\fi
}
\newcommand{\hatcurLBiIeccen}[1]{\ifnum#1=762 %
\hatcurLBiIeccenxxxxxA
\else
\ifnum#1=4643 %
\hatcurLBiIeccenxxxxxB
\else
??????\fi
\fi
}
\newcommand{\hatcurLBiigeccen}[1]{\ifnum#1=762 %
\hatcurLBiigeccenxxxxxA
\else
\ifnum#1=4643 %
\hatcurLBiigeccenxxxxxB
\else
??????\fi
\fi
}
\newcommand{\hatcurLBiiHeccen}[1]{\ifnum#1=762 %
\hatcurLBiiHeccenxxxxxA
\else
\ifnum#1=4643 %
\hatcurLBiiHeccenxxxxxB
\else
??????\fi
\fi
}
\newcommand{\hatcurLBiiieccen}[1]{\ifnum#1=762 %
\hatcurLBiiieccenxxxxxA
\else
\ifnum#1=4643 %
\hatcurLBiiieccenxxxxxB
\else
??????\fi
\fi
}
\newcommand{\hatcurLBiiIeccen}[1]{\ifnum#1=762 %
\hatcurLBiiIeccenxxxxxA
\else
\ifnum#1=4643 %
\hatcurLBiiIeccenxxxxxB
\else
??????\fi
\fi
}
\newcommand{\hatcurLBiiJeccen}[1]{\ifnum#1=762 %
\hatcurLBiiJeccenxxxxxA
\else
\ifnum#1=4643 %
\hatcurLBiiJeccenxxxxxB
\else
??????\fi
\fi
}
\newcommand{\hatcurLBiiKeccen}[1]{\ifnum#1=762 %
\hatcurLBiiKeccenxxxxxA
\else
\ifnum#1=4643 %
\hatcurLBiiKeccenxxxxxB
\else
??????\fi
\fi
}
\newcommand{\hatcurLBiikepeccen}[1]{\ifnum#1=762 %
\hatcurLBiikepeccenxxxxxA
\else
\ifnum#1=4643 %
\hatcurLBiikepeccenxxxxxB
\else
??????\fi
\fi
}
\newcommand{\hatcurLBiiMeccen}[1]{\ifnum#1=762 %
\hatcurLBiiMeccenxxxxxA
\else
\ifnum#1=4643 %
\hatcurLBiiMeccenxxxxxB
\else
??????\fi
\fi
}
\newcommand{\hatcurLBiireccen}[1]{\ifnum#1=762 %
\hatcurLBiireccenxxxxxA
\else
\ifnum#1=4643 %
\hatcurLBiireccenxxxxxB
\else
??????\fi
\fi
}
\newcommand{\hatcurLBiiReccen}[1]{\ifnum#1=762 %
\hatcurLBiiReccenxxxxxA
\else
\ifnum#1=4643 %
\hatcurLBiiReccenxxxxxB
\else
??????\fi
\fi
}
\newcommand{\hatcurLBiiSfoureccen}[1]{\ifnum#1=762 %
\hatcurLBiiSfoureccenxxxxxA
\else
\ifnum#1=4643 %
\hatcurLBiiSfoureccenxxxxxB
\else
??????\fi
\fi
}
\newcommand{\hatcurLBiiSoneeccen}[1]{\ifnum#1=762 %
\hatcurLBiiSoneeccenxxxxxA
\else
\ifnum#1=4643 %
\hatcurLBiiSoneeccenxxxxxB
\else
??????\fi
\fi
}
\newcommand{\hatcurLBiiSthreeeccen}[1]{\ifnum#1=762 %
\hatcurLBiiSthreeeccenxxxxxA
\else
\ifnum#1=4643 %
\hatcurLBiiSthreeeccenxxxxxB
\else
??????\fi
\fi
}
\newcommand{\hatcurLBiiStwoeccen}[1]{\ifnum#1=762 %
\hatcurLBiiStwoeccenxxxxxA
\else
\ifnum#1=4643 %
\hatcurLBiiStwoeccenxxxxxB
\else
??????\fi
\fi
}
\newcommand{\hatcurLBiiTeccen}[1]{\ifnum#1=762 %
\hatcurLBiiTeccenxxxxxA
\else
\ifnum#1=4643 %
\hatcurLBiiTeccenxxxxxB
\else
??????\fi
\fi
}
\newcommand{\hatcurLBiiueccen}[1]{\ifnum#1=762 %
\hatcurLBiiueccenxxxxxA
\else
\ifnum#1=4643 %
\hatcurLBiiueccenxxxxxB
\else
??????\fi
\fi
}
\newcommand{\hatcurLBiiVeccen}[1]{\ifnum#1=762 %
\hatcurLBiiVeccenxxxxxA
\else
\ifnum#1=4643 %
\hatcurLBiiVeccenxxxxxB
\else
??????\fi
\fi
}
\newcommand{\hatcurLBiizeccen}[1]{\ifnum#1=762 %
\hatcurLBiizeccenxxxxxA
\else
\ifnum#1=4643 %
\hatcurLBiizeccenxxxxxB
\else
??????\fi
\fi
}
\newcommand{\hatcurLBiJeccen}[1]{\ifnum#1=762 %
\hatcurLBiJeccenxxxxxA
\else
\ifnum#1=4643 %
\hatcurLBiJeccenxxxxxB
\else
??????\fi
\fi
}
\newcommand{\hatcurLBiKeccen}[1]{\ifnum#1=762 %
\hatcurLBiKeccenxxxxxA
\else
\ifnum#1=4643 %
\hatcurLBiKeccenxxxxxB
\else
??????\fi
\fi
}
\newcommand{\hatcurLBikepeccen}[1]{\ifnum#1=762 %
\hatcurLBikepeccenxxxxxA
\else
\ifnum#1=4643 %
\hatcurLBikepeccenxxxxxB
\else
??????\fi
\fi
}
\newcommand{\hatcurLBiMeccen}[1]{\ifnum#1=762 %
\hatcurLBiMeccenxxxxxA
\else
\ifnum#1=4643 %
\hatcurLBiMeccenxxxxxB
\else
??????\fi
\fi
}
\newcommand{\hatcurLBireccen}[1]{\ifnum#1=762 %
\hatcurLBireccenxxxxxA
\else
\ifnum#1=4643 %
\hatcurLBireccenxxxxxB
\else
??????\fi
\fi
}
\newcommand{\hatcurLBiReccen}[1]{\ifnum#1=762 %
\hatcurLBiReccenxxxxxA
\else
\ifnum#1=4643 %
\hatcurLBiReccenxxxxxB
\else
??????\fi
\fi
}
\newcommand{\hatcurLBiSfoureccen}[1]{\ifnum#1=762 %
\hatcurLBiSfoureccenxxxxxA
\else
\ifnum#1=4643 %
\hatcurLBiSfoureccenxxxxxB
\else
??????\fi
\fi
}
\newcommand{\hatcurLBiSoneeccen}[1]{\ifnum#1=762 %
\hatcurLBiSoneeccenxxxxxA
\else
\ifnum#1=4643 %
\hatcurLBiSoneeccenxxxxxB
\else
??????\fi
\fi
}
\newcommand{\hatcurLBiSthreeeccen}[1]{\ifnum#1=762 %
\hatcurLBiSthreeeccenxxxxxA
\else
\ifnum#1=4643 %
\hatcurLBiSthreeeccenxxxxxB
\else
??????\fi
\fi
}
\newcommand{\hatcurLBiStwoeccen}[1]{\ifnum#1=762 %
\hatcurLBiStwoeccenxxxxxA
\else
\ifnum#1=4643 %
\hatcurLBiStwoeccenxxxxxB
\else
??????\fi
\fi
}
\newcommand{\hatcurLBiTeccen}[1]{\ifnum#1=762 %
\hatcurLBiTeccenxxxxxA
\else
\ifnum#1=4643 %
\hatcurLBiTeccenxxxxxB
\else
??????\fi
\fi
}
\newcommand{\hatcurLBiueccen}[1]{\ifnum#1=762 %
\hatcurLBiueccenxxxxxA
\else
\ifnum#1=4643 %
\hatcurLBiueccenxxxxxB
\else
??????\fi
\fi
}
\newcommand{\hatcurLBiVeccen}[1]{\ifnum#1=762 %
\hatcurLBiVeccenxxxxxA
\else
\ifnum#1=4643 %
\hatcurLBiVeccenxxxxxB
\else
??????\fi
\fi
}
\newcommand{\hatcurLBizeccen}[1]{\ifnum#1=762 %
\hatcurLBizeccenxxxxxA
\else
\ifnum#1=4643 %
\hatcurLBizeccenxxxxxB
\else
??????\fi
\fi
}
\newcommand{\hatcurLCbsqeccen}[1]{\ifnum#1=762 %
\hatcurLCbsqeccenxxxxxA
\else
\ifnum#1=4643 %
\hatcurLCbsqeccenxxxxxB
\else
??????\fi
\fi
}
\newcommand{\hatcurLCdipeccen}[1]{\ifnum#1=762 %
\hatcurLCdipeccenxxxxxA
\else
\ifnum#1=4643 %
\hatcurLCdipeccenxxxxxB
\else
??????\fi
\fi
}
\newcommand{\hatcurLCdureccen}[1]{\ifnum#1=762 %
\hatcurLCdureccenxxxxxA
\else
\ifnum#1=4643 %
\hatcurLCdureccenxxxxxB
\else
??????\fi
\fi
}
\newcommand{\hatcurLCdurhreccen}[1]{\ifnum#1=762 %
\hatcurLCdurhreccenxxxxxA
\else
\ifnum#1=4643 %
\hatcurLCdurhreccenxxxxxB
\else
??????\fi
\fi
}
\newcommand{\hatcurLCdurhrshorteccen}[1]{\ifnum#1=762 %
\hatcurLCdurhrshorteccenxxxxxA
\else
\ifnum#1=4643 %
\hatcurLCdurhrshorteccenxxxxxB
\else
??????\fi
\fi
}
\newcommand{\hatcurLCdurshorteccen}[1]{\ifnum#1=762 %
\hatcurLCdurshorteccenxxxxxA
\else
\ifnum#1=4643 %
\hatcurLCdurshorteccenxxxxxB
\else
??????\fi
\fi
}
\newcommand{\hatcurLChatnetmAeccen}[1]{\ifnum#1=762 %
\hatcurLChatnetmAeccenxxxxxA
\else
\ifnum#1=4643 %
\hatcurLChatnetmAeccenxxxxxB
\else
??????\fi
\fi
}
\newcommand{\hatcurLChatnetmBeccen}[1]{\ifnum#1=762 %
\hatcurLChatnetmBeccenxxxxxA
\else
\ifnum#1=4643 %
\hatcurLChatnetmBeccenxxxxxB
\else
??????\fi
\fi
}
\newcommand{\hatcurLChatnetmCeccen}[1]{\ifnum#1=762 %
\hatcurLChatnetmCeccenxxxxxA
\else
??????\fi
}
\newcommand{\hatcurLChatnetmDeccen}[1]{\ifnum#1=762 %
\hatcurLChatnetmDeccenxxxxxA
\else
??????\fi
}
\newcommand{\hatcurLCiblendAeccen}[1]{\ifnum#1=762 %
\hatcurLCiblendAeccenxxxxxA
\else
\ifnum#1=4643 %
\hatcurLCiblendAeccenxxxxxB
\else
??????\fi
\fi
}
\newcommand{\hatcurLCiblendBeccen}[1]{\ifnum#1=762 %
\hatcurLCiblendBeccenxxxxxA
\else
\ifnum#1=4643 %
\hatcurLCiblendBeccenxxxxxB
\else
??????\fi
\fi
}
\newcommand{\hatcurLCiblendCeccen}[1]{\ifnum#1=762 %
\hatcurLCiblendCeccenxxxxxA
\else
??????\fi
}
\newcommand{\hatcurLCiblendDeccen}[1]{\ifnum#1=762 %
\hatcurLCiblendDeccenxxxxxA
\else
??????\fi
}
\newcommand{\hatcurLCimpeccen}[1]{\ifnum#1=762 %
\hatcurLCimpeccenxxxxxA
\else
\ifnum#1=4643 %
\hatcurLCimpeccenxxxxxB
\else
??????\fi
\fi
}
\newcommand{\hatcurLCingdureccen}[1]{\ifnum#1=762 %
\hatcurLCingdureccenxxxxxA
\else
\ifnum#1=4643 %
\hatcurLCingdureccenxxxxxB
\else
??????\fi
\fi
}
\newcommand{\hatcurLCPeccen}[1]{\ifnum#1=762 %
\hatcurLCPeccenxxxxxA
\else
\ifnum#1=4643 %
\hatcurLCPeccenxxxxxB
\else
??????\fi
\fi
}
\newcommand{\hatcurLCPprececcen}[1]{\ifnum#1=762 %
\hatcurLCPprececcenxxxxxA
\else
\ifnum#1=4643 %
\hatcurLCPprececcenxxxxxB
\else
??????\fi
\fi
}
\newcommand{\hatcurLCPshorteccen}[1]{\ifnum#1=762 %
\hatcurLCPshorteccenxxxxxA
\else
\ifnum#1=4643 %
\hatcurLCPshorteccenxxxxxB
\else
??????\fi
\fi
}
\newcommand{\hatcurLCqeccen}[1]{\ifnum#1=762 %
\hatcurLCqeccenxxxxxA
\else
\ifnum#1=4643 %
\hatcurLCqeccenxxxxxB
\else
??????\fi
\fi
}
\newcommand{\hatcurLCqshorteccen}[1]{\ifnum#1=762 %
\hatcurLCqshorteccenxxxxxA
\else
\ifnum#1=4643 %
\hatcurLCqshorteccenxxxxxB
\else
??????\fi
\fi
}
\newcommand{\hatcurLCrhoeccen}[1]{\ifnum#1=762 %
\hatcurLCrhoeccenxxxxxA
\else
\ifnum#1=4643 %
\hatcurLCrhoeccenxxxxxB
\else
??????\fi
\fi
}
\newcommand{\hatcurLCrprstareccen}[1]{\ifnum#1=762 %
\hatcurLCrprstareccenxxxxxA
\else
\ifnum#1=4643 %
\hatcurLCrprstareccenxxxxxB
\else
??????\fi
\fi
}
\newcommand{\hatcurLCTAeccen}[1]{\ifnum#1=762 %
\hatcurLCTAeccenxxxxxA
\else
\ifnum#1=4643 %
\hatcurLCTAeccenxxxxxB
\else
??????\fi
\fi
}
\newcommand{\hatcurLCTBeccen}[1]{\ifnum#1=762 %
\hatcurLCTBeccenxxxxxA
\else
\ifnum#1=4643 %
\hatcurLCTBeccenxxxxxB
\else
??????\fi
\fi
}
\newcommand{\hatcurLCTeccen}[1]{\ifnum#1=762 %
\hatcurLCTeccenxxxxxA
\else
\ifnum#1=4643 %
\hatcurLCTeccenxxxxxB
\else
??????\fi
\fi
}
\newcommand{\hatcurLCzetaeccen}[1]{\ifnum#1=762 %
\hatcurLCzetaeccenxxxxxA
\else
\ifnum#1=4643 %
\hatcurLCzetaeccenxxxxxB
\else
??????\fi
\fi
}
\newcommand{\hatcurPPaequiveccen}[1]{\ifnum#1=762 %
\hatcurPPaequiveccenxxxxxA
\else
\ifnum#1=4643 %
\hatcurPPaequiveccenxxxxxB
\else
??????\fi
\fi
}
\newcommand{\hatcurPPareccen}[1]{\ifnum#1=762 %
\hatcurPPareccenxxxxxA
\else
\ifnum#1=4643 %
\hatcurPPareccenxxxxxB
\else
??????\fi
\fi
}
\newcommand{\hatcurPPareleccen}[1]{\ifnum#1=762 %
\hatcurPPareleccenxxxxxA
\else
\ifnum#1=4643 %
\hatcurPPareleccenxxxxxB
\else
??????\fi
\fi
}
\newcommand{\hatcurPPfluxapdimeccen}[1]{\ifnum#1=762 %
\hatcurPPfluxapdimeccenxxxxxA
\else
\ifnum#1=4643 %
\hatcurPPfluxapdimeccenxxxxxB
\else
??????\fi
\fi
}
\newcommand{\hatcurPPfluxapeccen}[1]{\ifnum#1=762 %
\hatcurPPfluxapeccenxxxxxA
\else
\ifnum#1=4643 %
\hatcurPPfluxapeccenxxxxxB
\else
??????\fi
\fi
}
\newcommand{\hatcurPPfluxavgdimeccen}[1]{\ifnum#1=762 %
\hatcurPPfluxavgdimeccenxxxxxA
\else
\ifnum#1=4643 %
\hatcurPPfluxavgdimeccenxxxxxB
\else
??????\fi
\fi
}
\newcommand{\hatcurPPfluxavgeccen}[1]{\ifnum#1=762 %
\hatcurPPfluxavgeccenxxxxxA
\else
\ifnum#1=4643 %
\hatcurPPfluxavgeccenxxxxxB
\else
??????\fi
\fi
}
\newcommand{\hatcurPPfluxavglogeccen}[1]{\ifnum#1=762 %
\hatcurPPfluxavglogeccenxxxxxA
\else
\ifnum#1=4643 %
\hatcurPPfluxavglogeccenxxxxxB
\else
??????\fi
\fi
}
\newcommand{\hatcurPPfluxperidimeccen}[1]{\ifnum#1=762 %
\hatcurPPfluxperidimeccenxxxxxA
\else
\ifnum#1=4643 %
\hatcurPPfluxperidimeccenxxxxxB
\else
??????\fi
\fi
}
\newcommand{\hatcurPPfluxperieccen}[1]{\ifnum#1=762 %
\hatcurPPfluxperieccenxxxxxA
\else
\ifnum#1=4643 %
\hatcurPPfluxperieccenxxxxxB
\else
??????\fi
\fi
}
\newcommand{\hatcurPPgeccen}[1]{\ifnum#1=762 %
\hatcurPPgeccenxxxxxA
\else
\ifnum#1=4643 %
\hatcurPPgeccenxxxxxB
\else
??????\fi
\fi
}
\newcommand{\hatcurPPieccen}[1]{\ifnum#1=762 %
\hatcurPPieccenxxxxxA
\else
\ifnum#1=4643 %
\hatcurPPieccenxxxxxB
\else
??????\fi
\fi
}
\newcommand{\hatcurPPloggeccen}[1]{\ifnum#1=762 %
\hatcurPPloggeccenxxxxxA
\else
\ifnum#1=4643 %
\hatcurPPloggeccenxxxxxB
\else
??????\fi
\fi
}
\newcommand{\hatcurPPmeccen}[1]{\ifnum#1=762 %
\hatcurPPmeccenxxxxxA
\else
\ifnum#1=4643 %
\hatcurPPmeccenxxxxxB
\else
??????\fi
\fi
}
\newcommand{\hatcurPPmeeccen}[1]{\ifnum#1=762 %
\hatcurPPmeeccenxxxxxA
\else
\ifnum#1=4643 %
\hatcurPPmeeccenxxxxxB
\else
??????\fi
\fi
}
\newcommand{\hatcurPPmelongeccen}[1]{\ifnum#1=762 %
\hatcurPPmelongeccenxxxxxA
\else
\ifnum#1=4643 %
\hatcurPPmelongeccenxxxxxB
\else
??????\fi
\fi
}
\newcommand{\hatcurPPmeshorteccen}[1]{\ifnum#1=762 %
\hatcurPPmeshorteccenxxxxxA
\else
\ifnum#1=4643 %
\hatcurPPmeshorteccenxxxxxB
\else
??????\fi
\fi
}
\newcommand{\hatcurPPmlongeccen}[1]{\ifnum#1=762 %
\hatcurPPmlongeccenxxxxxA
\else
\ifnum#1=4643 %
\hatcurPPmlongeccenxxxxxB
\else
??????\fi
\fi
}
\newcommand{\hatcurPPmrcorreccen}[1]{\ifnum#1=762 %
\hatcurPPmrcorreccenxxxxxA
\else
\ifnum#1=4643 %
\hatcurPPmrcorreccenxxxxxB
\else
??????\fi
\fi
}
\newcommand{\hatcurPPmshorteccen}[1]{\ifnum#1=762 %
\hatcurPPmshorteccenxxxxxA
\else
\ifnum#1=4643 %
\hatcurPPmshorteccenxxxxxB
\else
??????\fi
\fi
}
\newcommand{\hatcurPPmtwosiglimeccen}[1]{\ifnum#1=762 %
\hatcurPPmtwosiglimeccenxxxxxA
\else
\ifnum#1=4643 %
\hatcurPPmtwosiglimeccenxxxxxB
\else
??????\fi
\fi
}
\newcommand{\hatcurPPperieccen}[1]{\ifnum#1=762 %
\hatcurPPperieccenxxxxxA
\else
\ifnum#1=4643 %
\hatcurPPperieccenxxxxxB
\else
??????\fi
\fi
}
\newcommand{\hatcurPPphiconjeccen}[1]{\ifnum#1=762 %
\hatcurPPphiconjeccenxxxxxA
\else
\ifnum#1=4643 %
\hatcurPPphiconjeccenxxxxxB
\else
??????\fi
\fi
}
\newcommand{\hatcurPPreccen}[1]{\ifnum#1=762 %
\hatcurPPreccenxxxxxA
\else
\ifnum#1=4643 %
\hatcurPPreccenxxxxxB
\else
??????\fi
\fi
}
\newcommand{\hatcurPPreeccen}[1]{\ifnum#1=762 %
\hatcurPPreeccenxxxxxA
\else
\ifnum#1=4643 %
\hatcurPPreeccenxxxxxB
\else
??????\fi
\fi
}
\newcommand{\hatcurPPrelongeccen}[1]{\ifnum#1=762 %
\hatcurPPrelongeccenxxxxxA
\else
\ifnum#1=4643 %
\hatcurPPrelongeccenxxxxxB
\else
??????\fi
\fi
}
\newcommand{\hatcurPPreshorteccen}[1]{\ifnum#1=762 %
\hatcurPPreshorteccenxxxxxA
\else
\ifnum#1=4643 %
\hatcurPPreshorteccenxxxxxB
\else
??????\fi
\fi
}
\newcommand{\hatcurPPrhoeccen}[1]{\ifnum#1=762 %
\hatcurPPrhoeccenxxxxxA
\else
\ifnum#1=4643 %
\hatcurPPrhoeccenxxxxxB
\else
??????\fi
\fi
}
\newcommand{\hatcurPPrlongeccen}[1]{\ifnum#1=762 %
\hatcurPPrlongeccenxxxxxA
\else
\ifnum#1=4643 %
\hatcurPPrlongeccenxxxxxB
\else
??????\fi
\fi
}
\newcommand{\hatcurPPrshorteccen}[1]{\ifnum#1=762 %
\hatcurPPrshorteccenxxxxxA
\else
\ifnum#1=4643 %
\hatcurPPrshorteccenxxxxxB
\else
??????\fi
\fi
}
\newcommand{\hatcurPPtcirceccen}[1]{\ifnum#1=762 %
\hatcurPPtcirceccenxxxxxA
\else
\ifnum#1=4643 %
\hatcurPPtcirceccenxxxxxB
\else
??????\fi
\fi
}
\newcommand{\hatcurPPteffeccen}[1]{\ifnum#1=762 %
\hatcurPPteffeccenxxxxxA
\else
\ifnum#1=4643 %
\hatcurPPteffeccenxxxxxB
\else
??????\fi
\fi
}
\newcommand{\hatcurPPthetaeccen}[1]{\ifnum#1=762 %
\hatcurPPthetaeccenxxxxxA
\else
\ifnum#1=4643 %
\hatcurPPthetaeccenxxxxxB
\else
??????\fi
\fi
}
\newcommand{\hatcurPPtinfalleccen}[1]{\ifnum#1=762 %
\hatcurPPtinfalleccenxxxxxA
\else
\ifnum#1=4643 %
\hatcurPPtinfalleccenxxxxxB
\else
??????\fi
\fi
}
\newcommand{\hatcurRVecceneccen}[1]{\ifnum#1=762 %
\hatcurRVecceneccenxxxxxA
\else
\ifnum#1=4643 %
\hatcurRVecceneccenxxxxxB
\else
??????\fi
\fi
}
\newcommand{\hatcurRVeccentwosiglimeccen}[1]{\ifnum#1=762 %
\hatcurRVeccentwosiglimeccenxxxxxA
\else
\ifnum#1=4643 %
\hatcurRVeccentwosiglimeccenxxxxxB
\else
??????\fi
\fi
}
\newcommand{\hatcurRVfitrmseccen}[1]{\ifnum#1=762 %
\hatcurRVfitrmseccenxxxxxA
\else
\ifnum#1=4643 %
\hatcurRVfitrmseccenxxxxxB
\else
??????\fi
\fi
}
\newcommand{\hatcurRVgammaeccen}[1]{\ifnum#1=762 %
\hatcurRVgammaeccenxxxxxA
\else
\ifnum#1=4643 %
\hatcurRVgammaeccenxxxxxB
\else
??????\fi
\fi
}
\newcommand{\hatcurRVheccen}[1]{\ifnum#1=762 %
\hatcurRVheccenxxxxxA
\else
\ifnum#1=4643 %
\hatcurRVheccenxxxxxB
\else
??????\fi
\fi
}
\newcommand{\hatcurRVjittereccen}[1]{\ifnum#1=762 %
\hatcurRVjittereccenxxxxxA
\else
\ifnum#1=4643 %
\hatcurRVjittereccenxxxxxB
\else
??????\fi
\fi
}
\newcommand{\hatcurRVjittertwosiglimeccen}[1]{\ifnum#1=762 %
\hatcurRVjittertwosiglimeccenxxxxxA
\else
\ifnum#1=4643 %
\hatcurRVjittertwosiglimeccenxxxxxB
\else
??????\fi
\fi
}
\newcommand{\hatcurRVkeccen}[1]{\ifnum#1=762 %
\hatcurRVkeccenxxxxxA
\else
\ifnum#1=4643 %
\hatcurRVkeccenxxxxxB
\else
??????\fi
\fi
}
\newcommand{\hatcurRVKeccen}[1]{\ifnum#1=762 %
\hatcurRVKeccenxxxxxA
\else
\ifnum#1=4643 %
\hatcurRVKeccenxxxxxB
\else
??????\fi
\fi
}
\newcommand{\hatcurRVKtwosiglimeccen}[1]{\ifnum#1=762 %
\hatcurRVKtwosiglimeccenxxxxxA
\else
\ifnum#1=4643 %
\hatcurRVKtwosiglimeccenxxxxxB
\else
??????\fi
\fi
}
\newcommand{\hatcurRVomegaeccen}[1]{\ifnum#1=762 %
\hatcurRVomegaeccenxxxxxA
\else
\ifnum#1=4643 %
\hatcurRVomegaeccenxxxxxB
\else
??????\fi
\fi
}
\newcommand{\hatcurRVrheccen}[1]{\ifnum#1=762 %
\hatcurRVrheccenxxxxxA
\else
\ifnum#1=4643 %
\hatcurRVrheccenxxxxxB
\else
??????\fi
\fi
}
\newcommand{\hatcurRVrkeccen}[1]{\ifnum#1=762 %
\hatcurRVrkeccenxxxxxA
\else
\ifnum#1=4643 %
\hatcurRVrkeccenxxxxxB
\else
??????\fi
\fi
}
\newcommand{\hatcurRVtroneeccen}[1]{\ifnum#1=762 %
\hatcurRVtroneeccenxxxxxA
\else
\ifnum#1=4643 %
\hatcurRVtroneeccenxxxxxB
\else
??????\fi
\fi
}
\newcommand{\hatcurRVtrtwoeccen}[1]{\ifnum#1=762 %
\hatcurRVtrtwoeccenxxxxxA
\else
\ifnum#1=4643 %
\hatcurRVtrtwoeccenxxxxxB
\else
??????\fi
\fi
}
\newcommand{\hatcurSMEiloggeccen}[1]{\ifnum#1=762 %
\hatcurSMEiloggeccenxxxxxA
\else
\ifnum#1=4643 %
\hatcurSMEiloggeccenxxxxxB
\else
??????\fi
\fi
}
\newcommand{\hatcurSMEiteffeccen}[1]{\ifnum#1=762 %
\hatcurSMEiteffeccenxxxxxA
\else
\ifnum#1=4643 %
\hatcurSMEiteffeccenxxxxxB
\else
??????\fi
\fi
}
\newcommand{\hatcurSMEivmaceccen}[1]{\ifnum#1=762 %
\hatcurSMEivmaceccenxxxxxA
\else
\ifnum#1=4643 %
\hatcurSMEivmaceccenxxxxxB
\else
??????\fi
\fi
}
\newcommand{\hatcurSMEivmiceccen}[1]{\ifnum#1=762 %
\hatcurSMEivmiceccenxxxxxA
\else
\ifnum#1=4643 %
\hatcurSMEivmiceccenxxxxxB
\else
??????\fi
\fi
}
\newcommand{\hatcurSMEivsineccen}[1]{\ifnum#1=762 %
\hatcurSMEivsineccenxxxxxA
\else
\ifnum#1=4643 %
\hatcurSMEivsineccenxxxxxB
\else
??????\fi
\fi
}
\newcommand{\hatcurSMEizfeheccen}[1]{\ifnum#1=762 %
\hatcurSMEizfeheccenxxxxxA
\else
\ifnum#1=4643 %
\hatcurSMEizfeheccenxxxxxB
\else
??????\fi
\fi
}
\newcommand{\hatcurSMEizfehshorteccen}[1]{\ifnum#1=762 %
\hatcurSMEizfehshorteccenxxxxxA
\else
\ifnum#1=4643 %
\hatcurSMEizfehshorteccenxxxxxB
\else
??????\fi
\fi
}
\newcommand{\hatcurXAveccen}[1]{\ifnum#1=762 %
\hatcurXAveccenxxxxxA
\else
\ifnum#1=4643 %
\hatcurXAveccenxxxxxB
\else
??????\fi
\fi
}
\newcommand{\hatcurXdisteccen}[1]{\ifnum#1=762 %
\hatcurXdisteccenxxxxxA
\else
\ifnum#1=4643 %
\hatcurXdisteccenxxxxxB
\else
??????\fi
\fi
}
\newcommand{\hatcurXdistredeccen}[1]{\ifnum#1=762 %
\hatcurXdistredeccenxxxxxA
\else
\ifnum#1=4643 %
\hatcurXdistredeccenxxxxxB
\else
??????\fi
\fi
}
\newcommand{\hatcurXEBVeccen}[1]{\ifnum#1=762 %
\hatcurXEBVeccenxxxxxA
\else
\ifnum#1=4643 %
\hatcurXEBVeccenxxxxxB
\else
??????\fi
\fi
}
\newcommand{\hatcurXsecdureccen}[1]{\ifnum#1=762 %
\hatcurXsecdureccenxxxxxA
\else
\ifnum#1=4643 %
\hatcurXsecdureccenxxxxxB
\else
??????\fi
\fi
}
\newcommand{\hatcurXsecingdureccen}[1]{\ifnum#1=762 %
\hatcurXsecingdureccenxxxxxA
\else
\ifnum#1=4643 %
\hatcurXsecingdureccenxxxxxB
\else
??????\fi
\fi
}
\newcommand{\hatcurXsecondaryeccen}[1]{\ifnum#1=762 %
\hatcurXsecondaryeccenxxxxxA
\else
\ifnum#1=4643 %
\hatcurXsecondaryeccenxxxxxB
\else
??????\fi
\fi
}
\newcommand{\hatcurXsecphaseeccen}[1]{\ifnum#1=762 %
\hatcurXsecphaseeccenxxxxxA
\else
\ifnum#1=4643 %
\hatcurXsecphaseeccenxxxxxB
\else
??????\fi
\fi
}
\newcommand{\hatcurxxxxxxA}{TOI~762\,A}
\newcommand{\hatcurbxxxxxxA}{TOI~762\,A\,b}
\newcommand{\hatcurcxxxxxxA}{TOI~762\,A\,c}

\newcommand{\hatcurplanetnumxxxxxxA}{762}

\newcommand{\hatcurCCtwomassshortxxxxxxA}{11041818-4749169}

\newcommand{\hatcurCCticxxxxxxA}{178709444}

\newcommand{\hatcurCCtoixxxxxxA}{762}

\newcommand{\hatcurCCgaiadrtwoshortxxxxxxA}{5362352744504000256}

\newcommand{\hatcurRVgammaabsxxxxxxA}{\hatcurRVgammaA{\hatcurplanetnumxxxxxxA}}                           

\newcommand{\hatcurRVgammarelxxxxxxA}{\hatcurRVgammaA{\hatcurplanetnumxxxxxxA}}                           

\newcommand{\hatcurCCtassvixxxxxxA}{\ensuremath{NULL\pm NULL}}                  

\newcommand{\hatcurSMEversionxxxxxxA}{i}                                       

\newcommand{\hatcurisoshortxxxxxxA}{YY}
\newcommand{\hatcurisofullxxxxxxA}{Yonsei-Yale (YY)}
\newcommand{\hatcurisocitexxxxxxA}{yi:2001}

\newcommand{\hatcurlumindxxxxxxA}{\arstar}

\newcommand{\hatcurjhkfilsetxxxxxxA}{ESO}

\newcommand{\hatcurSMEteffxxxxxxA}{\ifthenelse{\equal{\hatcurSMEversionxxxxxxA}{i}}{\hatcurSMEiteff{\hatcurplanetnumxxxxxxA}}{\hatcurSMEiiteff{\hatcurplanetnumxxxxxxA}}}
\newcommand{\hatcurSMEzfehxxxxxxA}{\ifthenelse{\equal{\hatcurSMEversionxxxxxxA}{i}}{\hatcurSMEizfeh{\hatcurplanetnumxxxxxxA}}{\hatcurSMEiizfeh{\hatcurplanetnumxxxxxxA}}}
\newcommand{\hatcurSMEzfehshortxxxxxxA}{\ifthenelse{\equal{\hatcurSMEversionxxxxxxA}{i}}{\hatcurSMEizfehshort{\hatcurplanetnumxxxxxxA}}{\hatcurSMEiizfehshort{\hatcurplanetnumxxxxxxA}}}
\newcommand{\hatcurSMEloggxxxxxxA}{\ifthenelse{\equal{\hatcurSMEversionxxxxxxA}{i}}{\hatcurSMEilogg{\hatcurplanetnumxxxxxxA}}{\hatcurSMEiilogg{\hatcurplanetnumxxxxxxA}}}
\newcommand{\hatcurSMEvsinxxxxxxA}{\ifthenelse{\equal{\hatcurSMEversionxxxxxxA}{i}}{\hatcurSMEivsin{\hatcurplanetnumxxxxxxA}}{\hatcurSMEiivsin{\hatcurplanetnumxxxxxxA}}}
\newcommand{\hatcurSMEvmacxxxxxxA}{\ifthenelse{\equal{\hatcurSMEversionxxxxxxA}{i}}{\hatcurSMEivmac{\hatcurplanetnumxxxxxxA}}{\hatcurSMEiivmac{\hatcurplanetnumxxxxxxA}}}
\newcommand{\hatcurSMEvmicxxxxxxA}{\ifthenelse{\equal{\hatcurSMEversionxxxxxxA}{i}}{\hatcurSMEivmic{\hatcurplanetnumxxxxxxA}}{\hatcurSMEiivmic{\hatcurplanetnumxxxxxxA}}}

\newcommand{\hatcurxxxxxxB}{TIC~46432937}
\newcommand{\hatcurbxxxxxxB}{TIC~46432937\,b}
\newcommand{\hatcurcxxxxxxB}{TIC~46432937\,c}

\newcommand{\hatcurplanetnumxxxxxxB}{4643}

\newcommand{\hatcurCCtwomassshortxxxxxxB}{05352856-1435504}

\newcommand{\hatcurCCticxxxxxxB}{46432937}

\newcommand{\hatcurCCtoixxxxxxB}{\ensuremath{\cdots}}

\newcommand{\hatcurCCgaiadrtwoshortxxxxxxB}{2984391358868786816}

\newcommand{\hatcurRVgammaabsxxxxxxB}{\hatcurRVgamma{\hatcurplanetnumxxxxxxB}}                           

\newcommand{\hatcurRVgammarelxxxxxxB}{\hatcurRVgamma{\hatcurplanetnumxxxxxxB}}                           

\newcommand{\hatcurCCtassvixxxxxxB}{\ensuremath{NULL\pm NULL}}                  

\newcommand{\hatcurSMEversionxxxxxxB}{i}                                       

\newcommand{\hatcurisoshortxxxxxxB}{YY}
\newcommand{\hatcurisofullxxxxxxB}{Yonsei-Yale (YY)}
\newcommand{\hatcurisocitexxxxxxB}{yi:2001}

\newcommand{\hatcurlumindxxxxxxB}{\arstar}

\newcommand{\hatcurjhkfilsetxxxxxxB}{ESO}

\newcommand{\hatcurSMEteffxxxxxxB}{\ifthenelse{\equal{\hatcurSMEversionxxxxxxB}{i}}{\hatcurSMEiteff{\hatcurplanetnumxxxxxxB}}{\hatcurSMEiiteff{\hatcurplanetnumxxxxxxB}}}
\newcommand{\hatcurSMEzfehxxxxxxB}{\ifthenelse{\equal{\hatcurSMEversionxxxxxxB}{i}}{\hatcurSMEizfeh{\hatcurplanetnumxxxxxxB}}{\hatcurSMEiizfeh{\hatcurplanetnumxxxxxxB}}}
\newcommand{\hatcurSMEzfehshortxxxxxxB}{\ifthenelse{\equal{\hatcurSMEversionxxxxxxB}{i}}{\hatcurSMEizfehshort{\hatcurplanetnumxxxxxxB}}{\hatcurSMEiizfehshort{\hatcurplanetnumxxxxxxB}}}
\newcommand{\hatcurSMEloggxxxxxxB}{\ifthenelse{\equal{\hatcurSMEversionxxxxxxB}{i}}{\hatcurSMEilogg{\hatcurplanetnumxxxxxxB}}{\hatcurSMEiilogg{\hatcurplanetnumxxxxxxB}}}
\newcommand{\hatcurSMEvsinxxxxxxB}{\ifthenelse{\equal{\hatcurSMEversionxxxxxxB}{i}}{\hatcurSMEivsin{\hatcurplanetnumxxxxxxB}}{\hatcurSMEiivsin{\hatcurplanetnumxxxxxxB}}}
\newcommand{\hatcurSMEvmacxxxxxxB}{\ifthenelse{\equal{\hatcurSMEversionxxxxxxB}{i}}{\hatcurSMEivmac{\hatcurplanetnumxxxxxxB}}{\hatcurSMEiivmac{\hatcurplanetnumxxxxxxB}}}
\newcommand{\hatcurSMEvmicxxxxxxB}{\ifthenelse{\equal{\hatcurSMEversionxxxxxxB}{i}}{\hatcurSMEivmic{\hatcurplanetnumxxxxxxB}}{\hatcurSMEiivmic{\hatcurplanetnumxxxxxxB}}}
\newcommand{\hatcur}[1]{\ifnum#1=762 %
\hatcurxxxxxxA
\else
\ifnum#1=4643 %
\hatcurxxxxxxB
\else
??????\fi
\fi
}
\newcommand{\hatcurb}[1]{\ifnum#1=762 %
\hatcurbxxxxxxA
\else
\ifnum#1=4643 %
\hatcurbxxxxxxB
\else
??????\fi
\fi
}
\newcommand{\hatcurc}[1]{\ifnum#1=762 %
\hatcurcxxxxxxA
\else
\ifnum#1=4643 %
\hatcurcxxxxxxB
\else
??????\fi
\fi
}
\newcommand{\hatcurCCgaiadrtwoshort}[1]{\ifnum#1=762 %
\hatcurCCgaiadrtwoshortxxxxxxA
\else
\ifnum#1=4643 %
\hatcurCCgaiadrtwoshortxxxxxxB
\else
??????\fi
\fi
}
\newcommand{\hatcurCCtassvi}[1]{\ifnum#1=762 %
\hatcurCCtassvixxxxxxA
\else
\ifnum#1=4643 %
\hatcurCCtassvixxxxxxB
\else
??????\fi
\fi
}
\newcommand{\hatcurCCtic}[1]{\ifnum#1=762 %
\hatcurCCticxxxxxxA
\else
\ifnum#1=4643 %
\hatcurCCticxxxxxxB
\else
??????\fi
\fi
}
\newcommand{\hatcurCCtoi}[1]{\ifnum#1=762 %
\hatcurCCtoixxxxxxA
\else
\ifnum#1=4643 %
\hatcurCCtoixxxxxxB
\else
??????\fi
\fi
}
\newcommand{\hatcurCCtwomassshort}[1]{\ifnum#1=762 %
\hatcurCCtwomassshortxxxxxxA
\else
\ifnum#1=4643 %
\hatcurCCtwomassshortxxxxxxB
\else
??????\fi
\fi
}
\newcommand{\hatcurisocite}[1]{\ifnum#1=762 %
\hatcurisocitexxxxxxA
\else
\ifnum#1=4643 %
\hatcurisocitexxxxxxB
\else
??????\fi
\fi
}
\newcommand{\hatcurisofull}[1]{\ifnum#1=762 %
\hatcurisofullxxxxxxA
\else
\ifnum#1=4643 %
\hatcurisofullxxxxxxB
\else
??????\fi
\fi
}
\newcommand{\hatcurisoshort}[1]{\ifnum#1=762 %
\hatcurisoshortxxxxxxA
\else
\ifnum#1=4643 %
\hatcurisoshortxxxxxxB
\else
??????\fi
\fi
}
\newcommand{\hatcurjhkfilset}[1]{\ifnum#1=762 %
\hatcurjhkfilsetxxxxxxA
\else
\ifnum#1=4643 %
\hatcurjhkfilsetxxxxxxB
\else
??????\fi
\fi
}
\newcommand{\hatcurlumind}[1]{\ifnum#1=762 %
\hatcurlumindxxxxxxA
\else
\ifnum#1=4643 %
\hatcurlumindxxxxxxB
\else
??????\fi
\fi
}
\newcommand{\hatcurplanetnum}[1]{\ifnum#1=762 %
\hatcurplanetnumxxxxxxA
\else
\ifnum#1=4643 %
\hatcurplanetnumxxxxxxB
\else
??????\fi
\fi
}
\newcommand{\hatcurRVgammaabs}[1]{\ifnum#1=762 %
\hatcurRVgammaabsxxxxxxA
\else
\ifnum#1=4643 %
\hatcurRVgammaabsxxxxxxB
\else
??????\fi
\fi
}
\newcommand{\hatcurRVgammarel}[1]{\ifnum#1=762 %
\hatcurRVgammarelxxxxxxA
\else
\ifnum#1=4643 %
\hatcurRVgammarelxxxxxxB
\else
??????\fi
\fi
}
\newcommand{\hatcurSMElogg}[1]{\ifnum#1=762 %
\hatcurSMEloggxxxxxxA
\else
\ifnum#1=4643 %
\hatcurSMEloggxxxxxxB
\else
??????\fi
\fi
}
\newcommand{\hatcurSMEteff}[1]{\ifnum#1=762 %
\hatcurSMEteffxxxxxxA
\else
\ifnum#1=4643 %
\hatcurSMEteffxxxxxxB
\else
??????\fi
\fi
}
\newcommand{\hatcurSMEversion}[1]{\ifnum#1=762 %
\hatcurSMEversionxxxxxxA
\else
\ifnum#1=4643 %
\hatcurSMEversionxxxxxxB
\else
??????\fi
\fi
}
\newcommand{\hatcurSMEvmac}[1]{\ifnum#1=762 %
\hatcurSMEvmacxxxxxxA
\else
\ifnum#1=4643 %
\hatcurSMEvmacxxxxxxB
\else
??????\fi
\fi
}
\newcommand{\hatcurSMEvmic}[1]{\ifnum#1=762 %
\hatcurSMEvmicxxxxxxA
\else
\ifnum#1=4643 %
\hatcurSMEvmicxxxxxxB
\else
??????\fi
\fi
}
\newcommand{\hatcurSMEvsin}[1]{\ifnum#1=762 %
\hatcurSMEvsinxxxxxxA
\else
\ifnum#1=4643 %
\hatcurSMEvsinxxxxxxB
\else
??????\fi
\fi
}
\newcommand{\hatcurSMEzfeh}[1]{\ifnum#1=762 %
\hatcurSMEzfehxxxxxxA
\else
\ifnum#1=4643 %
\hatcurSMEzfehxxxxxxB
\else
??????\fi
\fi
}
\newcommand{\hatcurSMEzfehshort}[1]{\ifnum#1=762 %
\hatcurSMEzfehshortxxxxxxA
\else
\ifnum#1=4643 %
\hatcurSMEzfehshortxxxxxxB
\else
??????\fi
\fi
}

\newcounter{planetcounter}

\newboolean{emulateapj}
\setboolean{emulateapj}{true}

\newboolean{rvtablelong}
\setboolean{rvtablelong}{true}

\newboolean{astroph}
\setboolean{astroph}{true}

\shortauthors{Hartman et al.}
\shorttitle{Transiting Giant Planets Around M Dwarfs}
\ifthenelse{\boolean{emulateapj}}{
    
}{
    
}

\begin{document}

\title{
TOI~762\,A\,\lowercase{b} and TIC~46432937\,\lowercase{b}: Two Giant Planets Transiting M Dwarf Stars
}

\correspondingauthor{Joel Hartman}
\email{jhartman@astro.princeton.edu}

\author[0000-0001-8732-6166]{Joel D. Hartman}
\affiliation{Department of Astrophysical Sciences, Princeton University, 4 Ivy Lane, Princeton, NJ 08544, USA}

\author[0000-0001-6023-1335]{Daniel Bayliss}
\affil{Dept. of Physics, University of Warwick, Gibbet Hill Road, Coventry CV4 7AL, UK}
\affil{Centre for Exoplanets and Habitability, University of Warwick, Gibbet Hill Road, Coventry CV4 7AL, UK}

\author[0000-0002-9158-7315]{Rafael Brahm}
\affil{Millennium Institute of Astrophysics (MAS), Nuncio Monse\~{n}or S\'otero Sanz 100, Providencia, Santiago, Chile}
\affil{Facultad de Ingenier\'ia y Ciencias, Universidad Adolfo Ib\'a\~{n}ez, Av. Diagonal las Torres 2640, Pe\~{n}alol\'en, Santiago, Chile}
\affil{Data Observatory Foundation, Santiago, Chile}

\author[0000-0001-7904-4441]{Edward M.\ Bryant}
\affil{Mullard Space Science Laboratory, University College London, Holmbury St Mary, Dorking, RH5 6NT, UK}

\author[0000-0002-5389-3944]{Andr\'es Jord\'an}
\affil{Millennium Institute of Astrophysics (MAS), Nuncio Monse\~{n}or S\'otero Sanz 100, Providencia, Santiago, Chile}
\affil{Facultad de Ingenier\'ia y Ciencias, Universidad Adolfo Ib\'a\~{n}ez, Av. Diagonal las Torres 2640, Pe\~{n}alol\'en, Santiago, Chile}
\affil{Data Observatory Foundation, Santiago, Chile}

\author[0000-0001-7204-6727]{G\'asp\'ar \'A. Bakos}
\affiliation{Department of Astrophysical Sciences, Princeton University, 4 Ivy Lane, Princeton, NJ 08544, USA}

\author[0000-0002-5945-7975]{Melissa J. Hobson}
\affil{Observatoire de Gen\`eve, D\'epartement d’Astronomie, Universit\'e de Gen\`eve, Chemin Pegasi 51b, 1290 Versoix, Switzerland}

\author[0000-0002-7444-5315]{Elyar Sedaghati}
\affil{European Southern Observatory (ESO), Av. Alonso de C\'ordova 3107, 763 0355 Vitacura, Santiago, Chile}


\author[0000-0001-9003-8894]{Xavier Bonfils}
\affil{Univ. Grenoble Alpes, CNRS, IPAG, F-38000 Grenoble, France}

\author{Marion Cointepas}
\affil{Univ. Grenoble Alpes, CNRS, IPAG, F-38000 Grenoble, France}
\affil{Observatoire de Gen\`eve, D\'epartement d’Astronomie, Universit\'e de Gen\`eve, Chemin Pegasi 51b, 1290 Versoix, Switzerland}

\author[0000-0003-3208-9815]{Jose Manuel Almenara}
\affil{Univ. Grenoble Alpes, CNRS, IPAG, F-38000 Grenoble, France}
\affil{Observatoire de Gen\`eve, D\'epartement d’Astronomie, Universit\'e de Gen\`eve, Chemin Pegasi 51b, 1290 Versoix, Switzerland}


\author[0000-0003-1464-9276]{Khalid Barkaoui}
\affil{Astrobiology Research Unit, Universit\'e de Li\`ege, All\'ee du 6 Ao\^ut 19C, B-4000 Li\`ege, Belgium}
\affil{Department of Earth, Atmospheric and Planetary Science, Massachusetts Institute of Technology, 77 Massachusetts Avenue, Cambridge, MA 02139, USA}
\affil{Instituto de Astrof\'isica de Canarias (IAC), Calle V\'ia L\'actea s/n, 38200, La Laguna, Tenerife, Spain}

\author{Mathilde Timmermans}
\affil{Astrobiology Research Unit, Universit\'e de Li\`ege, All\'ee du 6 Ao\^ut 19C, B-4000 Li\`ege, Belgium}

\author{George Dransfield}
\affil{School of Physics \& Astronomy, University of Birmingham, Edgbaston, Birmingham B15 2TT, UK}

\author[0000-0002-7008-6888]{Elsa Ducrot}
\affil{Paris Region Fellow, Marie Sklodowska-Curie Action}
\affil{AIM, CEA, CNRS, Universit\'e Paris-Saclay, Universit\'e de Paris, F-91191 Gif-sur-Yvette, France}

\author[0000-0002-9350-830X]{Sebasti\'an Z\'u\~{n}iga-Fern\'andez}
\affil{Astrobiology Research Unit, Universit\'e de Li\`ege, All\'ee du 6 Ao\^ut 19C, B-4000 Li\`ege, Belgium}

\author[0000-0003-0030-332X]{Matthew J. Hooton}
\affil{Cavendish Laboratory, JJ Thomson Avenue, Cambridge CB3 0HE, UK}

\author{Peter Pihlmann Pedersen}
\affil{Cavendish Laboratory, JJ Thomson Avenue, Cambridge CB3 0HE, UK}

\author[0000-0003-1572-7707]{Francisco J. Pozuelos}
\affil{Instituto de Astrof\'isica de Andaluc\'ia (IAA-CSIC), Glorieta de la Astronom\'ia s/n, 18008 Granada, Spain}

\author[0000-0002-5510-8751]{Amaury H. M. J. Triaud}
\affil{School of Physics \& Astronomy, University of Birmingham, Edgbaston, Birmingham B15 2TT, UK}

\author[0000-0003-1462-7739]{Micha\"el Gillon}
\affil{Astrobiology Research Unit, Universit\'e de Li\`ege, All\'ee du 6 Ao\^ut 19C, B-4000 Li\`ege, Belgium}

\author[0000-0001-8923-488X]{Emmanuel Jehin}
\affil{Space Sciences, Technologies and Astrophysics Research (STAR) Institute, Universit\'e de Li\`ege, All\'ee du 6 Ao\^ut 19C, B-4000 Li\`ege, Belgium}


\author[0000-0002-8961-0352]{William~C.~Waalkes}
\affiliation{Department of Physics and Astronomy, Dartmouth College, Hanover NH 03755, USA}

\author[0000-0002-3321-4924]{Zachory~K.~Berta-Thompson}
\affil{Department of Astrophysical \& Planetary Sciences, University of Colorado Boulder, Boulder CO 80309, USA}

\author[0000-0002-2532-2853]{Steve~B.~Howell}
\affil{NASA Ames Research Center, Moffett Field, CA 94035, USA} 

\author[0000-0001-9800-6248]{Elise Furlan}
\affiliation{NASA Exoplanet Science Institute, Caltech/IPAC, Mail Code 100-22, 1200 E. California Blvd., Pasadena, CA 91125, USA}

\author[0000-0003-2058-6662]{George~R.~Ricker}
\affiliation{Department of Physics and Kavli Institute for Astrophysics and Space Research, Massachusetts Institute of Technology, Cambridge, MA 02139, USA}

\author[0000-0001-6763-6562]{Roland~Vanderspek}
\affiliation{Department of Physics and Kavli Institute for Astrophysics and Space Research, Massachusetts Institute of Technology, Cambridge, MA 02139, USA}

\author[0000-0002-6892-6948]{Sara~Seager}
\affiliation{Department of Physics and Kavli Institute for Astrophysics and Space Research, Massachusetts Institute of Technology, Cambridge, MA 02139, USA}
\affiliation{Department of Earth, Atmospheric and Planetary Sciences, Massachusetts Institute of Technology, Cambridge, MA 02139, USA}
\affiliation{Department of Aeronautics and Astronautics, MIT, 77 Massachusetts Avenue, Cambridge, MA 02139, USA}

\author[0000-0002-4265-047X]{Joshua~N.~Winn}
\affiliation{Department of Astrophysical Sciences, Princeton University, 4 Ivy Lane, Princeton, NJ 08544, USA}

\author[0000-0002-4715-9460]{Jon~M.~Jenkins}
\affiliation{NASA Ames Research Center, Moffett Field, CA 94035, USA}

\author[0000-0003-2196-6675]{David Rapetti}
\affiliation{NASA Ames Research Center, Moffett Field, CA 94035, USA}
\affiliation{Research Institute for Advanced Computer Science, Universities Space Research Association, Washington, DC 20024, USA}

\author[0000-0001-6588-9574]{Karen A.\ Collins}
\affiliation{Center for Astrophysics \textbar \ Harvard \& Smithsonian, 60 Garden Street, Cambridge, MA 02138, USA}

\author[0000-0002-9003-484X]{David Charbonneau}
\affiliation{Center for Astrophysics \textbar \ Harvard \& Smithsonian, 60 Garden Street, Cambridge, MA 02138, USA}

\author[0000-0002-7754-9486]{Christopher~J.~Burke}
\affiliation{Department of Physics and Kavli Institute for Astrophysics and Space Research, Massachusetts Institute of Technology, Cambridge, MA 02139, USA}

\author{David~R.~Rodriguez}
\affiliation{Space Telescope Science Institute, 3700 San Martin Drive, Baltimore, MD, 21218, USA}


\begin{abstract}

\setcounter{footnote}{10}
We present the discovery of \hatcurb{762} and \hatcurb{4643}, two
giant planets transiting M dwarf stars.  Transits of both systems were
first detected from observations by the NASA {\em TESS} mission, and
the transiting objects are confirmed as planets through high-precision
radial velocity (RV) observations carried out with
VLT/ESPRESSO. \hatcurb{762} is a warm sub-Saturn with a mass
of \hatcurPPm{762}\,\mjup, a radius of \hatcurPPr{762}\,\rjup, and an
orbital period of \hatcurLCPshort{762}\,d. It transits a mid-M dwarf
star with a mass of \hatcurISOm{762}\,\msun\ and a radius of
\hatcurISOr{762}\,\rsun. The star \hatcur{762} has a resolved binary star companion TOI~762\,B that is separated from \hatcur{762} by 3\farcs2 ($\sim 319$\,AU) and has an estimated mass of $0.227 \pm 0.010$\,\msun. The planet \hatcurb{4643} is a warm
Super-Jupiter with a mass of \hatcurPPm{4643}\,\mjup\ and radius of
\hatcurPPr{4643}\,\rjup. The planet's orbital period is $P =
\hatcurLCPshort{4643}$\,d, and it undergoes grazing transits of its
early M dwarf host star, which has a mass of
\hatcurISOm{4643}\,\msun\ and a radius of
\hatcurISOr{4643}\,\rsun. 
\hatcurb{4643} is one of the
highest mass planets found to date transiting an M dwarf star. \hatcurb{4643} is also a promising target for atmospheric observations, having the highest Transmission Spectroscopy Metric or Emission Spectroscopy Metric value of any known warm Super-Jupiter (mass greater than 3.0\,\mjup, equilibrium temperature below 1000\,K).
\setcounter{footnote}{0}
\end{abstract}

\keywords{
    planetary systems ---
    stars: individual (
\setcounter{planetcounter}{1}
\hatcur{762},
\setcounter{planetcounter}{2}
\hatcur{4643},
\setcounter{planetcounter}{3}
) 
    techniques: spectroscopic, photometric
}


\section{Introduction}
\label{sec:introduction}

How do the properties of giant planets depend on the masses of their
host stars? This is a key open question and topic of current research
in the field of exoplanets. Efforts to address this question have been
limited by the relatively small number of low-mass and high-mass stars
that have been searched for planets compared to the much larger sample
of Solar-mass stars that have been searched to date by surveys like
{\em Kepler} \citep{borucki:2010} and {\em TESS} \citep{ricker:2015}.  

In order to expand the sample of giant planets known around
low-mass stars, we have been carrying out a program to follow-up
candidate transiting giant planets around M dwarf stars and late K
dwarf stars using the ESPRESSO instrument on the 8\,m Very Large
Telescope (VLT) at Paranal Observatory, in Chile. This facility has
proven to be very efficient at gathering high-precision RV
observations for faint stars, even down to $V \sim 16.5$. We have so
far published the confirmation of seven systems through this effort
\citep{hartman:2020:hats47484972,jordan:2022:hats74hats77,hobson:2023,almenara:2023},
and in this paper we publish the confirmation of two more such
objects.

The planets that we confirm in this paper were identified by the NASA
{\em TESS} mission, which has been carrying out a wide-field search
for transiting planets around bright stars since its launch in
2018. Thanks to its all-sky observing strategy, the mission has also
been successful at discovering the rare instances of transiting
giant planets ($\mpl > 0.1$\,\mjup, for discussion purposes) around M dwarf stars. A total of 17 of the 20
transiting giant planets that have been confirmed around M dwarf stars
so far were either first identified by {\em TESS}, or included {\em TESS} follow-up observations as part of the discovery \citep[]{bakos:2020:hats71,canas:2020,canas:2022,canas:2023:toi3984toi5293,gan:2022,gan:2023,hartman:2023:toi4201toi5344,hobson:2023,jordan:2022:hats74hats77,kagetani:2023,kanodia:2021,kanodia:2022,kanodia:2023,triaud:2023}. The three other known systems were discovered by {\em Kepler} \citep{johnson:2012}, HATSouth \citep{hartman:2015:hats6} and NGTS \citep{bayliss:2018}.

While {\em TESS} has been very successful at discovering giant planets
around M dwarfs, many of these objects were not identified through the
mission's primary transit search effort operated by the Science
Processing Operations Center \citep[SPOC;][]{jenkins:2016} at NASA
Ames Research Center (ARC), which focuses on the 2\,min cadence observations, but were instead identified through special efforts to produce
light curves for faint M dwarf stars from the Full Frame Images (FFIs) and to search them for transit
signals. One such effort is the {\em TESS} Faint-star search
\citep{kunimoto:2022}, which has identified some 3200 faint {\em TESS}
Objects of Interest (TOIs) to date, including 128 candidate giant planets transiting M dwarfs. Another example is \citet{bryant:2023}
who conducted a search for transiting giant planets around M dwarfs in
30\,min cadence {\em TESS} observations and identified 15 candidates,
including 7 that were not previously identified by other
projects. Both of the planets that we confirm in this paper were
included in the \citet{bryant:2023} sample, and one of these systems
(\hatcurb{4643}) was first identified by them.

In the following section we discuss the observations that are used to
confirm and characterize each planetary system. In
Section~\ref{sec:anal} we describe the analysis methods. In
Section~\ref{sec:discussion} we discuss the results.

\section{Observations}
\label{sec:obs}

\subsection{Initial Photometric Detection}

Both \hatcurb{762} and \hatcurb{4643} were first identified as
transiting planet candidates from observations gathered by the NASA
{\em TESS} mission. Table~\ref{tab:phfusummary} summarizes the {\em TESS} observations that are available for each system. Both targets were identified as candidates by an independent transit search performed by \citet{bryant:2023} with the aim of measuring the occurrence rates of giant planets with low-mass host stars. For this search the Box-fitting Least Squares \citep[BLS;][]{kovacs:2002:BLS} algorithm was utilised to search for giant planets transiting low-mass stars in light curves generated from the 30\,minute cadence Full-Frame-Images (FFIs) from Cycle~1 of the {\em TESS} mission by the TESS-SPOC team \citep{Caldwell_2020}. During Cycle~1 both \hatcurb{762} and \hatcurb{4643} were observed in a single sector: Sector 10 for \hatcurb{762} (2019 Mar 29--2019 Apr 22) and Sector 6 for \hatcurb{4643} (2018 Dec 15--2019 Jan 6). Following the BLS detection of these two candidates a number of automated checks were performed to investigate whether the transit-like signals could be the result of a number of different false positive scenarios, such as eclipsing binaries or variable stars. See \citet{bryant:2023} for more details on the analyses performed. For both \hatcurb{762} and \hatcurb{4643} we found no evidence that the transit-like signals were a result of a non-planetary scenario. An initial transit fitting analysis was performed on both candidates using the \textsc{batman} package \citep{batman:2015} to generate the transit models and using \textsc{emcee} \citep{emcee:2013} to perform the MCMC sampling. Both objects were identified as high likelihood giant planet candidates by this analysis and included in the sample reported in \citet{bryant:2023}.

\hatcurb{762} was independently identified as a candidate by the {\em TESS} Science Processing Operations Center \citep[SPOC;][]{jenkins:2016} at NASA ARC. The SPOC conducted a transit search of Sector 10 on 23 May 2019 with an adaptive, noise-compensating matched filter \citep{jenkins:2002,jenkins:2010,jenkins:2020}, producing a threshold crossing event (TCE) for which an initial limb-darkened transit model was fitted \citep{Li:DVmodelFit2019} and a suite of diagnostic tests were conducted to help make or break the planetary nature of the signal \citep{Twicken:DVdiagnostics2018}. The transit signature was also detected in a search of FFI data by the Quick Look Pipeline (QLP) at MIT \citep{huang:2020a,huang:2020b}. The TESS Science Office (TSO) reviewed the vetting information and issued an alert on 27 February 2020 \citep{guerrero:TOIs2021ApJS}. The signal was repeatedly recovered as additional observations were made in Sectors 36, 37, and 63, and the transit signature passed all the diagnostic tests presented in the Data Validation reports. The difference image centroiding test located the host star within $3\farcs18 \pm 2\farcs58$ of the source of the transit signal.

The light curves
used in the analysis of each system are summarized in
Table~\ref{tab:phfusummary}. For \hatcur{762} we make use of the 2\,min cadence TESS light curves corrected for systematics by SPOC using the Presearch Data Conditioning Simple Aperture Photometry (PDCSAP) method of \citet{Stumpe2012,Stumpe2014} and \citet{2012PASP..124.1000S}. For \hatcur{4643} we use the TESS-SPOC 30\,min and 10\,min cadence light curves, which were also corrected using the PDCSAP method. All of the {\em TESS} light curves were obtained from the Mikulski Archive for Space Telescopes (MAST) at the Space Telescope Science Institute (STScI).

\subsection{High Contrast Imaging}

\begin{figure}[!ht]
 {
 \centering
 \leavevmode
 \includegraphics[width={1.0\linewidth}]{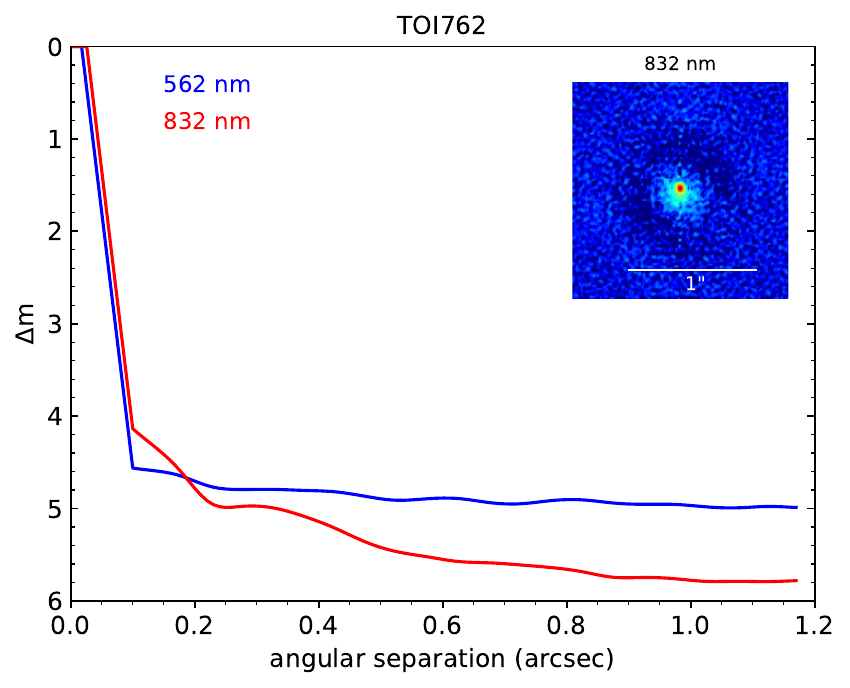}%
 }
\caption{
    $5\sigma$ magnitude contrast limits for any companions to \hatcur{762} based on Gemini/Zorro observations obtained in the 562\,nm and 832\,nm band-passes. The inset shows the reconstructed Gemini/Zorro image of \hatcur{762} in the 832\,nm band-pass.
\label{fig:toi762zorro}
}
    \end{figure}

\hatcur{762} was observed with the Zorro speckle imager on the
Gemini-South~8\,m telescope \citep{scott:2021}. Observations were
obtained in the $832 \pm 40$\,nm and $562 \pm 54$\,nm band-passes on
2020 Jan 11, and these were processed to generate reconstructed images
following \citet{howell:2011}. No companions to \hatcur{762} were
resolved through these observations within the field of view of Zorro,
with contrast limits of $\Delta m_{832} > 5.42$ and $\Delta m_{562} >
4.9$ beyond 0\farcs5. Figure~\ref{fig:toi762zorro} shows the 832\,nm
reconstructed image, and the resulting $5\sigma$ magnitude contrast
limits that we place on any companions to \hatcur{762}.

\startlongtable
    \begin{deluxetable*}{lccc}
\tablewidth{0pc}
\tabletypesize{\scriptsize}
\tablecaption{
    Sources within 10$\arcsec$ of \hatcur{762}
    \label{tab:toi762nbrs}
}
\tablehead{
   \multicolumn{1}{c}{Parameter} &
   \multicolumn{1}{c}{TOI~762\,A} &
   \multicolumn{1}{c}{TOI~762\,B} &
   \multicolumn{1}{c}{} 
}
\startdata
GAIA DR3 ID & 5362352744504000256 & 5362352744496315264 & 5362352744496318336 \\
$\Delta$R.A. (arcsec) & $\cdots$ & $-2.754 \pm 0.098$ & $5.501 \pm 0.092$ \\
$\Delta$Dec. (arcsec) & $\cdots$ & $-1.62 \pm 0.10$ & $4.003 \pm 0.092$ \\
$\mu_{\rm R.A.}$ (\masy) & $\hatcurCCpmra{762}$ & $-157.37 \pm 0.13$ & $-5.15 \pm 0.12$ \\
$\mu_{\rm Dec.}$ (\masy) & $\hatcurCCpmdec{762}$ & $-24.38 \pm 0.12$ & $1.02 \pm 0.11$ \\
$\pi$ (mas) & $\hatcurCCparallax{762}$ & $9.79 \pm 0.14$ & $0.40 \pm 0.13$ \\
$G$ (mag) & $\hatcurCCgaiamG{762}$ & $18.1928 \pm 0.0031$ & $18.3444 \pm 0.0030$ \\
$BP - RP$ (mag)\tablenotemark{a} & $2.7548 \pm 0.0059$ & $3.554 \pm 0.093$ & $0.994 \pm 0.020$ \\
$\mstar$ (\msun) & $\hatcurISOm{762}$ & $0.227 \pm 0.010$ & $\cdots$ \\
\enddata 
\tablenotetext{a}{We caution that the $BP$ and $RP$ photometry for each source appears to be contaminated by the other neighboring sources as indicated by the high values of phot\_bp\_rp\_excess\_factor.}
    \end{deluxetable*}

There are two resolved sources within 10\arcsec\ of \hatcur{762} that
are listed in the {\em Gaia} DR3 catalog (see also
Table~\ref{tab:toi762nbrs}). Gaia DR3 5362352744496318336 is located
4\farcs7 to the northeast from \hatcur{762} with $\Delta G =
3.42$\,mag, $\Delta BP = 2.29$\,mag, and $\Delta RP = 4.05$\,mag
relative to \hatcur{762}. The parallax and proper motion of this
object differ significantly from the values for \hatcur{762}
indicating that the two sources are not physically
associated. Gaia~DR3~5362352744496315264 is located 3\farcs2 to the
southwest of \hatcur{762} and does appear to be physically bound to
\hatcur{762} with $\pi = 9.79 \pm 0.14$\,mas, ${\rm pmRA} = -157.37
\pm 0.13$\,mas\,yr$^{-1}$ and ${\rm pmDE} = -24.38 \pm
0.12$\,mas\,yr$^{-1}$ compared to $\pi = 10.118 \pm 0.023$\,mas, ${\rm
  pmRA} = -159.174 \pm 0.020$\,mas\,yr$^{-1}$ and ${\rm pmDE} =
-24.780 \pm 0.020$\,mas\,yr$^{-1}$ for \hatcur{762}.

Gaia~DR3~5362352744496315264 has
$\Delta G = 3.26$\,mag, $\Delta BP = 3.73$\,mag and $\Delta RP =
2.93$\,mag relative to \hatcur{762}. This object has been identified as a wide binary companion to \hatcur{762} in both the SUPERWIDE catalog \citep{hartman:2020:superwide} and the catalog of \citet{elbadry:2021}. We refer to this star as TOI~762\,B. Assuming this is a bound companion, we estimate that it is a late-M dwarf star with a mass of $0.227 \pm 0.010$\,\msun, at a projected physical separation of $\sim 319$\,AU from \hatcur{762}. The mass estimate is based on comparing the absolute $G$ magnitude of the source to version 1.2 of
the MIST theoretical stellar evolution models
\citep{paxton:2011,paxton:2013,paxton:2015,choi:2016,dotter:2016}, assuming the age, distance, metallicity and redenning to the source are the same as for \hatcur{762}. The inferred 2MASS and WISE magnitudes for the star are $J = 14.057 \pm 0.086$\,mag, $H = 13.475 \pm 0.076$\,mag, $K_{S} = 13.177 \pm 0.085$\,mag, $W_{1} = 12.954 \pm 0.095$\,mag, $W_{2} = 12.71 \pm 0.10$\,mag, $W_{3} = 12.45 \pm 0.11$\,mag, and $W_{4} = 12.33 \pm 0.11$\,mag. The measured relative proper motion difference between TOI~762\,A and TOI~762\,B of $\Delta {\rm pm} = 1.85 \pm 0.13$\,mas\,yr$^{-1}$ is less than the value of $2.9$\,mas\,yr$^{-1}$ that would be expected if the binary star system has a circular, face-on orbit, and suggests that the orbit is inclined and/or eccentric. We note that TOI~762\,B is resolved from \hatcur{762} in the spectroscopic observations (Section~\ref{sec:specobs}) which reveal a RV variation for \hatcur{762} that is in phase with the transit ephemeris. Based on this we conclude that \hatcur{762} is the source of the transit signal.

The closer of the two neighbors is within the photometric aperture of
the follow-up light curves for \hatcur{762}, and would also
have been unresolved in the 2MASS and WISE photometry of this
source. The source is also close enough to be of concern for the {\em Gaia} $BP$ and $RP$ measurements. Using the relations in \citet{riello:2021} for the expected value and scatter of phot\_bp\_rp\_excess\_factor as functions of the $BP-RP$ color and $G$ magnitude for an isolated source, we find that \hatcur{762} has a value that is 3.3$\sigma$ greater than expected. This indicates that these measurements may be contaminated, however correcting for this contamination may be difficult to do accurately. For this reason we exclude the $BP$ and $RP$ photometry from the analysis of this system. We account for the contamination in the follow-up light curves
as described in Section~\ref{sec:transitmodel}. To account for the
contamination in the 2MASS and WISE photometry we use MIST v 1.2 to
estimate the apparent magnitude of the neighbor in these band-passes
given the absolute $G$ magnitude, and assuming the age, metallicity,
distance, and redenning to the source have the values that we
determine for \hatcur{762}. We then subtract the flux contribution of
the neighbor from the observed 2MASS and WISE magnitudes, accounting
for the uncertainty on the flux of the neighbor. These corrected
magnitudes are listed in Table~\ref{tab:stellarobserved}, and are also the values
that we include in our analysis of the system
(Section~\ref{sec:transitmodel}). 

There are no nearby stars within 10\arcsec\ of \hatcur{4643} listed in the {\em Gaia} DR3 catalog. Ground-based high-spatial-resolution imaging is not available for this target, so stellar companions within 1\arcsec\ cannot be ruled out.

\subsection{Ground-based Photometric Follow-up}\label{sec:phfu}

Ground-based photometric follow-up observations of \hatcur{762} were
obtained through the {\em TESS} Follow-up Program \citep[TFOP;][]{collins:2019}. Photometric follow-up observations of \hatcur{4643} were also obtained. The
observations that have been carried out for these systems are listed in
Table~\ref{tab:phfusummary}. The data are shown in
Figures~\ref{fig:toi762} and~\ref{fig:tic4643lc2}, and are made
available in Table~\ref{tab:phfu}. Here we briefly discuss each
facility that was used.

\subsubsection{LCOGT~1\,m}

A transit of \hatcurb{762} was observed using a 1\,m telescope at the Siding Spring Observatory, Australia station of the Las Cumbres Observatory Global Telescope (LCOGT) network \citep{brown:2013:lcogt}. The observations were gathered through an $I_{C}$ band-pass on 2019 June 28 using a SINISTRO imager. The images have a pixel scale of 0\farcs389\,px$^{-1}$, with an estimated average point-spread function (PSF) full-width at half-maximum (FWHM) of $\sim 2\arcsec$. The data were reduced to ensemble-corrected light curves using the {\sc AstroImageJ} package \citep{collins:2017} with a photomeric aperture radius of 14 pixels. 

A separate photometric measurement was performed on the LCOGT images using a 2 pixel (0\farcs8) radius aperture to confirm that the transits are due to \hatcur{762} rather than the neighbor TOI~762\,B. This analysis revealed that \hatcur{762} is indeed the source of the $\sim 30$\,ppt transit events.

\subsubsection{ExTrA~0.6\,m}

A total of six transits of \hatcurb{762} and six transits of \hatcurb{4643} were monitored using the Exoplanets in Transits and their Atmospheres (ExTrA) facility \citep{bonfils:2015} at La Silla Observatory in Chile. Several of the transits were simultaneously observed by more than one of the three 0.6\,m telescopes in the facility, and for \hatcurb{4643} we made multi-band light curves from the observations, leading to a total of 14 separate transit light curves of \hatcurb{762} and 64 separate light curves of \hatcurb{4643} from this facility. The facility performs spectro-photometric observations over a wavelength range of 0.85\,$\mu$m to 1.55\,$\mu$m using a NIRvana 640 LN camera which is fed with optical fibers from the three telescopes. Observations of \hatcur{762} were gathered with the 4\arcsec\ diameter aperture fibers, except for on the night of 2021 Mar 13 when the 8\arcsec\ fibers were used. All transits of \hatcur{4643} were observed with 8\arcsec\ diameter aperture fibers. Both targets were observed with the low resolution mode of the spectrograph ($R \sim 20$). Band-pass-integrated, comparison-star-corrected light curves were produced following the method of \citet{cointepas:2021}. For \hatcur{762} we made use of light curves integrated over the full band-pass of the instrument, while for \hatcur{4643}, where the grazing transits make the solution more sensitive to limb darkening, we produced and analyzed separate $Z$, $Y$, $J$ and $H$-band light curves from each transit observation. 

\subsubsection{TRAPPIST-South~0.6\,m}

Two transits of \hatcurb{762} were observed using the southern 0.6\,m TRAnsiting Planets and PlanetesImals Small Telescope \citep[TRAPPIST-South;][]{Jehin2011,Gillon2011,Barkaoui2019_TN} at La Silla Observatory. The first transit on 2023 February 2 was observed through an $I+z$ filter, while the second transit on 2023 April 16 was observed through a $z^{\prime}$ filter. Observations were obtained at a pixel scale of 0\farcs64. For the first transit the estimated PSF FWHM was 2\farcs6 and a photometric aperture of 3\farcs84 was used, while for the second transit the estimated PSF FWHM was 1\farcs4 and a photometric aperture of 4\farcs48 was used. Scheduling of the observations was performed using the tools of \citet{jensen:2013}, while ensemble-corrected light curves were derived from the observations following the methods of \citet{garcia:2022}.

\subsubsection{SPECULOOS-South~1.0\,m}

A transit of \hatcurb{762} was observed on 2023 April 16 using three of the 1.0\,m telescopes in the Search for habitable Planets EClipsing ULtra-cOOl Stars Southern observatory \citep[SPECULOOS-South][]{delrez:2018,sebastian:2021,Burdanov2022} at Paranal Observatory in Chile. Observations were gathered through $g^{\prime}$, $r^{\prime}$ and $z^{\prime}$ filters. The observations had a pixel scale of 0\farcs35 and an estimated PSF FWHM of 1\farcs6. An aperture of radius 2\farcs45 was used to extract the photometry. Scheduling of the observations was performed following \citet{sebastian:2021}, while the data were reduced to ensemble-corrected light curves following \citet{murray:2020} and \citet{garcia:2021,garcia:2022}.

\subsection{Spectroscopic Observations}\label{sec:specobs}

Time series spectroscopy was obtained for both \hatcur{762} and
\hatcur{4643} using the ESPRESSO instrument \citep{pepe:2021} on the
8\,m Very Large Telescope (VLT) at Paranal Observatory in Chile. We
obtained five observations of \hatcur{762} in Period 104 between 2019
Dec 1 and 2019 Dec 24, and three observations in Period 110 between
2022 Dec 25 and 2022 Dec 30.  We allow for an instrumental offset
between the two periods in fitting the observations of
\hatcur{762}. For \hatcur{4643}, we obtained eight observations all in
Period 110, between 2022 Nov 24 and 2023 Feb 22. Observations were
gathered in HR mode (using a single Unit Telescope, and a
spectroscopic resolution of $R \equiv \Delta\lambda/\lambda \sim
140,000$; the fiber aperture is 1\arcsec\ in this mode, and the wavelength coverage is from $3770$\,\AA\ to 7910\,\AA). A sky fiber was placed 7\arcsec\ from the target fiber, however there is no difference in the RVs derived from the sky-subtracted spectra compared to the non-sky-subtracted spectra due to the fact that moon contamination was minimal during the observations. An exposure time of 1800\,s was used for the observations. For \hatcur{762} the peak S/N varied from 23 to 30, while for \hatcur{4643} the peak S/N was between 41 and 64. 

The data were reduced to Radial Velocity (RV) measurements in the Solar System
Barycentric frame using the ESPRESSO DRS pipeline
\citep[v2.3.5][]{sosnowska:2015,modigliani:2020} in the EsoReflex
environment \citep{freudling:2013}. The ESPRESSO DRS calculates the RVs from individual spectra by measuring the cross correlation function (CCF) for each slice separately, using a template (stellar model) matching closest to the spectral type of the star. It then adds all these CCFs for the different orders, ignoring those orders that are severely affected by telluric contamination. It then fits a Gaussian function to the final CCF, where the center of the Gaussian is the RV and the width of the Gaussian represents the precision of the RV measurements. 

\begin{figure*}[!ht]
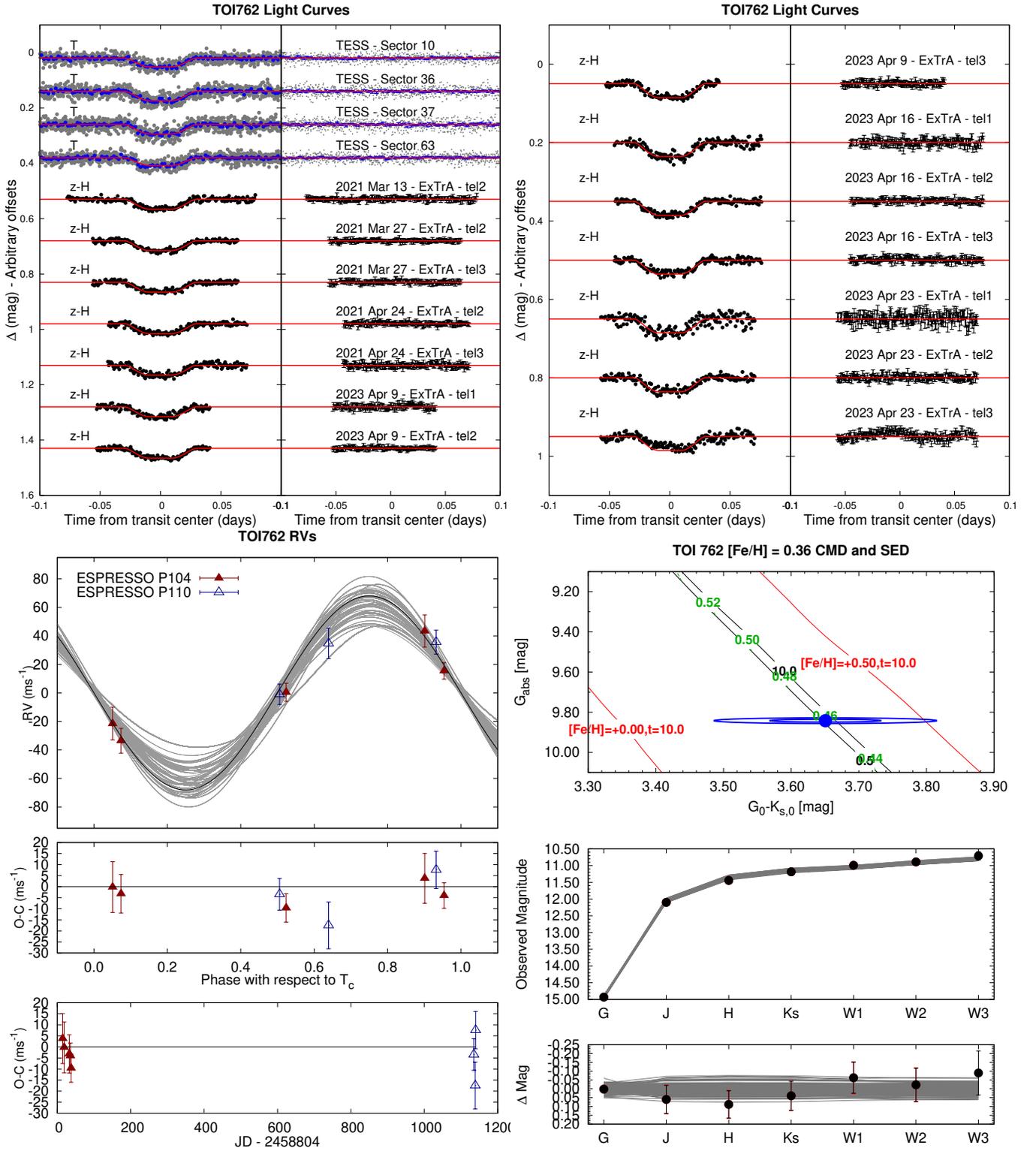

 {
 \centering
 \leavevmode
 \includegraphics[width={1.0\linewidth}]{\hatcurhtr{762}-banner}
}
 {
 \centering
 \leavevmode
 \includegraphics[width={0.5\linewidth}]{\hatcurhtr{762}-lc}%
 \hfil
 \includegraphics[width={0.5\linewidth}]{\hatcurhtr{762}-lc2}%
 }
 {
 \centering
 \leavevmode
 \includegraphics[width={0.5\linewidth}]{\hatcurhtr{762}-rv}%
 \hfil
 \includegraphics[width={0.5\linewidth}]{\hatcurhtr{762}-iso-gk-gabs-isofeh-SED}%
 }                        
\caption{ Observations incorporated into the analysis of the
  transiting planet system \hatcur{762}. Additional light curves are shown in Fig.~\ref{fig:toi762lc2}. {\em Top:} Transit light
  curves with best-fitted model (maximum likelihood) overplotted.
    The dates, filters and instruments used are indicated. The term $z-H$ used for the ExTrA light curves refers to the full band-pass not the color. (Caption continued on next page.) 
\label{fig:toi762}
}
    \end{figure*}

Both systems show significant
RV variations that are in phase with the transit ephemerides, and of
amplitudes that indicate the transiting components are of planetary
mass in both cases. The RV measurements for each system are
shown in Figures~\ref{fig:toi762} and~\ref{fig:tic4643} and are listed
in Table~\ref{tab:rvs}.

\startlongtable
    \begin{deluxetable*}{llrrrr}
\tablewidth{0pc}
\tabletypesize{\scriptsize}
\tablecaption{
    Summary of photometric observations
    \label{tab:phfusummary}
}
\tablehead{
    \multicolumn{1}{c}{Instrument/Field\tablenotemark{a}} &
    \multicolumn{1}{c}{Date(s)} &
    \multicolumn{1}{c}{\# Images\tablenotemark{b}} &
    \multicolumn{1}{c}{Cadence\tablenotemark{c}} &
    \multicolumn{1}{c}{Filter} &
    \multicolumn{1}{c}{Precision\tablenotemark{d}} \\
    \multicolumn{1}{c}{} &
    \multicolumn{1}{c}{} &
    \multicolumn{1}{c}{} &
    \multicolumn{1}{c}{(sec)} &
    \multicolumn{1}{c}{} &
    \multicolumn{1}{c}{(mmag)}
}
\startdata
\sidehead{\textbf{\hatcur{762}}}
~~~~TESS/Sector 10 & 2019 Mar--2019 Apr & 13,754 & 120 & $T$ & 11.7 \\
~~~~TESS/Sector 36 & 2021 Mar--2021 Apr & 15,491 & 120 & $T$ & 13.3 \\
~~~~TESS/Sector 37 & 2021 Apr & 15,061 & 120 & $T$ & 13.9 \\
~~~~TESS/Sector 63 & 2023 Mar--2023 Apr & 17,456 & 120 & $T$ & 12.5 \\
~~~~LCOGT~1.0\,m & 2019 Jun 28 & 109 & 97 & $I_{C}$ & 3.1 \\
~~~~ExTrA - tel2 & 2021 Mar 13 & 214 & 62 & 0.85--1.55\,$\mu$m & 5.9 \\
~~~~ExTrA - tel2 & 2021 Mar 27 & 168 & 62 & 0.85--1.55\,$\mu$m & 4.7 \\
~~~~ExTrA - tel3 & 2021 Mar 27 & 164 & 62 & 0.85--1.55\,$\mu$m & 5.5 \\
~~~~ExTrA - tel2 & 2021 Apr 24 & 159 & 62 & 0.85--1.55\,$\mu$m & 5.7 \\
~~~~ExTrA - tel3 & 2021 Apr 24 & 160 & 62 & 0.85--1.55\,$\mu$m & 7.5 \\
~~~~TRAPPIST-South & 2023 Feb  2 & 214 & 83 & I$+z$ & 3.5 \\
~~~~ExTrA - tel1 & 2023 Apr  9 & 129 & 62 & 0.85--1.55\,$\mu$m & 7.3 \\
~~~~ExTrA - tel2 & 2023 Apr  9 & 131 & 62 & 0.85--1.55\,$\mu$m & 4.8 \\
~~~~ExTrA - tel3 & 2023 Apr  9 & 131 & 62 & 0.85--1.55\,$\mu$m & 5.0 \\
~~~~SPECULOOS-South & 2023 Apr 16 & 97 & 210 & $g^{\prime}$ & 3.2 \\
~~~~SPECULOOS-South & 2023 Apr 16 & 169 & 85 & $r^{\prime}$ & 3.2 \\
~~~~SPECULOOS-South & 2023 Apr 16 & 565 & 36 & $z^{\prime}$ & 2.9 \\
~~~~ExTrA - tel1 & 2023 Apr 16 & 162 & 62 & 0.85--1.55\,$\mu$m & 7.4 \\
~~~~ExTrA - tel2 & 2023 Apr 16 & 167 & 62 & 0.85--1.55\,$\mu$m & 4.4 \\
~~~~ExTrA - tel3 & 2023 Apr 16 & 169 & 62 & 0.85--1.55\,$\mu$m & 5.4 \\
~~~~TRAPPIST-South & 2023 Apr 16 & 141 & 111 & $z^{\prime}$ & 3.0 \\
~~~~ExTrA - tel1 & 2023 Apr 23 & 173 & 62 & 0.85--1.55\,$\mu$m & 11.2 \\
~~~~ExTrA - tel2 & 2023 Apr 23 & 175 & 62 & 0.85--1.55\,$\mu$m & 6.1 \\
~~~~ExTrA - tel3 & 2023 Apr 23 & 174 & 62 & 0.85--1.55\,$\mu$m & 8.7 \\
\sidehead{\textbf{\hatcur{4643}}}
~~~~TESS/Sector 6 & 2018 Dec -- 2019 Jan & 959 & 1800 & $T$ & 1.2 \\
~~~~TESS/Sector 32 & 2020 Nov -- 2020 Dec & 3467 & 600 & $T$ & 2.4 \\
~~~~ExTrA - tel1 & 2023 Oct 10 & 141 & 62 & $Z$ & 7.7 \\
~~~~ExTrA - tel1 & 2023 Oct 10 & 141 & 62 & $Y$ & 4.7 \\
~~~~ExTrA - tel1 & 2023 Oct 10 & 140 & 62 & $J$ & 4.3 \\
~~~~ExTrA - tel1 & 2023 Oct 10 & 141 & 62 & $H$ & 12.6 \\
~~~~ExTrA - tel2 & 2023 Oct 10 & 140 & 62 & $Z$ & 7.0 \\
~~~~ExTrA - tel2 & 2023 Oct 10 & 140 & 62 & $Y$ & 5.0 \\
~~~~ExTrA - tel2 & 2023 Oct 10 & 140 & 62 & $J$ & 4.3 \\
~~~~ExTrA - tel2 & 2023 Oct 10 & 141 & 62 & $H$ & 12.6 \\
~~~~ExTrA - tel3 & 2023 Oct 10 & 141 & 62 & $Z$ & 7.1 \\
~~~~ExTrA - tel3 & 2023 Oct 10 & 141 & 62 & $Y$ & 5.9 \\
~~~~ExTrA - tel3 & 2023 Oct 10 & 141 & 62 & $J$ & 5.3 \\
~~~~ExTrA - tel3 & 2023 Oct 10 & 140 & 62 & $H$ & 22.8 \\
~~~~ExTrA - tel1 & 2023 Nov 5 & 165 & 62 & $Z$ & 9.9 \\
~~~~ExTrA - tel1 & 2023 Nov 5 & 165 & 62 & $Y$ & 5.6 \\
~~~~ExTrA - tel1 & 2023 Nov 5 & 165 & 62 & $J$ & 5.2 \\
~~~~ExTrA - tel1 & 2023 Nov 5 & 163 & 62 & $H$ & 14.7 \\
~~~~ExTrA - tel2 & 2023 Nov 5 & 163 & 62 & $Z$ & 7.5 \\
~~~~ExTrA - tel2 & 2023 Nov 5 & 165 & 62 & $Y$ & 5.4 \\
~~~~ExTrA - tel2 & 2023 Nov 5 & 165 & 62 & $J$ & 5.3 \\
~~~~ExTrA - tel2 & 2023 Nov 5 & 165 & 62 & $H$ & 14.7 \\
~~~~ExTrA - tel3 & 2023 Nov 5 & 165 & 62 & $Z$ & 8.9 \\
~~~~ExTrA - tel3 & 2023 Nov 5 & 165 & 62 & $Y$ & 5.7 \\
~~~~ExTrA - tel3 & 2023 Nov 5 & 164 & 62 & $J$ & 6.0 \\
~~~~ExTrA - tel3 & 2023 Nov 5 & 165 & 62 & $H$ & 18.8 \\
~~~~ExTrA - tel1 & 2023 Nov 18 & 269 & 62 & $Z$ & 8.0 \\
~~~~ExTrA - tel1 & 2023 Nov 18 & 268 & 62 & $Y$ & 4.9 \\
~~~~ExTrA - tel1 & 2023 Nov 18 & 269 & 62 & $J$ & 4.6 \\
~~~~ExTrA - tel1 & 2023 Nov 18 & 270 & 62 & $H$ & 12.2 \\
~~~~ExTrA - tel2 & 2023 Nov 18 & 269 & 62 & $Z$ & 7.6 \\
~~~~ExTrA - tel2 & 2023 Nov 18 & 269 & 62 & $Y$ & 4.8 \\
~~~~ExTrA - tel2 & 2023 Nov 18 & 268 & 62 & $J$ & 4.5 \\
~~~~ExTrA - tel2 & 2023 Nov 18 & 270 & 62 & $H$ & 11.8 \\
~~~~ExTrA - tel3 & 2023 Nov 18 & 269 & 62 & $Z$ & 8.6 \\
~~~~ExTrA - tel3 & 2023 Nov 18 & 268 & 62 & $Y$ & 4.9 \\
~~~~ExTrA - tel3 & 2023 Nov 18 & 269 & 62 & $J$ & 5.2 \\
~~~~ExTrA - tel3 & 2023 Nov 18 & 269 & 62 & $H$ & 14.8 \\
~~~~ExTrA - tel1 & 2023 Dec 1 & 221 & 62 & $Z$ & 8.1 \\
~~~~ExTrA - tel1 & 2023 Dec 1 & 221 & 62 & $Y$ & 5.8 \\
~~~~ExTrA - tel1 & 2023 Dec 1 & 220 & 62 & $J$ & 5.0 \\
~~~~ExTrA - tel1 & 2023 Dec 1 & 220 & 62 & $H$ & 11.8 \\
~~~~ExTrA - tel2 & 2023 Dec 1 & 219 & 62 & $Z$ & 8.0 \\
~~~~ExTrA - tel2 & 2023 Dec 1 & 219 & 62 & $Y$ & 5.6 \\
~~~~ExTrA - tel2 & 2023 Dec 1 & 220 & 62 & $J$ & 4.7 \\
~~~~ExTrA - tel2 & 2023 Dec 1 & 220 & 62 & $H$ & 15.1 \\
~~~~ExTrA - tel3 & 2023 Dec 1 & 220 & 62 & $Z$ & 9.0 \\
~~~~ExTrA - tel3 & 2023 Dec 1 & 220 & 62 & $Y$ & 5.9 \\
~~~~ExTrA - tel3 & 2023 Dec 1 & 220 & 62 & $J$ & 5.9 \\
~~~~ExTrA - tel3 & 2023 Dec 1 & 221 & 62 & $H$ & 16.4 \\
~~~~ExTrA - tel1 & 2023 Dec 14 & 312 & 62 & $Z$ & 8.3 \\
~~~~ExTrA - tel1 & 2023 Dec 14 & 313 & 62 & $Y$ & 5.3 \\
~~~~ExTrA - tel1 & 2023 Dec 14 & 314 & 62 & $J$ & 4.8 \\
~~~~ExTrA - tel1 & 2023 Dec 14 & 313 & 62 & $H$ & 15.8 \\
~~~~ExTrA - tel3 & 2023 Dec 14 & 314 & 62 & $Z$ & 8.2 \\
~~~~ExTrA - tel3 & 2023 Dec 14 & 314 & 62 & $Y$ & 5.6 \\
~~~~ExTrA - tel3 & 2023 Dec 14 & 311 & 62 & $J$ & 5.8 \\
~~~~ExTrA - tel3 & 2023 Dec 14 & 313 & 62 & $H$ & 15.9 \\
~~~~ExTrA - tel1 & 2024 Jan 6 & 141 & 62 & $Z$ & 8.6 \\
~~~~ExTrA - tel1 & 2024 Jan 6 & 141 & 62 & $Y$ & 5.2 \\
~~~~ExTrA - tel1 & 2024 Jan 6 & 141 & 62 & $J$ & 5.0 \\
~~~~ExTrA - tel1 & 2024 Jan 6 & 140 & 62 & $H$ & 12.7 \\
~~~~ExTrA - tel3 & 2024 Jan 6 & 141 & 62 & $Z$ & 8.7 \\
~~~~ExTrA - tel3 & 2024 Jan 6 & 140 & 62 & $Y$ & 5.7 \\
~~~~ExTrA - tel3 & 2024 Jan 6 & 141 & 62 & $J$ & 6.2 \\
~~~~ExTrA - tel3 & 2024 Jan 6 & 140 & 62 & $H$ & 16.4 \\
\enddata 
\tablenotetext{a}{ For {\em TESS} data we list the Sector from which the observations are taken.
}
\tablenotetext{b}{ Excluding any outliers or other data not included in the modelling. }
\tablenotetext{c}{ The median time between consecutive images rounded
  to the nearest second. Due to factors such as weather, the
  day--night cycle, guiding and focus corrections the cadence is only
  approximately uniform over short timescales.  } 
\tablenotetext{d}{
  The RMS of the residuals from the best-fit model. Note that in the case of {\em TESS} observations the transit may appear artificially shallower due to over-filtering and/or blending from unresolved neighbors. As a result the S/N of the transit may be less than what would be calculated from $\rpl/\rstar$ and the RMS estimates given here. }
    \end{deluxetable*}

%
%
    \begin{deluxetable*}{llrrrrl}
\tablewidth{0pc}
\tablecaption{
    Light curve data for \hatcur{762} and \hatcur{4643}\label{tab:phfu}.
}
\tablehead{
    \colhead{Object\tablenotemark{a}} &
    \colhead{BJD\tablenotemark{b}} & 
    \colhead{Mag\tablenotemark{c}} & 
    \colhead{\ensuremath{\sigma_{\rm Mag}}} &
    \colhead{Mag(orig)\tablenotemark{d}} & 
    \colhead{Filter} &
    \colhead{Instrument}
}
\startdata
TOI-762 & $ 2458595.24514 $ & $   0.01246 $ & $   0.01259 $ & $ \cdots $ & $ T$ & TESS/Sec10\\
TOI-762 & $ 2458577.88699 $ & $  -0.00409 $ & $   0.01237 $ & $ \cdots $ & $ T$ & TESS/Sec10\\
TOI-762 & $ 2458588.30220 $ & $  -0.01125 $ & $   0.01251 $ & $ \cdots $ & $ T$ & TESS/Sec10\\
TOI-762 & $ 2458581.35921 $ & $  -0.01476 $ & $   0.01231 $ & $ \cdots $ & $ T$ & TESS/Sec10\\
TOI-762 & $ 2458591.77437 $ & $  -0.02359 $ & $   0.01222 $ & $ \cdots $ & $ T$ & TESS/Sec10\\
TOI-762 & $ 2458574.41615 $ & $   0.01873 $ & $   0.01261 $ & $ \cdots $ & $ T$ & TESS/Sec10\\
TOI-762 & $ 2458595.24653 $ & $  -0.01866 $ & $   0.01228 $ & $ \cdots $ & $ T$ & TESS/Sec10\\
TOI-762 & $ 2458577.88838 $ & $  -0.01855 $ & $   0.01225 $ & $ \cdots $ & $ T$ & TESS/Sec10\\
TOI-762 & $ 2458588.30358 $ & $  -0.00049 $ & $   0.01260 $ & $ \cdots $ & $ T$ & TESS/Sec10\\
TOI-762 & $ 2458581.36059 $ & $  -0.01694 $ & $   0.01228 $ & $ \cdots $ & $ T$ & TESS/Sec10\\
\enddata
\tablenotetext{a}{
    Either \hatcur{762} or \hatcur{4643}.
}
\tablenotetext{b}{
    Barycentric Julian Dates in this paper are reported on the
    Barycentric Dynamical Time (TDB) system.  
} \tablenotetext{c}{
    The out-of-transit level has been subtracted. For observations
    made with TESS these magnitudes have been corrected for trends
    {\em prior} to fitting the transit model. For
    observations made with follow-up instruments (anything other than
    ``TESS'' in the ``Instrument'' column), the
    magnitudes have been corrected for a quadratic trend in time
    fit simultaneously with the transit.
}
\tablenotetext{d}{
    Raw magnitude values without correction for the quadratic trend in
    time, or for trends correlated with the seeing. These are only
    reported for the follow-up observations.
}
\tablecomments{
    This table is available in a machine-readable form in the online
    journal.  A portion is shown here for guidance regarding its form
    and content.
}
    \end{deluxetable*}

%
%

%
%
\addtocounter{figure}{-1}
    \begin{figure*}[!ht]
\caption{
    (Caption continued from previous page.)
For {\em TESS} we phase-fold the data, and plot the un-binned observations in grey, with the phase-binned values overplotted in blue. The residuals for each light curve are shown on the right-hand-side in the same order as the original light curves.  The error bars represent the photon
    and background shot noise, plus the readout noise. Note that these
    uncertainties are scaled in the fitting procedure to achieve a
    reduced $\chi^2$ of unity, but the uncertainties shown in the plot
    have not been scaled.
{\em Bottom Left:}
High-precision RVs phased with respect to the mid-transit time. 
The top panel shows the phased measurements together with the best-fit (maximum likelihood) model with no eccentricity. The gray-scale curves show 1000 models randomly selected from the MCMC posterior distribution generated when the eccentricity is allowed to vary.
The center-of-mass velocity has been subtracted. The middle panel shows the phase-folded velocity $O\!-\!C$ residuals.
The error bars include the estimated jitter, which is varied as a free parameter in the fitting. The bottom panel shows the $O\!-\!C$ residuals as a function of time.
{\em Bottom Right:} Color-magnitude diagram (CMD) and spectral energy distribution (SED). The top panel shows the absolute $G$ magnitude vs.\ the de-reddened $G - K_{S}$ color compared to
  theoretical isochrones (black lines) and stellar evolution tracks
  (green lines) from the MIST models interpolated at
  the best-estimate value for the metallicity of the host. The age
  of each isochrone is listed in black in Gyr, while the mass of each
  evolution track is listed in green in solar masses. The solid red lines show isochrones at higher and lower metallicities than the best-estimate value, with the metallicity and age in Gyr of each isochrone labelled on the plot. The filled
  blue circles show the measured reddening- and distance-corrected
  values from {\em Gaia} DR3 and 2MASS, while the blue lines indicate
  the $1\sigma$ and $2\sigma$ confidence regions, including the
  estimated systematic errors in the photometry. The middle panel shows the SED as measured via broadband photometry through the listed filters. Here we plot the observed magnitudes without correcting for distance or extinction. Overplotted are 200 model SEDs randomly selected from the MCMC posterior distribution produced through the global analysis (gray lines). 
The model makes use of the predicted absolute magnitudes in each bandpass from the MIST isochrones, the distance to the system (constrained largely via {\em Gaia} DR3) and extinction (constrained from the SED with a prior coming from the {\sc mwdust} 3D Galactic extinction model).  
The bottom panel shows the $O\!-\!C$ residuals from the best-fit model SED. The errors listed in the catalogs for the broad-band photometry measurements are shown with black lines, while the errors including an assumed 0.02\,mag systematic uncertainty, which is added in quadrature to the listed uncertainties, are shown with red lines. These latter uncertainties are what we use in the fit.
\label{fig:toi762:labcontinue}}
    \end{figure*}

    \begin{figure}[!ht]
 {
 \centering
 \leavevmode
 \includegraphics[width={1.0\linewidth}]{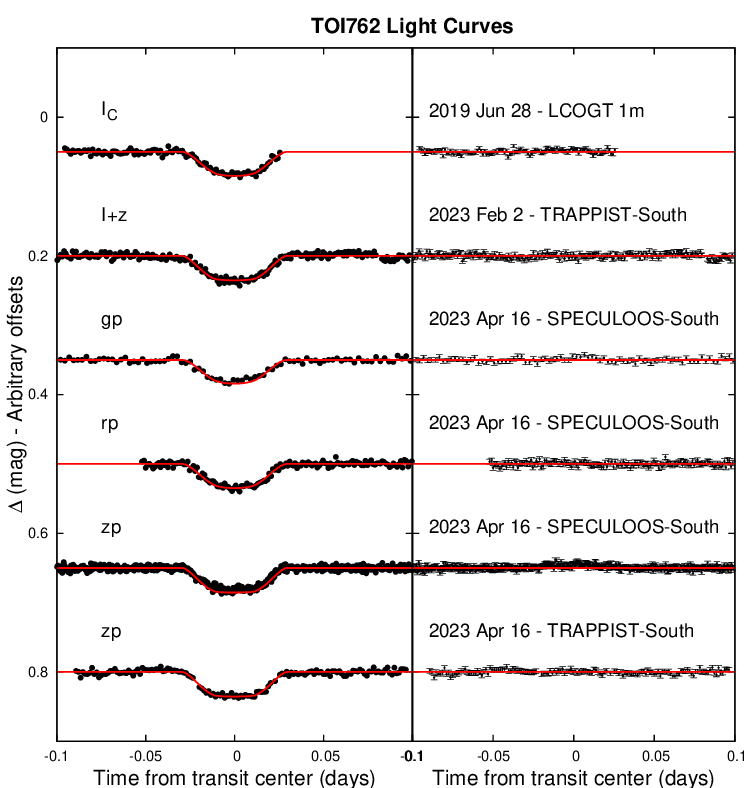}%
 }
\caption{
    Additional follow-up light curves of \hatcur{762}, shown as described in Fig.~\ref{fig:toi762}.
\label{fig:toi762lc2}
}
    \end{figure}

%
%
    \begin{figure*}[!ht]
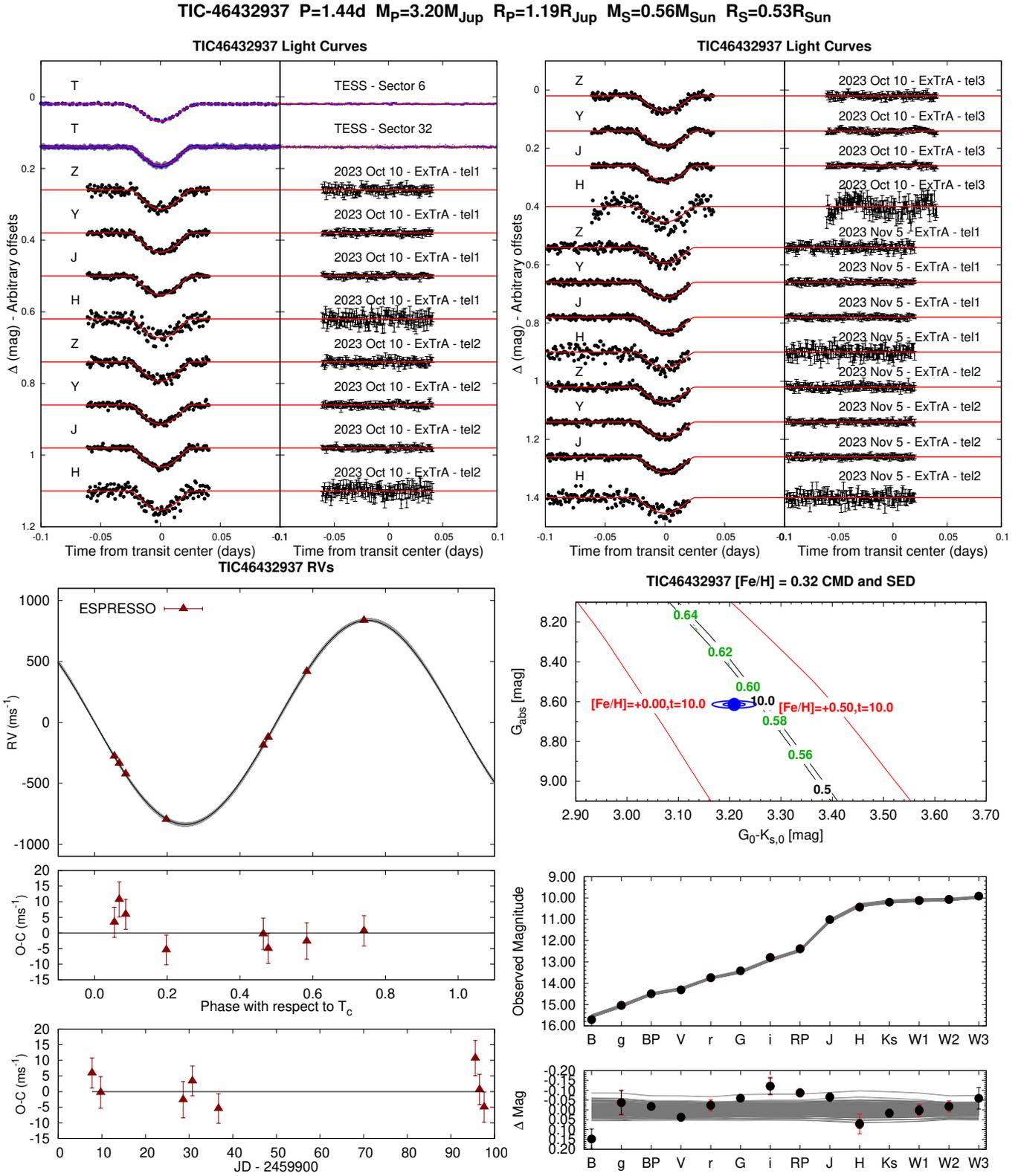

 {
 \centering
 \leavevmode
 \includegraphics[width={1.0\linewidth}]{\hatcurhtr{4643}-banner}
}
  {
 \centering
 \leavevmode
 \includegraphics[width={0.5\linewidth}]{\hatcurhtr{4643}-lc}%
 \hfil
 \includegraphics[width={0.5\linewidth}]{\hatcurhtr{4643}-lc2}%
 }
 {
 \centering
 \leavevmode
 \includegraphics[width={0.5\linewidth}]{\hatcurhtr{4643}-rv}%
 \hfil
 \includegraphics[width={0.5\linewidth}]{\hatcurhtr{4643}-iso-gk-gabs-isofeh-SED}%
 }                        
\caption{
    Same as Figure~\ref{fig:toi762}, here we show the observations of \hatcur{4643} together with our best-fit model. Additional follow-up light curves of \hatcur{4643} are shown in Figure~\ref{fig:tic4643lc2}.
\label{fig:tic4643}
}
    \end{figure*}

    \begin{figure*}[!ht]
 {
 \centering
 \leavevmode
 \includegraphics[width={0.5\linewidth}]{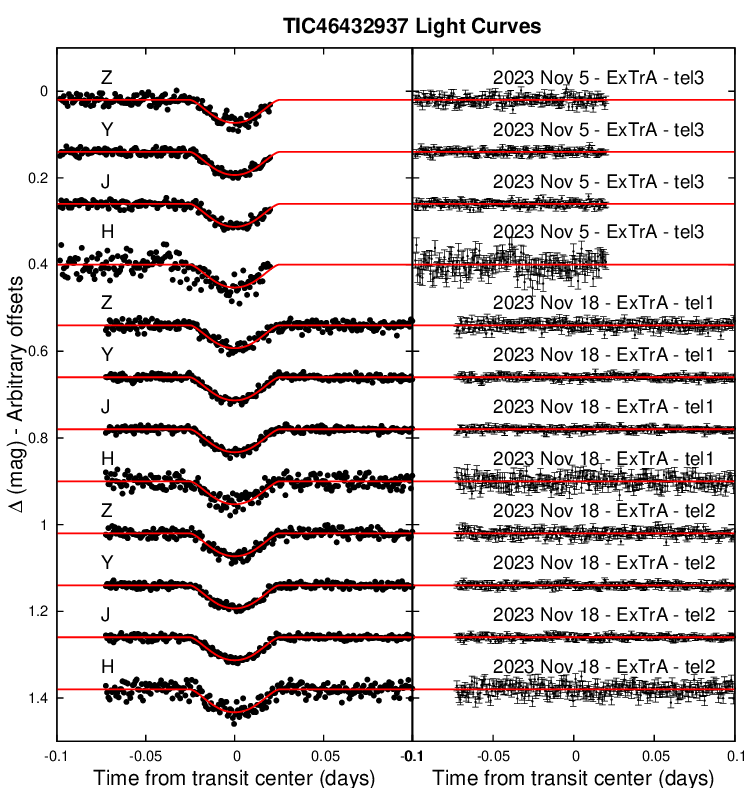}%
 \hfil
 \includegraphics[width={0.5\linewidth}]{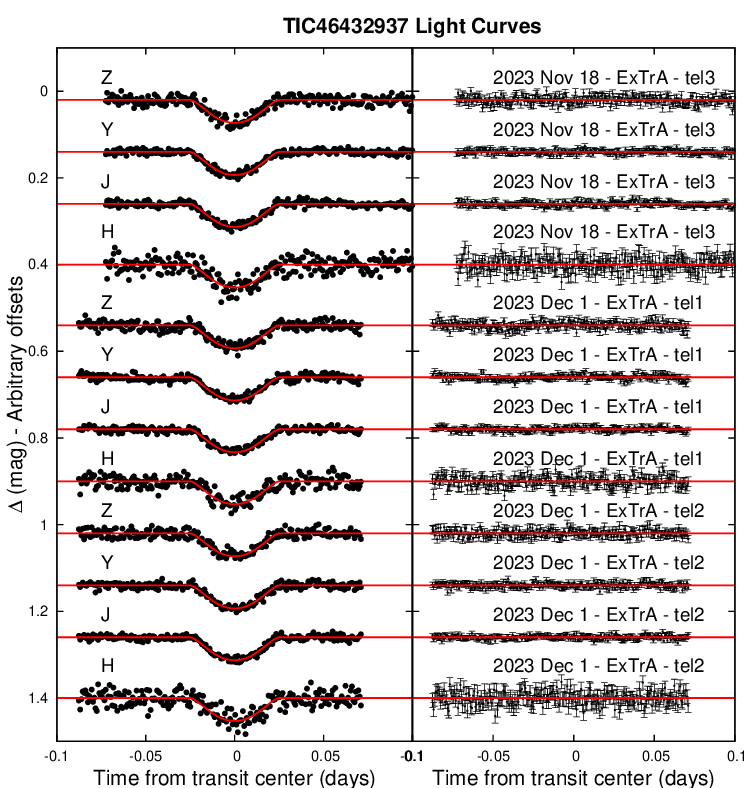}%
 }
 {
 \centering
 \leavevmode
 \includegraphics[width={0.5\linewidth}]{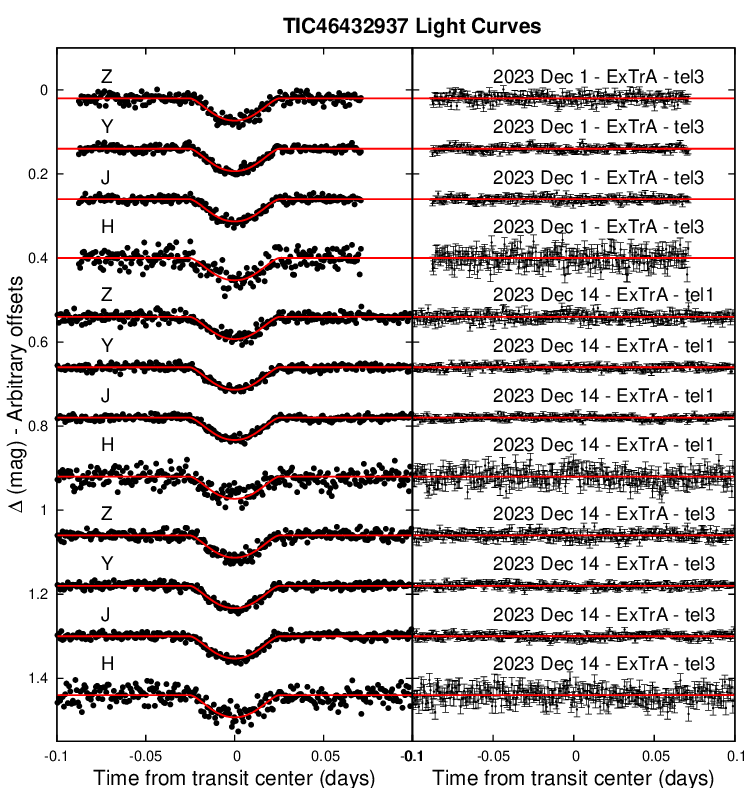}%
 \hfil
 \includegraphics[width={0.5\linewidth}]{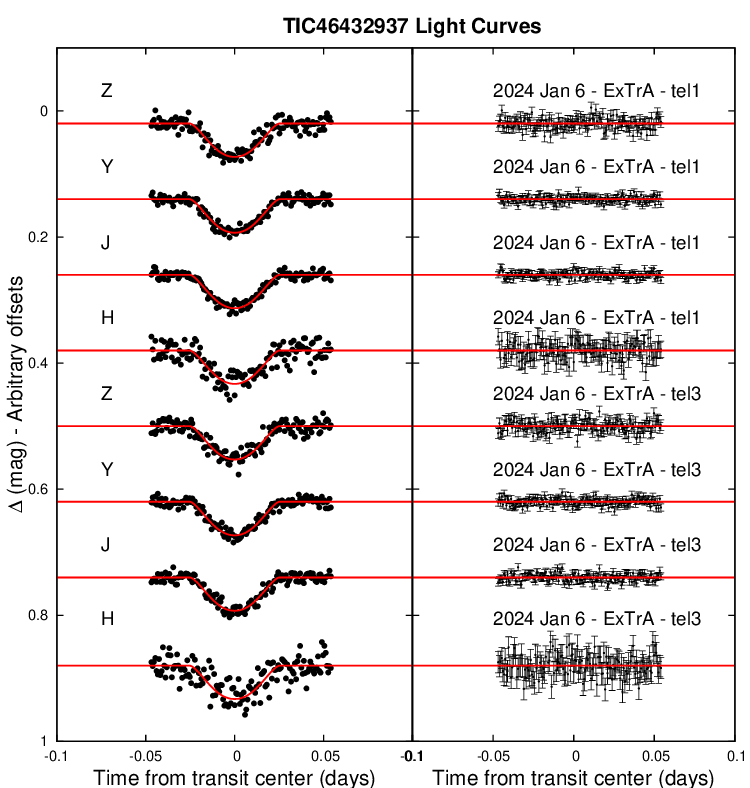}%
 }
\caption{
    Additional follow-up light curves of \hatcur{4643}, shown as described in Fig.~\ref{fig:toi762}.
\label{fig:tic4643lc2}
}
    \end{figure*}


\clearpage

\section{Analysis}
\label{sec:anal}

\subsection{Derivation of Stellar Atmospheric Parameters}
\label{sec:atmosphericparams}

We used the machine-learning based ODUSSEAS package
\citep{antoniadiskarnavas:2020} to measure the photospheric effective
temperature \teffstar\ and metallicity \feh\ (assuming Solar-scaled
abundances) of both systems from the ESPRESSO spectra. The tool measures pseudo-equivalent widths for thousands of lines and compares them to a training set of reference M dwarf stars observed by HARPS. Although the HARPS spectra have lower resolution than the ESPRESSO spectra, the tools has been shown to work effectively on high resolution ESPRESSO spectra as well \citep{lillobox:2020}. The analysis
was performed using the ``wide error'' mode, taking into consideration
the intrinsic uncertainties of the reference parameters in the machine
learning process in addition to the output machine learning model errors. We measure $\teffstar = \hatcurSMEteff{762}$\,K and
$\feh = \hatcurSMEzfeh{762}$ for \hatcur{762} and $\teffstar =
\hatcurSMEteff{4643}$\,K and $\feh = \hatcurSMEzfeh{4643}$ for
\hatcur{4643}.

\subsection{Stellar Activity and Galactic Kinematics}
\label{sec:activity}

We checked the {\em TESS} light curves of \hatcur{762} for evidence of
stellar rotational variability or flaring events. No
significant out-of-transit variability is observed. A total of $6$
bright outliers ($> 5\sigma$) are present in the out-of-transit {\em
  TESS} light curves of \hatcur{762}. In each case the outlier is
isolated to a single point, so if it is due to a flare, the flare
would have been shorter than 120\,s in duration. It is worth noting
though that no $5\sigma$ faint outliers are detected. We
also checked the publicly available ASAS-SN
\citep{hart:2023,shappee:2014} light curve \hatcur{762} and see no
evidence for periodic variability or stellar flares.

For \hatcur{4643} a periodic signal is present in the out-of-transit
{\em TESS} light curves from Sectors 6 and 32. We find a period of $P
= 5.88 \pm 0.54$\,day with a signal-to-noise ratio of $35.2$ as
measured in the Generalized Lomb-Scargle periodogram \citep{zechmeister:2009}. The
estimated uncertainty on the period is the half-width at half-maximum
of the peak in the periodogram. The peak-to-peak variation in the
phase-binned light curve is $\sim 1$\,mmag. The formal false alarm
probability is vanishingly small, however this does not account for
the possibility of systematic variations in the light curve due to
uncorrected instrumental effects. We checked the Zwicky Transient
Facility \citep[ZTF;][]{masci:2019} DR20 light curve of \hatcur{4643}, but
see no evidence for periodic variability. However, the scatter in the
ZTF light curve of this source is $\sim 0.02$\,mag, and the $\sim
1$\,mmag amplitude signal seen in {\em TESS} would not be detectable
if it is present in the ZTF data. No variability is detected in the
publicly available ASAS-SN light curve of \hatcur{4643} either, which
also has too high a scatter to permit detection of a signal comparable
to that seen in the {\em TESS} data. No candidate flare events or
significant bright outliers are seen in the {\em TESS} light curve of
\hatcur{4643}. While bright outliers are seen in the ZTF and ASAS-SN
light curves, the number of faint outliers is comparable. 

If the $P =
5.88$ day signal corresponds to the rotation period of \hatcur{4643},
the rotation period would be faster than the bulk of early M dwarf stars
with $0.9 < G - RP < 1.1$ seen in the {\em Kepler} sample \citep{mcquillan:2014}. While this may suggest a young age for \hatcur{4643},
existing gyrochronology relations are not suitable for M dwarf stars
\citep{popinchalk:2021}. It is also unclear how a relatively massive $\sim
3$\,\mjup\ planet on a short $P = 1.44$\,day orbit might affect the
rotational evolution of an early M dwarf. For these reasons we do not
attempt to estimate an age for \hatcur{4643} based on the possible
rotation period.

As an additional check on the ages of the systems, we used the
systemic radial velocities that we measured for each star with
ESPRESSO, together with the position, parallax, and proper motions from
{\em Gaia} DR3 to compute the $U$, $V$, and $W$ space velocities. We
follow the convention that $U$ increases toward the Galactic center,
$V$ increases in the direction of Galactic rotation, and $W$ increases
toward the North Galactic Pole. These velocities are then corrected to
be relative to the local standard of rest (LSR) by adding the Solar
peculiar velocities of $(U_{\odot},$ $V_{\odot},$ $W_{\odot})$ $=$ $(11.10,$
$12.24,$ $7.25)$\,\kms\ from \citet{schonrich:2010}. This yields
$(U_{LSR},$ $V_{LSR},$ $W_{LSR})$ $=$ $(-35.56,$ $-58.28,$ $-22.96)$\,\kms\ and
$(-73.82,$ $-33.11,$ $-31.21)$\,\kms, for \hatcur{762} and \hatcur{4643},
respectively. We use these velocities to compute the relative
probabilities of each star being a member of the Galactic thin and
thick disks, or of the Galactic halo, following
\citet{bensby:2014}. We find that both objects have a 70\% probability
of being members of the thin disk, a 30\% probability of being members
of the thick disk, and negligible probabilities of being members of
the Galactic halo. While each object is more likely than not to be in
the thin disk, both stars exhibit fairly high space motion compared to
typical thin disk members (both stars have at least one component of
their space velocity that is different from the mean value for thin
disk stars by more than $2\sigma$), which is consistent with both
objects being older main sequence stars.

\subsection{Joint Stellar and Planet Modeling}
\label{sec:transitmodel}

We carried out a joint analysis of the available observations to
determine the stellar and planetary parameters of each system
following the method of
\citet{hartman:2019:hats6069} and \citet{bakos:2020:hats71}. For each object we
performed a simultaneous fit to all light curves, RV measurements, the
observed Spectral-Energy Distribution (SED) as traced by the available
catalog broad-band photometry, the spectroscopic \teffstar\ and \feh,
and the astrometric parallax from {\em Gaia}~DR3 \citep{gaiadr3}. The
catalog photometry, spectroscopic parameters, and parallax values that
we included in the fit for each system are listed in
Table~\ref{tab:stellarobserved}.  The light curves were modeled using
the semi-analytic \citet{mandel:2002} model with quadratic limb
darkening coefficients that were allowed to vary in the fit, but with
priors based on the stellar atmospheric parameters in the theoretical
tabulations by \citet{claret:2012,claret:2013,claret:2018}. For the {\em TESS} observations of \hatcur{4643}, which have exposure times of 30\,min or 10\,min, we integrated the model over the exposure time. The RVs
were fit assuming the star follows a Keplerian orbit around the system
barycenter. The stellar parameters and SED are constrained to follow
the MIST stellar evolution model
while allowing for systematic errors in the stellar physical
parameters following the procedure of \citet{hartman:2023:toi4201toi5344}. We adopt systematic errors of $0.08$\,dex, 2.4\%, 5\% and 0.021\,mag on the metallicity, effective temperature, stellar mass, and bolometric magnitudes in the stellar evolution models, respectively \citep{tayar:2022}. We impose
a prior on the line of sight extinction $A_{V}$ using the MWDUST 3D
Galactic extinction model \citep{bovy:2016} and assume an $R_{V} =
3.1$ extinction law.  The fit is performed twice for each system,
first assuming a circular orbit for the planet, and second allowing
for a non-zero eccentricity. The fit is carried out through a
differential evolution Markov Chain Monte Carlo procedure (see \citealp{hartman:2019:hats6069} for a full list of parameters and priors) using visual inspection to confirm that the chains are converged and well-mixed, and to set the burn-in period.

To account for dilution in the light curves of \hatcur{762} due
  to the neighboring star TOI~762~B, we include for each light curve a
  parameter that specifies the fraction of the flux in the light curve
  that comes from the transit hosting star when out of transit (with
  the remaining fraction assumed to come from a constant source). This
  parameter is allowed to vary in the fit independently for each light
  curve.  We place a prior and uncertainty on each of these dilution
  parameters by calculating the expected flux from TOI~762~B that
  would contaminate the aperture, assuming that TOI~762~B has the same
  distance, reddening, age, and metallicity as inferred for TOI~762~A,
  and using the $G$ magnitude together with the MIST isochrones to
  infer the mass of TOI~762~B and its expected magnitude in various
  passbands. We note that the resulting dilution values are all
  consistent with the priors, to within the uncertainties. No
  systematic trend with wavelength is observed in the residuals.

The observations are consistent with circular orbits for both systems. For \hatcurb{762} we place a 95\% confidence upper limit on the eccentricity of $e\hatcurRVeccentwosiglimeccen{762}$, while for \hatcurb{4643} the upper limit is $e\hatcurRVeccentwosiglimeccen{4643}$. We therefore adopt the parameters derived for each system assuming a circular orbit. Note that in the joint fitting that we perform, the constraints on the eccentricities of these systems come not just from the RVs, but also from the combination of the transit observations and the theoretical stellar evolution models that constrain the allowed combinations of stellar mass, radius, and metallicity. For M dwarf host stars the constraints on mass and radius from comparing the observations to stellar evolution models are much tighter than for higher mass stars, which can lead to a much tighter constraint on the eccentricity than might be allowed by the RVs alone \citep[e.g.,][]{hartman:2015:hats6}. This appears to be the case for \hatcurb{762}, for which the RVs alone provide a much less stringent constraint on the eccentricity than is achieved through our joint analysis of the data. For \hatcurb{4643} on the other hand, the eccentricity constraint appears to come primarily from the RVs. Here the large semi-amplitude of the orbital variation caused by the massive planet and the well-sampled phase curve enable a strong constraint on the orbital eccentricity.

We find that \hatcurb{4643} exhibits grazing transits which in some cases can lead to a strong degeneracy between the impact parameter of the planet and the planet-to-star radius ratio. In such cases it may only be possible to provide a lower bound on the planetary radius \citep[e.g., HATS-23\,b;][]{bento:2017}. We confirmed that the Markov Chains for \hatcur{4643} are well-converged and display a clear upper-limit on both the impact parameter and the planet-to-star radius ratio, so we provide best-estimates for these parameters together with two-sided uncertainties.

Figures~\ref{fig:toi762}--\ref{fig:tic4643lc2} compare the best-fit
models to the observational data. The adopted stellar parameters are
listed in Table~\ref{tab:stellarderived}, while the adopted planetary
parameters are listed in Table~\ref{tab:planetparam} and the limb darkening coefficients are listed in Table~\ref{tab:ldparams}. In both cases we list the parameters determined assuming circular orbits.

\subsection{Ruling Out Blended Stellar Eclipsing Binary Scenarios}
\label{sec:blendmodel}

A line-of-sight blend of three stars, including two that are
eclipsing, is a relatively common false positive that can mimic the
photometric transit and radial velocity signals produced by a
transiting giant planet system \citep[e.g.,][]{torres:2004}. In order to rule out such
an explanation for the observations of \hatcur{762} or \hatcur{4643}
we carried out a blend analysis following the methods of
\citet{hartman:2019:hats6069}. Here we model the photometric and astrometric data as a
blend of three or more stars using the MIST version~1.2 stellar evolution
models to constrain the physical properties of the stars. We consider
both a hierarchical triple star system, where the two fainter stars in
the system are eclipsing, and a line-of-sight blend between an
eclipsing binary star system and a brighter, physically unrelated,
foreground star.

For \hatcur{4643} the analysis was performed before the photometric follow-up observations were available, and only the {\em TESS} light curves are included in this case. Because the parameters that were measured for the system when using only the {\em TESS} observations are fully consistent with the parameters found when incorporating the follow-up light curves from ExTrA, and as we discuss below we can already rule out blend scenarios with only the {\em TESS} data, we do not expect the conclusions from the blend analysis to change when incorporating the follow-up light curves. We therefore did not repeat the blend analysis for this system after obtaining the light curves from ExTrA.

For the hierarchical triple system the parameters that we vary in the
fit include the times of two reference transit events, the age of the
stellar system, the masses of the three stellar companions, the
distance to the system, the metallicity of the system, the impact
parameter of the eclipsing pair, and the limb darkening coefficients
of the primary star in the eclipsing pair. For the line-of-sight blend
we include all of these parameters, as well as the age, metallicity
and distance of the blended foreground star.  We compare both of these
scenarios to a model consisting of a transiting planet around a single
star for which we vary the system age, metallicity and distance, the mass
and limb darkening coefficients of the host star, the impact parameter
of the system, and the planet-to-star radius ratio.  

For TOI~762 we also account for the 3\farcs2 resolved stellar
companents TOI~762\,A and TOI~762\,B in modelling the observations, and attempt to model the
brighter target TOI~762\,A, for which RV variations consistent with a planetary
companion were detected, as a blended system of three stars. Thus for
TOI~762 there are four stars considered in the blend scenarios
that we investigate, and we include the mass of the 3\farcs2 neighbor TOI~762\,B
as a parameter in the fit, and assume that this neighbor has the same
age, metallicity and distance as the brightest star in the blended
object TOI~762\,A.

We find that for both TOI~762 and \hatcur{4643} the transiting
planet scenario provides a significantly better fit to the photometric
and astrometric data, with lower $\chi^2$, than the blended stellar
eclipsing binary systems that we considered. For TOI~762 we find
that the best-fit transiting planet model has $\Delta \chi^2 = -305$
compared to the best-fit blended eclipsing binary model, while for
\hatcur{4643} we find $\Delta \chi^2 = -287$. As the transiting planet
scenarios also require fewer free parameters, we can confidently rule
out a blended stellar eclipsing binary system for both objects. The best-fit blend scenarios exhibit secondary eclipses that are ruled out by the {\em TESS} observations, as well as distances and implied apparent magnitudes for the blended objects that are inconsistent with the {\em Gaia} parallax and measured apparent magnitudes. We
therefore consider both objects to be confirmed transiting planet systems.



\section{Discussion}
\label{sec:discussion}

%
%
    \begin{figure*}[!ht]
 {
 \centering
 \leavevmode
 \includegraphics[width={0.5\linewidth}]{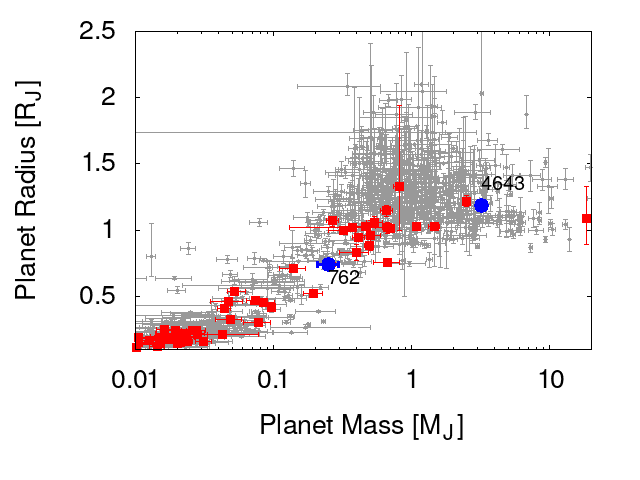}%
 \hfil
 \includegraphics[width={0.5\linewidth}]{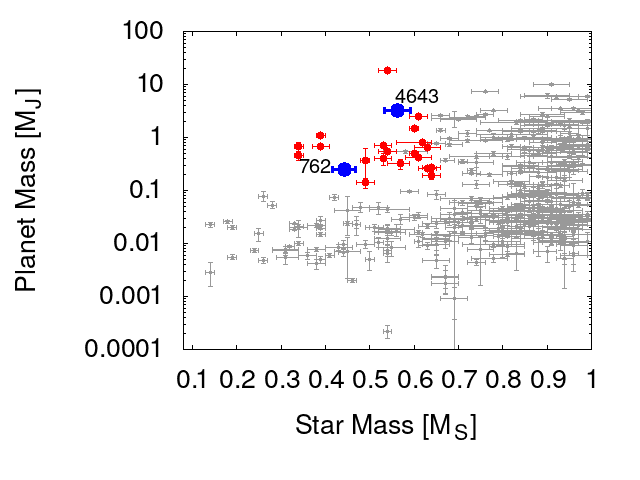}
 }
\caption{
    {\em Left:} Planet radius vs.\ mass. The two new planet discoveries are indicated (we abbreviate \hatcurb{4643} as 4643). Small grey points show all transiting planets with measured masses and radii around K and earlier type stars from the NASA Exoplanet Archive, while red points show transiting planets with masses and radii around M dwarfs. {\em Right:} Planet mass vs.\ host star mass for transiting planets around sub-solar-mass stars. The red points in this case indicate giant planets transiting M dwarfs, while the grey points indicate other planets.
\label{fig:plmassradplmassstmass}
}
    \end{figure*}

    \begin{figure*}[!ht]
 {
 \centering
 \leavevmode
 \includegraphics[width={0.5\linewidth}]{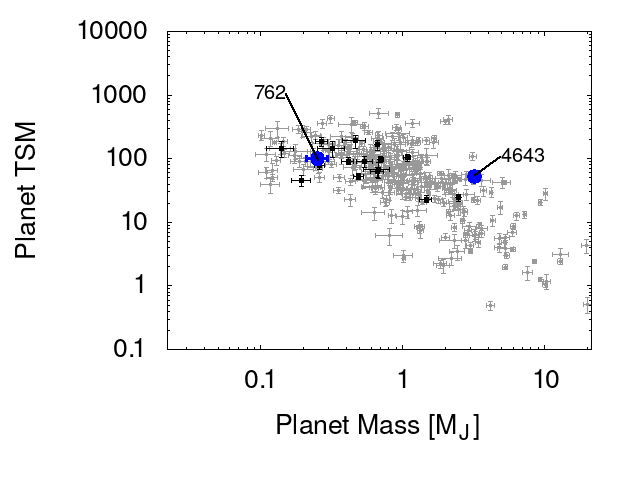}%
 \hfil
 \includegraphics[width={0.5\linewidth}]{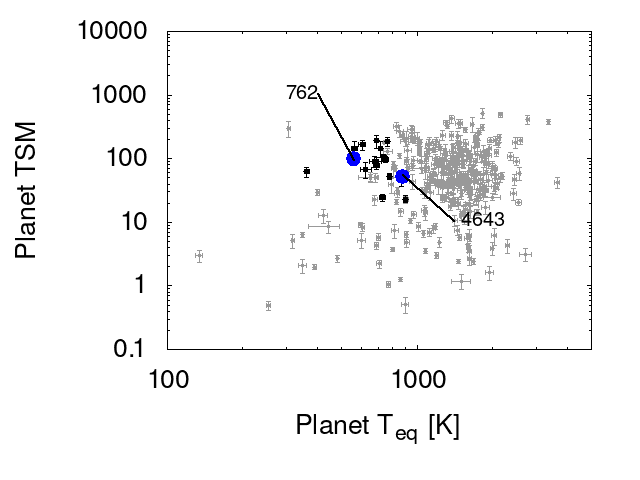}
 }
 {
 \centering
 \leavevmode
 \includegraphics[width={0.5\linewidth}]{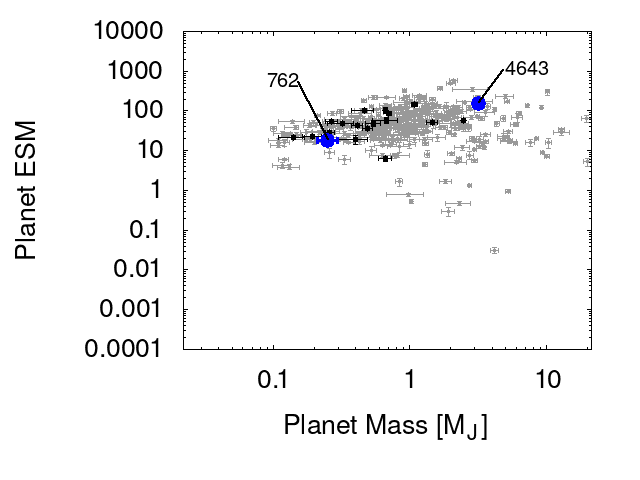}%
 \hfil
 \includegraphics[width={0.5\linewidth}]{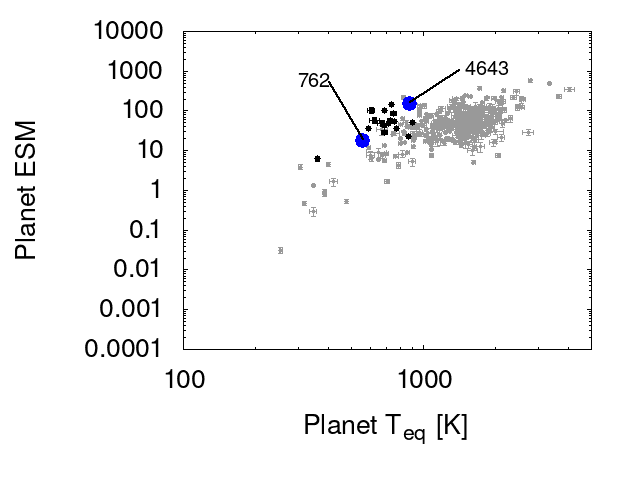}
 }
\caption{
    {\em Top Left:} Transmission Spectroscopy Metric (TSM) vs.\ planet mass. \hatcurb{762} and \hatcurb{4643} are
    indicated. Small gray points show all transiting planets with measured masses $> 0.1$\,\mjup\ and radii $> 0.5$\,\rjup\ from the NASA Exoplanet
    Archive, while small black points denote planets with M dwarf host stars. For all panels we restrict the plots to planets for which the plotted quantities have uncertainties of less than 30\%. {\em Top Right:} TSM vs.\ planet T$_{\rm eq}$ computed assuming zero albedo and full redistribution of heat. The symbols are the same as in the left plot. {\em Bottom Left:} Emission Spectroscopy Metric (ESM) vs.\ planet mass. {\em Bottom Right:} ESM vs.\ planet T$_{\rm eq}$. \hatcurb{4643} has the highest value of TSM or ESM for a transiting warm Super-Jupiter with $\mpl > 3.0$\,\mjup\ and $T_{\rm eq} < 1000$\,K.
\label{fig:tsm}
}
    \end{figure*}

In this paper we presented the discovery of two transiting giant planets that orbit M dwarf stars. Fig.~\ref{fig:plmassradplmassstmass} compares these two new systems to other confirmed transiting planet systems listed in the \citet{ps}\footnote{Accessed on 2024-01-17 at 12:40}. We find that with $\mstar = \hatcurISOm{762}$\,\msun, \hatcur{762} is one of the lowest mass stars known to host a transiting giant planet (taking $\mpl > 0.1$\,\mjup\ as our definition of such a planet; it is also one of the lowest mass stars known to host a giant planet in general), while with $\mpl = \hatcurPPm{4643}$\,\mjup, \hatcurb{4643} is one of the highest mass planets known to transit an M dwarf star. Stars of lower mass than \hatcur{762} with giant planets include TOI-4860 \citep[$0.336$\,\msun;][]{triaud:2023}, TOI-519 \citep[$0.335$\,\msun;][]{kagetani:2023}, TOI-5205 \citep[$0.392$\,\msun;][]{kanodia:2023}, and TOI-3235 \citep[$0.370$\,\msun;][]{hobson:2023}. The only transiting object orbiting an M dwarf star listed on the NASA Exoplanet Archive with a higher mass than \hatcurb{4643} is TOI-1278\,b \citep{artigau:2021}, but with a mass of $18.5\pm0.5$\,\mjup, this object may be a brown dwarf rather than a planet. The next highest mass planet is TOI-4201\,b \citep[$2.48$\,\mjup;][]{gan:2023,hartman:2023:toi4201toi5344,delamer:2023}.

\paragraph{Stellar Companion to \hatcur{762}} We find that \hatcur{762} has a resolved stellar companion TOI~762\,B. The angular separation of 3\farcs2 between the two stars corresponds to a current projected physical separation of 319\,AU. We estimate that TOI~762\,B has a mass of $0.227 \pm 0.010$\,\msun, which is 45\% the mass of \hatcur{762}. The eccentricity and orbital period of TOI~762\,B are not known. 

If the eccentricity is sufficiently high, then TOI~762\,B might have induced high-eccentricity migration for the planet \hatcurb{762} via the Kozai-Lidov mechanism \citep{kozai:1962,lidov:1962,naoz:2016}. To get a rough sense for the time-scale of this mechanism we use eq.~7 of \citet{kiseleva:1998} assuming that the current projected physical separation corresponds to the semi-major axis of the orbit of the binary star system. We find a time-scale of $\sim 100$\,Myr if TOI~762\,B has a very high eccentricity of 0.95, or $\sim 3.5$\,Gyr if the eccentricity is close to zero. 

Four other M dwarfs that host transiting giant planets also have known wide stellar binary companions (TOI~3984\,A, \citealp{canas:2023:toi3984toi5293}; TOI~5293\,A, \citealp{canas:2023:toi3984toi5293}; TOI~3714, \citealp{canas:2022}; and HATS-74\,A, \citealp{jordan:2022:hats74hats77}). Thus, approximately 20\% of the sample of M dwarfs with transiting giant planets are known to have resolved stellar binary companions. The overall stellar multiplicity fraction for M dwarf stars is estimated to be $46\pm5$\% \citep{susemiehl:2022}, however, a meaningful comparison to the rate for giant planet host stars will require a careful correction for observational completeness. Such a study will likely require a much larger sample of giant planet-hosting M dwarfs to enable a statistically significant result. 

\citet{ngo:2016} has compared the stellar multiplicity rate for FGK stars that host hot Jupiters to the rate for field FGK stars. They find that the fraction of hot Jupiter systems with stellar companions between 50 and 2000\,AU is approximately 2.9 times the field star companion fraction. But, they also find that in the majority of cases the stellar binary companions could not drive high-eccentricity migration through the Kozai-Lidov effect. They conclude that for FGK systems, binarity is likely correlated with the formation of hot Jupiters, but the binary companions themselves do not physically drive their migration.

\paragraph{High mass of \hatcurb{4643}} The high mass of \hatcurb{4643} given its low stellar host mass of \hatcurISOm{4643}\,\msun\ poses a challenge for theories of planet formation and evolution. \citet{gan:2023} carried out interior structure modelling of the planet TOI-4201\,b, which has a similar radius to \hatcurb{4643}, similar host star metallicity and similar planet equilibrium temperature, but somewhat lower planet mass and somewhat larger host star mass. They find that TOI-4201\,b requires a low planet bulk metal content which appears to be at odds with the high metallicity inferred for the host star. Given the higher planet mass of \hatcurb{4643}, the comparably high metallicity of its host star, and the comparable planet radius and equilibrium temperature, we anticipate that a similar analysis would result in the same conclusion for \hatcurb{4643} as \citet{gan:2023} arrived at for TOI-4201\,b---i.e., that the planet mass/radius/equilibrium temperature requires a low planet bulk metal content that is at odds with the high host star metallicity. \citet{gan:2023} considered a variety of potential heating sources that might explain the discrepancy, including tidal heating \citep{leconte:2010}, a gas giant merger \citep{li:2010,liu:2015}, and a more quiescent process involving an embryo being captured by a gas giant during inward migration \citep{lin:1996}. The fact that two super-Jupiter-mass planets have now been found around M dwarf stars with somewhat high radii, suggesting low bulk metal content, indicates that if there is an additional heating mechanism, it may be common for short-period giant planets around M dwarf stars. With only two planets, however, we cannot draw any definite conclusions at this point.

\paragraph{Grazing Transits of \hatcurb{4643}} We find that \hatcurb{4643} exhibits grazing transit events. Two other
giant planets around M dwarfs have been found with grazing transits:
NGTS~1\,b \citep{bayliss:2018}, and the super-massive planet or brown
dwarf TOI~1278\,B \citep{artigau:2021}. The relatively large values of
$R_{p}/R_{\star}$ for giant planets orbiting M dwarfs enhances the
probability of finding these systems in grazing configurations
compared to giant planets transiting Sun-like stars. The deep V-shaped
transit events exhibited by these systems are often considered the
hallmark of stellar eclipsing binaries, so it is important to keep in
mind that transiting planets can have light curves of this form as
well when searching for giant planets transiting M dwarf stars.

Because grazing transit events do not have second or third contact
points (the times when a planet starts and stops being fully in front
of its host star, respectively), there is less information content in
a grazing transit than in a full transit event. This can lead to
degeneracies between the impact parameter and the planet-to-star
radius ratio, resulting in larger uncertainties on the planetary
radius and other parameters for these systems. High-precision
observations, such as from {\em TESS}, can help break this degeneracy and enable an accurate
measurement of the planetary radius in spite of the grazing
transits. We find that this is the case for \hatcurb{4643}, for which we find that the radius is measured to 2.5\% uncertainty despite the grazing events. 

One minor advantage of a grazing system is that secular
variations in the orbits of these planets, due for example to the
presence of exterior planets in the system, may be detectable with
higher signal-to-noise than for full transits \citep{ribas:2008}.  An example of this is K2-146\,c, a planet whose orbit has been seen to vary between grazing and fully transiting 
 configurations \citep{hamann:2019}. While no timing variations have been detected yet for \hatcurb{4643}, the grazing transits make this an interesting target for continued transit timing observations going forward.

\paragraph{Prospects for Atmospheric Characterization} Transiting giant planets orbiting M dwarf stars can be useful objects for atmospheric characterization due to their relatively deep transits. We computed the Transmission Spectroscopy Metric and Emission Spectroscopy Metric \citep[TSM and ESM, respectively;][]{kempton:2018} for the two new planetary systems presented here, finding values of ${\rm TSM} = 101 \pm 21$, and ${\rm ESM} = 18.3 \pm 1.5$ for \hatcurb{762} and ${\rm TSM} = 52.3 \pm 4.8$, and ${\rm ESM} = 152 \pm 11$ for \hatcurb{4643}. These are compared to other transiting giant planets from the NASA Exoplanet Archive in Figure~\ref{fig:tsm}. While neither of the newly discovered planets has an especially high TSM value compared to the overall sample of transiting giant planets, their values are both greater than the median value of ${\rm TSM} = 47.4$ for the full sample of confirmed warm transiting giant planets with equilibrium temperatures $T_{\rm eq} < 1000$\,K. When comparing against planets of $\mpl > 3.0$\,\mjup, the TSM of \hatcurb{4643} does stand out as being greater than the value for all but four of these planets (i.e., it has a greater TSM value than 92\% of all known transiting planets with $\mpl > 3.0$\,\mjup). These four planets (HAT-P-70\,b, ${\rm TSM} = 57.8$, \citealp{zhou:2019}; MASCARA-4\,b, ${\rm TSM} = 107$, \citealp{dorval:2020}; TOI-1431\,b, ${\rm TSM} = 109$, \citealp{addison:2021}; and HIP~65\,A\,b, ${\rm TSM} = 780$, \citealp{nielsen:2020}) all have equilibrium temperatures greater than $1400$\,K, which is significantly higher than that of \hatcurb{4643}. The next highest TSM value for a planet with $\mpl > 3.0$\,\mjup\ and $T_{\rm eq} < 1000$\,K is 13.1 for the planet HAT-P-20\,b \citep{bakos:2011:hat20hat23}, which is a factor of four smaller than the value for \hatcurb{4643}. \hatcurb{4643} also has a high value of ESM for a giant planet with $T_{\rm eq} < 1000$\,K. The only giant planet that has $T_{\rm eq} < 1000$\,K and a higher ESM value than \hatcurb{4643} is WASP-80\,b \citep{triaud:2013}, with ${\rm ESM} = 216$. WASP-80\,b also has a significantly lower mass of $0.538 \pm 0.035$\,\mjup. Thus \hatcurb{4643} presents a particularly good opportunity for studying the atmosphere of a warm, high-mass planet. The frequent transits of this $P = \hatcurLCPshort{4643}$\,d system should also facilitate scheduling such observations.



\acknowledgements 

We thank the anonymous referee for their careful reading this paper, and helpful comments that have improved the quality of the work.
JH and GB acknowledge funding from NASA grant 80NSSC22K0315. 
AJ and RB acknowledge support from ANID -- Millennium  Science
Initiative -- ICN12\_009. R.B. acknowledges support from FONDECYT
Project 11200751. AJ acknowledges support from FONDECYT project 1210718.
We acknowledge funding from the European Research Council under the ERC Grant Agreement n. 337591-ExTrA.
This research has made use of the Exoplanet Follow-up Observation Program (ExoFOP; DOI: 10.26134/ExoFOP5) website, which is operated by the California Institute of Technology, under contract with the National Aeronautics and Space Administration under the Exoplanet Exploration Program.
This paper made use of data collected by the TESS mission and are publicly available from the Mikulski Archive for Space Telescopes (MAST) operated by the Space Telescope Science Institute (STScI). Funding for the TESS mission is provided by NASA’s Science Mission Directorate. The specific observations from MAST analyzed in this paper can be accessed from \dataset[DOI]{https://doi.org/10.17909/e3cp-9907}.
We acknowledge the use of public TESS data from pipelines at the TESS Science Office and at the TESS Science Processing Operations Center.
Resources supporting this work were provided by the NASA High-End Computing (HEC) Program through the NASA Advanced Supercomputing (NAS) Division at Ames Research Center for the production of the SPOC data products.
This research has made use of the NASA Exoplanet Archive, which is operated by the California Institute of Technology, under contract with the National Aeronautics and Space Administration under the Exoplanet Exploration Program.
The contributions at the Mullard Space Science Laboratory by EMB have been supported by STFC through the consolidated grant ST/W001136/1.
The postdoctoral fellowship of KB is funded by F.R.S.-FNRS grant T.0109.20 and by the Francqui Foundation.
This publication benefits from the support of the French Community of Belgium in the context of the FRIA Doctoral Grant awarded to MT.
MG is F.R.S.-FNRS Research Director and EJ is F.R.S.-FNRS Senior Research Associate. 
F.J.P acknowledges financial support from the grant CEX2021-001131-S funded by MCIN/AEI/ 10.13039/501100011033 and through projects PID2019-109522GB-C52 and PID2022-137241NB-C43. 
Based on data collected by the SPECULOOS-South Observatory at the ESO Paranal Observatory in Chile.The ULiege's contribution to SPECULOOS has received funding from the European Research Council under the European Union's Seventh Framework Programme (FP/2007-2013) (grant Agreement n$^\circ$ 336480/SPECULOOS), from the Balzan Prize and Francqui Foundations, from the Belgian Scientific Research Foundation (F.R.S.-FNRS; grant n$^\circ$ T.0109.20), from the University of Liege, and from the ARC grant for Concerted Research Actions financed by the Wallonia-Brussels Federation. This work is supported by a grant from the Simons Foundation (PI Queloz, grant number 327127).
Based on data collected by the TRAPPIST-South telescope at the ESO La Silla Observatory. TRAPPIST is funded by the Belgian Fund for Scientific Research (Fond National de la Recherche Scientifique, FNRS) under the grant PDR T.0120.21, with the participation of the Swiss National Science Fundation (SNF). 
DR was supported by NASA under award number NNA16BD14C for NASA Academic Mission Services.
%

\facilities{TESS, LCOGT, TRAPPIST, Gaia, ExTrA, SPECULOOS, VLT:ESPRESSO, Exoplanet Archive, IRSA, ZTF, ASAS-SN}

\software{VARTOOLS \citep{hartman:2016:vartools}, MWDUST \citep{bovy:2016}, Astropy \citep{astropy:2013,astropy:2018}, ESPRESSO DRS pipeline v2.3.5 \citep{sosnowska:2015,modigliani:2020}, EsoReflex \citep{freudling:2013}, AstroImageJ \citep{collins:2017}}


\bibliographystyle{aasjournal}
\bibliography{hatsbib}

\clearpage

%
%
    \begin{deluxetable*}{lccl}
\tablewidth{0pc}
\tabletypesize{\tiny}
\tablecaption{
    Astrometric, Spectroscopic and Photometric parameters for \hatcur{762} and \hatcur{4643}
    \label{tab:stellarobserved}
}
\tablehead{
    \multicolumn{1}{c}{} &
    \multicolumn{1}{c}{\bf TOI~762\,A} &
    \multicolumn{1}{c}{\bf TIC~46432937} &
    \multicolumn{1}{c}{} \\
    \multicolumn{1}{c}{~~~~~~~~Parameter~~~~~~~~} &
    \multicolumn{1}{c}{Value}                     &
    \multicolumn{1}{c}{Value}                     &
    \multicolumn{1}{c}{Source}
}
\startdata
\noalign{\vskip -3pt}
\sidehead{Astrometric properties and cross-identifications}
~~~~2MASS-ID\dotfill               & \hatcurCCtwomassshort{762} & \hatcurCCtwomassshort{4643} & \\
~~~~TIC-ID\dotfill                 & \hatcurCCtic{762} & \hatcurCCtic{4643} & \\
~~~~TOI-ID\dotfill                 & \hatcurCCtoi{762} & \hatcurCCtoi{4643} & \\
~~~~GAIA~DR2-ID\dotfill                 & \hatcurCCgaiadrtwoshort{762} & \hatcurCCgaiadrtwoshort{4643} & \\
~~~~R.A. (J2000)\dotfill            & \hatcurCCra{762}    & \hatcurCCra{4643}    & GAIA DR3\\
~~~~Dec. (J2000)\dotfill            & \hatcurCCdec{762}   & \hatcurCCdec{4643}   & GAIA DR3\\
~~~~$\mu_{\rm R.A.}$ (\masy)              & \hatcurCCpmra{762} & \hatcurCCpmra{4643} & GAIA DR3\\
~~~~$\mu_{\rm Dec.}$ (\masy)              & \hatcurCCpmdec{762} & \hatcurCCpmdec{4643} & GAIA DR3\\
~~~~parallax (mas)              & \hatcurCCparallax{762} & \hatcurCCparallax{4643} & GAIA DR3\\
\sidehead{Spectroscopic properties}
~~~~$\teffstar$ (K)\dotfill         & \hatcurSMEiteff{762} & \hatcurSMEteff{4643} & see Section~\ref{sec:atmosphericparams}\\
~~~~$\feh$\dotfill                  & \hatcurSMEizfeh{762} & \hatcurSMEizfeh{4643} & see Section~\ref{sec:atmosphericparams}\\
~~~~$\gamma_{\rm RV}$ (\ms)\dotfill& $\hatcurRVgammaabs{762}$ & $\hatcurRVgammaabs{4643}$ & ESPRESSO  \\
\sidehead{Photometric properties\tablenotemark{a}}
~~~~$G$ (mag)\tablenotemark{b}\dotfill               & \hatcurCCgaiamG{762} & \hatcurCCgaiamG{4643} & GAIA DR3 \\
~~~~$BP$ (mag)\tablenotemark{b}\dotfill               & $\cdots$ & \hatcurCCgaiamBP{4643} & GAIA DR3 \\
~~~~$RP$ (mag)\tablenotemark{b}\dotfill               & $\cdots$ & \hatcurCCgaiamRP{4643} & GAIA DR3 \\
~~~~$B$ (mag)\dotfill               & $\cdots$ & \hatcurCCtassmB{4643} & APASS\tablenotemark{c} \\
~~~~$V$ (mag)\dotfill               & $\cdots$ & \hatcurCCtassmv{4643} & APASS\tablenotemark{c} \\
~~~~$g$ (mag)\dotfill               & $\cdots$ & \hatcurCCtassmg{4643} & APASS\tablenotemark{c} \\
~~~~$r$ (mag)\dotfill               & $\cdots$ & \hatcurCCtassmr{4643} & APASS\tablenotemark{c} \\
~~~~$i$ (mag)\dotfill               & $\cdots$ & \hatcurCCtassmi{4643} & APASS\tablenotemark{c} \\
~~~~$J$ (mag)\tablenotemark{d}\dotfill               & \hatcurCCtwomassJmag{762} & \hatcurCCtwomassJmag{4643} & 2MASS           \\
~~~~$H$ (mag)\tablenotemark{d}\dotfill               & \hatcurCCtwomassHmag{762} & \hatcurCCtwomassHmag{4643} & 2MASS           \\
~~~~$K_s$ (mag)\tablenotemark{d}\dotfill             & \hatcurCCtwomassKmag{762} & \hatcurCCtwomassKmag{4643} & 2MASS           \\
~~~~$W1$ (mag)\tablenotemark{e}\dotfill             & \hatcurCCWonemag{762} & \hatcurCCWonemag{4643} & WISE           \\
~~~~$W2$ (mag)\tablenotemark{e}\dotfill             & \hatcurCCWtwomag{762} & \hatcurCCWtwomag{4643} & WISE           \\
~~~~$W3$ (mag)\tablenotemark{e}\dotfill             & \hatcurCCWthreemag{762} & \hatcurCCWthreemag{4643} & WISE           \\
\enddata
\tablenotetext{a}{
    We only include in the table catalog magnitudes that were included in our analysis of each system.
}
\tablenotetext{b}{
    The listed uncertainties for the Gaia DR3 photometry are taken from the catalog. For the analysis we assume an additional systematic uncertainty of 0.02\,mag for all bandpasses.
}
\tablenotetext{c}{
    From APASS DR6 as
    listed in the UCAC 4 catalog \citep{zacharias:2013:ucac4}. Although these measurements are also available for \hatcur{762}, we do not include them in the analysis or list them here because the degree to which these measurements are contaminated by flux from the 3\farcs2 and 4\farcs7 neighbors is difficult to determine.
}
\tablenotetext{d}{
    From the 2MASS catalog \citep{skrutskie:2006}. For \hatcur{762} we subtracted an estimate for the flux contribution from TOI~762\,B.
}
\tablenotetext{e}{
    From the 2021 Feb 16 ALLWISE Data release of the WISE mission \citep{cutri:2021}. For \hatcur{762} we subtracted an estimate for the flux contribution from TOI~762\,B.
}
    \end{deluxetable*}

\startlongtable
\tabletypesize{\scriptsize}
    \begin{deluxetable*}{lrrrrl}
\tablewidth{0pc}
\tablecaption{
    ESPRESSO radial velocities for \hatcur{762} and \hatcur{4643}.
    \label{tab:rvs}
}
\tablehead{
    \colhead{System} &
    \colhead{BJD} &
    \colhead{RV\tablenotemark{a}} &
    \colhead{\ensuremath{\sigma_{\rm RV}}\tablenotemark{b}} &
    \colhead{Phase} &
    \colhead{Instrument\tablenotemark{c}}\\
    \colhead{} &
    \colhead{\hbox{(2,450,000$+$)}} &
    \colhead{(\ms)} &
    \colhead{(\ms)} &
    \colhead{(\ms)} &
    \colhead{}
}
\startdata
TOI-762 & $ 8818.82472 $ & $    47.93 $ & $    11.30 $ & $   0.901 $ & ESPRESSO/P104 \\
TOI-762 & $ 8822.81705 $ & $   -17.07 $ & $    11.50 $ & $   0.051 $ & ESPRESSO/P104 \\
TOI-762 & $ 8836.78307 $ & $   -29.07 $ & $     8.70 $ & $   0.074 $ & ESPRESSO/P104 \\
TOI-762 & $ 8839.83846 $ & $    19.93 $ & $     5.80 $ & $   0.954 $ & ESPRESSO/P104 \\
TOI-762 & $ 8841.81658 $ & $     4.93 $ & $     6.40 $ & $   0.524 $ & ESPRESSO/P104 \\
TOI-762 & $ 9938.80622 $ & $     3.43 $ & $     7.14 $ & $   0.506 $ & ESPRESSO/P110 \\
TOI-762 & $ 9942.74132 $ & $    39.13 $ & $    10.60 $ & $   0.639 $ & ESPRESSO/P110 \\
TOI-762 & $ 9943.75858 $ & $    40.13 $ & $     8.42 $ & $   0.932 $ & ESPRESSO/P110 \\
TIC-46432937 & $ 9907.75857 $ & $  -423.38 $ & $     1.95 $ & $   0.086 $ & ESPRESSO \\
TIC-46432937 & $ 9909.74414 $ & $  -187.38 $ & $     2.49 $ & $   0.464 $ & ESPRESSO \\
TIC-46432937 & $ 9928.64212 $ & $   417.62 $ & $     3.81 $ & $   0.584 $ & ESPRESSO \\
TIC-46432937 & $ 9930.76066 $ & $  -277.38 $ & $     1.94 $ & $   0.054 $ & ESPRESSO \\
TIC-46432937 & $ 9936.72836 $ & $  -797.38 $ & $     1.81 $ & $   0.197 $ & ESPRESSO \\
TIC-46432937 & $ 9995.60029 $ & $  -336.38 $ & $     3.53 $ & $   0.068 $ & ESPRESSO \\
TIC-46432937 & $ 9996.56963 $ & $   836.62 $ & $     2.08 $ & $   0.741 $ & ESPRESSO \\
TIC-46432937 & $ 9997.63070 $ & $  -122.38 $ & $     1.97 $ & $   0.478 $ & ESPRESSO \\
\enddata
\tablenotetext{a}{
    The zero-point of these velocities is arbitrary. An overall offset
    $\gamma_{\rm rel}$ fitted to the orbit has been subtracted for each system.
}
\tablenotetext{b}{
    Internal errors excluding the component of astrophysical jitter
    allowed to vary in the fit.
}
\tablenotetext{c}{
    For \hatcur{762} we distinguish between the ESPRESSO observations obtained during P104 and those obtained during P110 for which we allow independent zero-point offsets in the fit.
}
    \end{deluxetable*}

%
%
    \begin{deluxetable*}{lcc}
\tablewidth{0pc}
\tabletypesize{\footnotesize}
\tablecaption{
    Adopted derived stellar parameters for \hatcur{762} and \hatcur{4643}.
    \label{tab:stellarderived}
}
\tablehead{
    \multicolumn{1}{c}{} &
    \multicolumn{1}{c}{\bf TOI~762\,A} &
    \multicolumn{1}{c}{\bf TIC~46432937} \\
    \multicolumn{1}{c}{~~~~~~~~Parameter~~~~~~~~} &
    \multicolumn{1}{c}{Value}                     &
    \multicolumn{1}{c}{Value}                     
}
\startdata
~~~~$\mstar$ ($\msun$)\dotfill      & \hatcurISOmlong{762} & \hatcurISOmlong{4643} \\
~~~~$\rstar$ ($\rsun$)\dotfill      & \hatcurISOrlong{762} & \hatcurISOrlong{4643} \\
~~~~$\loggstar$ (cgs)\dotfill       & \hatcurISOlogg{762} & \hatcurISOlogg{4643} \\
~~~~$\rhostar$ (\gcmc)\dotfill       & \hatcurLCrho{762} & \hatcurLCrho{4643} \\
~~~~$\lstar$ ($\lsun$)\dotfill      & \hatcurISOlum{762} & \hatcurISOlum{4643} \\
~~~~$\teffstar$ (K)\dotfill      &  \hatcurISOteff{762} &  \hatcurISOteff{4643} \\
~~~~$\feh$\dotfill      &  \hatcurISOzfeh{762} &  \hatcurISOzfeh{4643} \\
~~~~Age (Gyr)\dotfill               & \hatcurISOage{762} & \hatcurISOage{4643} \\
~~~~$A_{V}$ (mag)\dotfill               & \hatcurXAv{762} & \hatcurXAv{4643} \\
~~~~Distance (pc)\dotfill           & \hatcurXdistred{762} & \hatcurXdistred{4643} \\
\enddata
\tablecomments{
The listed parameters are those determined through the joint differential evolution Markov Chain analysis, including systematic errors in the stellar evolution models, described in Section~\ref{sec:transitmodel}. For all systems the RV observations are consistent with a circular orbit, and we assume a fixed circular orbit in generating the parameters listed here. 
}
    \end{deluxetable*}

%
    \begin{deluxetable*}{lcc}
\tabletypesize{\tiny}
\tablecaption{Adopted orbital and planetary parameters for \hatcurb{762}, and \hatcurb{4643}\label{tab:planetparam}}
\tablehead{
    \multicolumn{1}{c}{} &
    \multicolumn{1}{c}{\bf TOI~762\,A\,b} &
    \multicolumn{1}{c}{\bf TIC~46432937\,b} \\
    \multicolumn{1}{c}{~~~~~~~~~~~~~~~Parameter~~~~~~~~~~~~~~~} &
    \multicolumn{1}{c}{Value} &
    \multicolumn{1}{c}{Value}
}
\startdata
\noalign{\vskip -3pt}
\sidehead{\Lc{} parameters}
~~~$P$ (days)             \dotfill    & $\hatcurLCP{762}$ & $\hatcurLCP{4643}$ \\
~~~$T_c$ (${\rm BJD\_{}TDB}$)    
      \tablenotemark{a}   \dotfill    & $\hatcurLCT{762}$ & $\hatcurLCT{4643}$ \\
~~~$T_{14}$ (days)
      \tablenotemark{a}   \dotfill    & $\hatcurLCdur{762}$ & $\hatcurLCdur{4643}$ \\
~~~$T_{12} = T_{34}$ (days)
      \tablenotemark{a}   \dotfill    & $\hatcurLCingdur{762}$ & $\cdots$ \\
~~~$\phi_{\rm occ}$ (phase) 
       \tablenotemark{b}  \dotfill    & $\hatcurXsecphaseeccen{762}$ & $\hatcurXsecphaseeccen{4643}$ \\
~~~$T_{c,{\rm occ}}$ (${\rm BJD\_{}TDB}$) 
       \tablenotemark{b}  \dotfill    & $\hatcurXsecondaryeccen{762}$ & $\hatcurXsecondaryeccen{4643}$ \\
~~~$T_{14,{\rm occ}}$ (days) 
       \tablenotemark{b}  \dotfill    & $\hatcurXsecdureccen{762}$ & $\hatcurXsecdureccen{4643}$ \\
~~~$\arstar$              \dotfill    & $\hatcurPPar{762}$ & $\hatcurPPar{4643}$ \\
~~~$\zrstar$ \tablenotemark{c}             \dotfill    & $\hatcurLCzeta{762}$\phn& $\hatcurLCzeta{4643}$\phn\\
~~~$\rpl/\rstar$          \dotfill    & $\hatcurLCrprstar{762}$& $\hatcurLCrprstar{4643}$\\
~~~$b^2$                  \dotfill    & $\hatcurLCbsq{762}$& $\hatcurLCbsq{4643}$\\
~~~$b \equiv a \cos i/\rstar$
                          \dotfill    & $\hatcurLCimp{762}$& $\hatcurLCimp{4643}$\\
~~~$i$ (deg)              \dotfill    & $\hatcurPPi{762}$\phn& $\hatcurPPi{4643}$\phn\\
\sidehead{RV parameters}
~~~$K$ (\ms)              \dotfill    & $\hatcurRVK{762}$\phn\phn& $\hatcurRVK{4643}$\phn\phn\\
~~~$e$ \tablenotemark{d}               \dotfill    & $\hatcurRVeccentwosiglimeccen{762}$ & $\hatcurRVeccentwosiglimeccen{4643}$ \\
~~~RV jitter ESPRESSO 1\tablenotemark{e} (\ms)        \dotfill    & $\hatcurRVjitterA{762}$& $\hatcurRVjitter{4643}$\\
~~~RV jitter ESPRESSO 2\tablenotemark{e} (\ms)        \dotfill    & $\hatcurRVjitterB{762}$& $\cdots$\\
\sidehead{Planetary parameters}
~~~$\mpl$ ($\mjup$)       \dotfill    & $\hatcurPPmlong{762}$& $\hatcurPPmlong{4643}$\\
~~~$\rpl$ ($\rjup$)       \dotfill    & $\hatcurPPrlong{762}$& $\hatcurPPrlong{4643}$\\
~~~$C(\mpl,\rpl)$
    \tablenotemark{f}     \dotfill    & $\hatcurPPmrcorr{762}$& $\hatcurPPmrcorr{4643}$\\
~~~$\rhopl$ (\gcmc)       \dotfill    & $\hatcurPPrho{762}$& $\hatcurPPrho{4643}$\\
~~~$\log g_p$ (cgs)       \dotfill    & $\hatcurPPlogg{762}$& $\hatcurPPlogg{4643}$\\
~~~$a$ (AU)               \dotfill   & $\hatcurPParel{762}$& $\hatcurPParel{4643}$\\
~~~$T_{\rm eq}$ (K)        \dotfill   & $\hatcurPPteff{762}$& $\hatcurPPteff{4643}$\\
~~~$\Theta$ \tablenotemark{g} \dotfill & $\hatcurPPtheta{762}$& $\hatcurPPtheta{4643}$\\
~~~$\log_{10}\langle F \rangle$ (cgs) \tablenotemark{h}
                          \dotfill    & $\hatcurPPfluxavglog{762}$& $\hatcurPPfluxavglog{4643}$\\
\enddata
\tablecomments{
For both systems we adopt a model in which the orbit is assumed to be circular. Except where noted otherwise, the listed parameters are calculated assuming circular orbits. See the discussion in Section~\ref{sec:transitmodel}.
}
\tablenotetext{a}{
    Times are in Barycentric Julian Date calculated on the Barycentric Dynamical Time (TDB) system.
    \ensuremath{T_c}: Reference epoch of
    mid transit that minimizes the correlation with the orbital
    period.
    \ensuremath{T_{14}}: total transit duration, time
    between first to last contact;
    \ensuremath{T_{12}=T_{34}}: ingress/egress time, time between first
    and second, or third and fourth contact.
}
\tablenotetext{b}{
    Inferred timing of occultation events calculated from the fit where the eccentricity is allowed to vary. Occultations have not been observed for either system. Times are in Barycentric Julian Date calculated on the Barycentric Dynamical Time (TDB) system.
    \ensuremath{\phi_{\rm occ}}: orbital phase of the occultation. Phase 0 refers to the time of mid transit.
    \ensuremath{T_{c,{\rm occ}}}: Reference epoch of
    mid occultation.
    \ensuremath{T_{14,{\rm occ}}}: total occultation duration.
}
\tablenotetext{c}{
   Reciprocal of the half duration of the transit used as a jump parameter in our MCMC analysis in place of $\arstar$. It is related to $\arstar$ by the expression $\zrstar = \arstar(2\pi(1+e\sin\omega))/(P\sqrt{1-b^2}\sqrt{1-e^2})$ \citep{bakos:2010:hat11}.
}
\tablenotetext{d}{
    The 95\% confidence upper limit on the eccentricity determined
    when $\sqrt{e}\cos\omega$ and $\sqrt{e}\sin\omega$ are allowed to
    vary in the fit.
}
\tablenotetext{e}{
    Term added in quadrature to the formal RV uncertainties for each
    instrument. This is treated as a free parameter in the fitting
    routine. The two values listed for \hatcurb{762} are for the ESPRESSO P104 and P110 data, respectively.
}
\tablenotetext{f}{
    Correlation coefficient between the planetary mass \mpl\ and radius
    \rpl\ estimated from the posterior parameter distribution.
}
\tablenotetext{g}{
    The Safronov number is given by $\Theta = \frac{1}{2}(V_{\rm
    esc}/V_{\rm orb})^2 = (a/\rpl)(\mpl / \mstar )$
    \citep[see][]{hansen:2007}.
}
\tablenotetext{h}{
    Incoming flux per unit surface area, averaged over the orbit.
}
    \end{deluxetable*}

%
    \begin{deluxetable*}{lcc}
\tabletypesize{\tiny}
\tablecaption{Adopted limb darkening coefficients for \hatcurb{762}, and \hatcurb{4643}\label{tab:ldparams}}
\tablehead{
    \multicolumn{1}{c}{} &
    \multicolumn{1}{c}{\bf TOI~762\,A\,b} &
    \multicolumn{1}{c}{\bf TIC~46432937\,b} \\
    \multicolumn{1}{c}{~~~~~~~~~~~~~~~Parameter~~~~~~~~~~~~~~~} &
    \multicolumn{1}{c}{Value} &
    \multicolumn{1}{c}{Value}
}
\startdata
\noalign{\vskip -3pt}
~~~$c_1,g$                  \dotfill    & $\hatcurLBig{762}$& $\cdots$\\
~~~$c_2,g$                  \dotfill    & $\hatcurLBiig{762}$& $\cdots$\\
~~~$c_1,r$                  \dotfill    & $\hatcurLBir{762}$& $\cdots$\\
~~~$c_2,r$                  \dotfill    & $\hatcurLBiir{762}$& $\cdots$\\
~~~$c_1,i$                  \dotfill    & $\hatcurLBii{762}$ & $\cdots$\\
~~~$c_2,i$                  \dotfill    & $\hatcurLBiii{762}$ & $\cdots$\\
~~~$c_1,zs$                  \dotfill   & $\hatcurLBiz{762}$& $\cdots$\\
~~~$c_2,zs$                  \dotfill   & $\hatcurLBiiz{762}$& $\cdots$\\
~~~$c_1,I+z$                  \dotfill    & $\hatcurLBiI{762}$ & $\cdots$\\
~~~$c_2,I+z$                  \dotfill    & $\hatcurLBiiI{762}$ & $\cdots$\\
~~~$c_1,Z$                  \dotfill    & $\cdots$ & $\hatcurLBiz{4643}$\\
~~~$c_2,Z$                  \dotfill    & $\cdots$ & $\hatcurLBiiz{4643}$\\
~~~$c_1,Y$                  \dotfill    & $\cdots$ & $\hatcurLBiI{4643}$\\
~~~$c_2,Y$                  \dotfill    & $\cdots$ & $\hatcurLBiiI{4643}$\\
~~~$c_1,J$                  \dotfill    & $\cdots$ & $\hatcurLBiJ{4643}$\\
~~~$c_2,J$                  \dotfill    & $\cdots$ & $\hatcurLBiiJ{4643}$\\
~~~$c_1,H$                  \dotfill    & $\cdots$ & $\hatcurLBiH{4643}$\\
~~~$c_2,H$                  \dotfill    & $\cdots$ & $\hatcurLBiiH{4643}$\\
~~~$c_1,z$--$H$                  
\dotfill    & $\hatcurLBiJ{762}$ & $\cdots$\\
~~~$c_2,z$--$H$                  \dotfill    & $\hatcurLBiiJ{762}$ & $\cdots$\\
~~~$c_1,T$                  \dotfill    & $\hatcurLBiT{762}$& $\hatcurLBiT{4643}$\\
~~~$c_2,T$                  \dotfill    & $\hatcurLBiiT{762}$& $\hatcurLBiiT{4643}$\\
\enddata
\tablecomments{
For all systems we adopt a model in which the orbit is assumed to be circular. See the discussion in Section~\ref{sec:transitmodel}. The values listed are for a quadratic law. The limb darkening parameters were
    directly varied in the fit, using the tabulations from
    \cite{claret:2012,claret:2013,claret:2018} to place Gaussian prior
    constraints on their values, assuming a prior uncertainty of $0.2$
    for each coefficient. 
}
    \end{deluxetable*}

\end{document}